%% file: main.tex
\documentclass[12pt,letterpaper]{article}
\usepackage{natbib}
\usepackage[margin=0.9in]{geometry}
\usepackage{setspace}
\usepackage{titling}
\usepackage{authblk}

\usepackage[colorlinks=true,citecolor=BrickRed,linkcolor=BrickRed,urlcolor=BrickRed]{hyperref}

\input{macros}

\title{{\Large Transporting causal effects from a randomized trial without ``transportability:'' a case study of political advertising during U.S.\ elections}}
\date{\vspace{-1.5cm}}

\author[1]{Xinran Miao}
\author[1,2]{Jiwei Zhao}
\author[1]{Hyunseung Kang}
\affil[1]{Department of Statistics, University of Wisconsin-Madison}
\affil[2]{Department of Biostatistics and Medical Informatics, University of Wisconsin-Madison}

\begin{document}
\maketitle
\vspace{-1cm}

\begin{abstract}
During the 2020 U.S.\ presidential election, \citet{aggarwal20232} conducted a large-scale randomized experiment to evaluate a digital ad campaign against Trump in five battleground states. While the study found no effect on voter turnout, it's unclear whether this null result generalizes to other battleground states, notably Georgia, which played a unique role in the 2020 election and differs from the battleground states.

Inspired by the study, we present a transfer learning framework to estimate treatment effects in a target population (e.g., Georgia) based on a randomized experiment from a source population (e.g., the five battleground states). Our framework is based on a sensitivity analysis that allows for violation of transportability, a popular yet impractical assumption which requires all differences between the source and target populations to be characterized by observed variables. Under our framework, we propose two estimators of the target treatment effect: a simple regression estimator with bootstrap, which we recommend for practitioners in this field, and an estimator based on the efficient influence function. Importantly, both estimators allow for  covariates to differ between the target and the source populations, another common scenario in practice. We also propose a new, sample splitting approach to calibrate the sensitivity parameter. We apply our framework to estimate the effect of the ad campaign on voter turnout in Georgia during the 2020 election. Our findings indicate that small departures from transportability can lead to dramatically different ad effects across counties of Georgia. The direction of the effects is largely driven by racial composition: counties with higher White and lower Black percents tend to show positive effects, while counties with higher Latinx percents tend to show negative effects.
\end{abstract}


\newpage


\section{Introduction}
\subsection{Motivation: generalizing  randomized experiments of political ad campaigns} \label{sec:intro.motivation}
In recent years, political campaigns have used randomized  experiments to
evaluate political ads (e.g., \citet{gerber2011large,kalla2018minimal,aggarwal20232}).
For example, during the 2020 U.S.\ presidential election, \citet{aggarwal20232} conducted a large-scale, randomized controlled trial (RCT) among $1,999,282$ registered voters from five battleground states: Pennsylvania (PA), Wisconsin (WI), Michigan (MI), North Carolina (NC) and Arizona (AZ). They found that their negative, digital ad campaign against President Donald Trump during the 2020 U.S.\ presidential election was ineffective in changing voter turnout.

The main empirical question we address in this paper is as follows: would \citet{aggarwal20232}'s negative ad campaign against Trump  be ineffective in Georgia during the 2020 U.S.\ presidential election? Georgia was another important battleground state during the 2020 election and was unique among the battleground states for several reasons. First, President Joe Biden won the state by the narrowest margin of any state
and became the first Democrat to carry the state since 1992. Second, compared with other battleground states, Georgia presented distinct voter demographics, notably having the highest share of Black voters among battleground states \citep{pew2020race_in_GA}.
Third, even after accounting for voter demographics (e.g., age, race, gender), there were  substantial measured and unmeasured variations between Georgia and the other battleground states such as health insurance coverage, education attainment, immigrant status, and income \citep{pew2020black}.
These unmeasured differences constitute a violation of the transportability assumption, a widely-adopted assumption that requires all differences between populations be captured by observed covariates, thereby rendering most existing methods insufficient.

We introduce a framework that relaxes the transportability assumption
by combining transfer learning with sensitivity analysis and allowing for both measured and unmeasured differences between the populations.
Specifically,
transfer learning uses the overlapping measurements about voters in Georgia and \citet{aggarwal20232}'s experiment (e.g., gender, age group, race) to adjust for measurable differences between the two populations.
Sensitivity analysis quantifies any unmeasured differences between populations  (e.g., health insurance coverage, socioeconomic status, changes in electoral context) with interpretable sensitivity parameters.
Crucially, we also propose a new, data-driven procedure to calibrate/benchmark the magnitude of sensitivity parameters based on sample splitting and design sensitivity \citep{rosenbaum_designsens_2004,rosenbaum2020design}.

Applying our framework, we estimate the effect of \citet{aggarwal20232}'s  negative, digital ad campaign against Trump on voter turnout among roughly 3.9 million registered voters in Georgia during the 2020 U.S.\ presidential election. With a county-by-county analysis of the treatment effect, we show that if Georgia differed from the five battleground states in the experiment only in terms of measured voter demographics, then
\citet{aggarwal20232}'s ad campaign against Trump would remain ineffective  in all counties of Georgia.
But, if there are slight, unmeasured differences between Georgia and the five battleground states, then the ad effects can be significant, with the direction of the effect varying noticeably by race.

\subsection{Related work and our contributions}\label{subsec:related.work}
Our work builds upon several works on generalizing or transporting treatment effects from a source population to a target population under a sensitivity analysis framework \citep{nguyen2017sensitivity,
colnet2021generalizing,dahabreh2023sensitivity,zeng2025efficient,duong2023sensitivity,ek2023externally,huang2024sensitivity}.
Specifically, we work under the exponential tilting sensitivity model \citep{robins2000sensitivity}, which has been used in works on generalizability and transportability
\citep{dahabreh2022global}.
We briefly remark that while our motivating example concerns a binary outcome (i.e., voted or not), the exponentially tilting sensitivity model applies to continuous outcomes, as elaborated in Section \ref{sec:sens_model}.
Building on this sensitivity model, we make the following new contributions.
\begin{itemize}
    \item[(a)] We allow
    the source data
    to have more covariates than the target data.
    Not only does this setup arise in our motivating application, but it is also common when the source data is derived from a randomized experiment where detailed covariate information about study units is collected.
    Related setups have been studied for similar motivations but with different emphases; see \citet{egami2021covariate} for covariate selection and \citet{zeng2025efficient} for efficient and minimax estimation.
    We remark that  most existing work requires that the same covariates are measured between the source and target populations (e.g., \citet{stuart2011use,tipton2013improving,dahabreh2019sensitivity,nie2021covariate,huang2024sensitivity}), and our framework explicitly relaxes this requirement.
    \item[(b)] We propose a simple regression estimator with nonparametric, percentile bootstrap for estimation and inference of the treatment effect in the target population; see Section \ref{subsec:OR}. Notably, while qualitatively suggested by several prior works (e.g., \citet{ackerman2021generalizing,lee2023improving,chen2026confidence}), we formally show one theoretically correct approach to conduct bootstrap-based inference for transfer learning. We also recommend this analysis pipeline for practitioners
    because of its simplicity, theoretically attractive properties (e.g., consistency, asymptotic normality), and the estimator based on the efficient influence function (EIF) is not doubly robust; see the next bullet point.
    \item[(c)] We also propose an estimator based on the EIF. This result extends the novel results in \citet{zeng2025efficient} to the setting where sensitivity parameters are present. While this estimator is optimal from a semiparametric efficiency perspective, it is more complex, requires estimating four nuisance functions and is not doubly robust; see Section \ref{subsec:eif}.
    \item[(d)] For either procedure (b) or (c), we propose a simple calibration procedure to generate interpretable, ``reference'' magnitudes of the sensitivity parameter.
    Unlike existing methods for calibration based on omitting a measured covariate
(e.g., \citet{hsu2013calibrating,cinelli2020making,ek2023externally,huang2024sensitivity}), our calibration procedure uses the same covariates for both calibration and sensitivity analysis. Our procedure is inspired by a clever idea underlying design sensitivity \citep{rosenbaum_designsens_2004}
     and sample splitting where we create a data-driven ``favorable'' situation \citep[Chapter 15]{rosenbaum2020design} by splitting the source data in a particular way; see Section \ref{sec:calibrate}.
  \end{itemize}

While the listed contributions are directly motivated from the statistical challenges in generalizing \citet{aggarwal20232}'s experiment on political advertising, our work is applicable in other contexts, notably in generalizing the results of a randomized trial to a target population with (1) mis-matching covariates and (2) unmeasurable differences between the source and the target populations.
Some examples include generalizing treatment effects from trial-eligible patients in a randomized, double-blind, placebo control clinical study to all eligible patients (e.g., \citet{goldstein2019outcome,lyden2023transportability,anderson2020generalizability,heirali2025eligibility}),
and
 inferring the impact of a multi-site developmental program on a new site based on data from existing sites \citep{nie2021covariate,huang2024sensitivity}.
More broadly, our analysis pipeline centered on sensitivity analysis is useful to practitioners who want a simple, theoretically valid approach for generalization or transportation tasks without assuming transportability.


\section{Transfer learning between elections}\label{sec:setting}
\subsection{Setup: Observed data}

Suppose we collect $\ns$ independent and identically distributed (i.i.d.) samples from a source population. For each study unit $i \in \Is = \{1,\ldots,\ns\}$ in the source data, we observe the following:
\begin{align*}
 \text{Source Data: }   \{\O_i = (\X_i,A_i,Y_i,S_i=1), i\in\Is\}.
\end{align*}
The variable $\X_i \in \calX$ is the pre-treatment covariate (e.g., voter demographics), $A_i$ is the binary treatment indicator (e.g., assigned to  ad campaign against Trump or not in the experiment), $Y_i$ is the binary outcome (e.g., voted or not), and $S_i$ indicates whether unit $i$ is from the source sample (i.e., $S_i=1$) or not (i.e., $S_i=0$).
In our data analysis, the source data is from \citet{aggarwal20232}.

\begin{figure}
    \centering
    \includegraphics[width=1\linewidth]{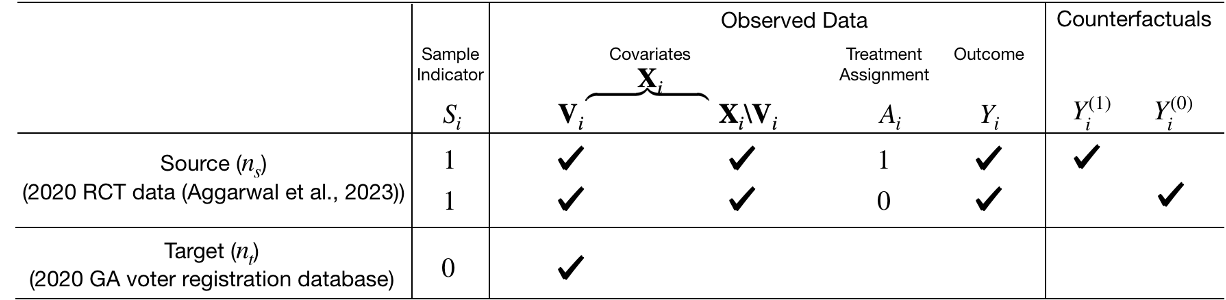}
    \caption{A visualization of the data setup. $\checkmark$ or filled entries indicate that the data is observed.}
    \label{fig:data}
\end{figure}

Independently from the source data, we also collect $\nt$ i.i.d.\ samples from the target population. For each study unit $i \in \It = \{\ns+1,\ldots,\ns+\nt=n\}$ from the target data, we observe the following\footnote{For notational convenience, we overload the notation $\O_i$ to represent the observed data from unit $i$.
If the data is from the source, $\O_i = (\X_i$, $A_i$, $Y_i, S_i=1)$ and if the data is from the target, $\O_i = (\V_i, S_i=0)$.}:
\begin{align*}
 \text{Target Data: }    \{\O_i = (\V_i,S_i=0), i\in\It\}.
\end{align*}
The variable $\V_i \in \calV \subseteq \calX$
is a subset of the covariates in $\cal{X}$.
In our data analysis, the target data consists of
registered voters from Georgia's voter registration database and $\V_i$ is voter $i$'s demographic information from the database (e.g., age group, gender, race). Because $\V_i$ is present in both the source and target data, we refer to it as the shared covariate. Figure \ref{fig:data} summarizes our data setup.

We make some remarks about  the setup.
First, we allow $\calV \subseteq \calX$ because, as far as we are aware of, there is no publicly available dataset of Georgia voters that measured the same attributes as the source data from the five battleground states.
In general, it's uncommon in practice for the target data to contain the same covariates as the source data, and allowing for mismatches in the covariates is more realistic than assuming identical covariates; see \citet{egami2021covariate} and \citet{zeng2025efficient} who echoed a similar sentiment. 
Second, while we focus on binary outcomes $Y_i$ due to our data analysis, our framework generalizes to a continuous outcome; see Section A of the Supplementary Materials.

\subsection{Setup: Causal estimands and nuisance functions} \label{subsec:setup}

We use the counterfactual framework to define causal effects.
Let $\Ya_i$ be the counterfactual outcome of unit $i$ when the treatment is, possibly contrary to fact, set to $a \in \{0,1\}$. The causal estimand of interest, denoted as $\theta$, is the average treatment effect in the target population (TATE):
\begin{align*}
\theta = \theta_1-\theta_0,\text{ where } \theta_a=\E\left[\Ya_i \mid S_i=0\right] \text{ and }a \in \{0,1\}.
\end{align*}
In our data analysis, $\theta$ is the average effect of \citet{aggarwal20232}'s ad campaign against Trump on voter turnout among 2020 voters in Georgia; see Section \ref{subsubsec:interpretation.tate} for a more detailed discussion of this estimand.
We remark that for a binary outcome, other measures of treatment effects are possible, such as the risk ratio $\theta_1/\theta_0$ and the odds ratio $[\theta_1/(1-\theta_1)][\theta_0/(1-\theta_0)]^{-1}$. While we focus on mean differences (i.e.\ $\theta_1 - \theta_0$),
our results are derived for $\theta_a, a\in \{0,1\}$ and thus, can be extended to cover the risk ratio and the odds ratio;  see \cite{ye2023robust} for an example.

We define the following functions, often referred to as nuisance functions.
\bit
\item The propensity score in the source population: $\pi(\x) = \P(A_i=1\mid\X_i=\x,S_i=1)$, $\x\in\calX$.
\item The outcome regression functions in the source population for each level of treatment $a \in \{0,1\}$:
\begin{itemize}
\item With all covariates $\X_i$: $
\mu_a(\x) = \E(Y_i\mid\X_i = \x,A_i=a,S_i=1)$, $\x\in\calX$.
\item With the shared covariates $\V_i$: $\rho_a(\v) = \E(\mu_a(\X_i) \mid \V_i=\v,S_i=1)$, $\v\in\calV$.
\end{itemize}
\item The ratio of probability densities of $\V_i$ between the two populations: $w(\v) = p_{\V_i\mid S_i=0}(\v\mid S_i=0)/p_{\V_i\mid S_i=1}(\v\mid S_i=1)$ where $p_{\V_i\mid S_i=s}(\cdot)$ is the conditional density of $\V_i$ given $S_i=s$ for $s=0,1$.
\eit
We conclude by defining the following notations for order and convergences. For two real sequences of numbers $b_n$ and $d_n$, we denote $b_n=O(d_n)$ if $|b_n|\leq C|d_n|$ for a constant $C$ and denote $b_n\asymp d_n$ if $b_n=O(d_n)$ and $d_n=O(b_n)$. We use $\to_p$ to denote convergence in probability and $\to_d$ to denote convergence in distribution. For a sequence of random variables $Z_n$ and a real sequence of numbers $b_n$, we denote $Z_n=o_p(b_n)$ if $Z_n/b_n\to_p0$. For a measurable and integrable function $f(\cdot)$, we denote its $L_2$ norm by $\lVert f(\O_i) \rVert=\sqrt{\E\{f^2(\O_i)\}}$.


\subsection{Causal identification}

To identify the TATE, it's common to make two sets of assumptions \citep{stuart2011use,tipton2013improving,nguyen2017sensitivity,dahabreh2023sensitivity,zeng2025efficient, huang2024overlap,huang2024sensitivity}. The first set of assumptions ensures the identification of the average treatment effect (ATE) in the source population with the source data.
\begin{assumption}[Identification of the ATE in the Source Population]\label{assump:source}\
\bnum
    \item[a.] (Stable Unit Treatment Value Assumption, SUTVA, \citet{rubin1980randomization}): $Y_i=Y_i^{(A_i)}$ if  $S_i = 1$.
    \item[b.] (Strong Ignorability, \citet{rosenbaum1983central}):  $\Yone_i, \Yzero_i \indep A_i \mid\X_i,S_i=1$ and $0 < \pi(\x) < 1$ for  $\x\in\calX$.
\enum
\end{assumption}
\noindent
Assumption \ref{assump:source}a rules out interference \citep{cox1958planning} and different versions of treatment. Assumption \ref{assump:source}b ensures all confounders are measured, which is satisfied if the source data is from a randomized controlled trial.
The plausibility of Assumption \ref{assump:source} in our data application is discussed in Section \ref{subsubsec:plausible.assump}.

The second set of assumptions ensures that we can generalize or transport the identified ATE from the source population to the target population.
\begin{assumption}[Positivity of Selection]\label{assump:pos.s}
$\P(S_i=1\mid\V_i=\v)>0$ for $\v\in\calV$; $\P(S_i=0)>0$.
\end{assumption}
\begin{assumption}[Transportability]
    \label{assump:trans}
    $\Yone_i, \Yzero_i \indep S_i \mid\V_i$.
\end{assumption}
\noindent
The first part of Assumption \ref{assump:pos.s} will be violated if there are some values of the shared covariates $\V_i$ that are only observed in the target population.
The second part of Assumption \ref{assump:pos.s} excludes the case where the target sample size is much smaller than the source sample size.
Because both parts of Assumption \ref{assump:pos.s} depend solely on observable quantities, Assumption \ref{assump:pos.s} can be checked with data; see Figure \ref{fig:demo.source.target} for an example from our data analysis.

Assumption \ref{assump:trans}, referred to as transportability, states that conditional on the shared covariates $\V_i$, the distributions of the potential outcomes are identical between the source and the target populations. The assumption is violated if the distribution of the potential outcomes differ between the source and the target populations after adjusting for $\V_i$.

For example, if $\V_i$ only contains gender, Assumption \ref{assump:trans} will be violated if within each political party, voter turnout under treatment or control is different between Georgia and the five battleground states.

Unfortunately, unlike Assumption \ref{assump:pos.s}, Assumption \ref{assump:trans} depends on counterfactual quantities and cannot be checked from data. Also, unlike strong ignorability in Assumption \ref{assump:source}, we are not aware of a feasible experimental design to guarantee Assumption \ref{assump:trans} in electoral contexts.\footnote{
A study design that satisfies Assumption \ref{assump:trans} is to randomize the selection of study units into the source data
\citep{tipton2013improving,tipton2017design}.
In our data analysis, this design implies \citet{aggarwal20232} randomized voters to be either in their study among five battleground states or be a voter in Georgia. We believe that this design is impractical.} This is the main motivation to embed sensitivity analysis within transfer learning so that we do not rely on Assumption \ref{assump:trans}.

Under Assumptions \ref{assump:source}-\ref{assump:trans}, the TATE can be identified \citep{zeng2025efficient}:
\begin{align}
    \theta &= \E \left[\E\left\{ \mu_1(\X_i)-\mu_0(\X_i)\mid\V_i,S_i=1\right\}\mid S_i=0\right].\label{eq:trans.identify}
\end{align}
In words, $\theta$ is identified by first averaging the conditional average treatment (CATE) effect in the source population (i.e., $\mu_1(\X_i)-\mu_0(\X_i)$) over the shared covariates $\V_i$ (i.e., the inner expectation in equation \eqref{eq:trans.identify}) and second, averaging this quantity among units in the target population (i.e., the outer expectation in equation \eqref{eq:trans.identify}). For efficient and minimax estimation of $\theta$ under Assumptions \ref{assump:source}-\ref{assump:trans}, see \citet{zeng2025efficient}.

\section{Sensitivity analysis of transportability} \label{sec:sens_model}
As discussed above, suppose transportability (i.e., Assumption \ref{assump:trans}) no longer holds even after conditioning on $\V_i$ and we measure the departure from it by the sensitivity parameter $\Gamma_a \in \mathbb{R}^+$
for each treatment level $a \in \{0,1\}$. Specifically, for a given $\v \in \calV$, the parameter $\Gamma_a$ is defined as the odds ratio of counterfactual outcomes  between the target and source populations:
\begin{align}\label{eq:tilt.binary}
    \Gamma_a &= \frac{{\rm ODD}_a(\v,0)}{{\rm ODD}_a(\v,1)},
    \ {\rm ODD}_a(\v,s) = \frac{\P(\Ya_i=1\mid\V_i=\v,S_i=s)}{\P(\Ya_i=0\mid\V_i=\v,S_i=s)},\  s\in\{0,1\},
    \v\in\calV.
\end{align}
When $\Gamma_a=1$, the conditional distributions of $\Ya_i$ given $\V_i$ are identical between the source and target populations and Assumption \ref{assump:trans} holds.
As $\Gamma_a$ moves away from $1$,  the degree of violating transportability increases.
For example, in our data analysis, $\Gamma_1=1.01$ means that the counterfactual odd of voting in Georgia is $1.01$ times that in the five battleground states among  voters who, possibly contrary to fact, get the negative ads from \citet{aggarwal20232}'s ad campaign. Similarly, $\Gamma_1 = 0.99$ means the counterfactual odd of voting in GA is $0.99$ times that in the five battleground states when a registered voter, possibly contrary to fact, gets the negative ad campaign.

Similar to other sensitivity analyses, the sensitivity parameter $\Gamma_a$ is not identifiable. Instead, investigators identify and estimate the TATE for a given $\Gamma_a$ and in doing so, study the sensitivity of the TATE when transportability is violated. The following result shows that
for a given $\Gamma_a \in \mathbb{R}^+$, the expected counterfactual outcome under treatment level $a \in \{0,1\}$ is identified.
\begin{theorem}[Identification of TATE Under Sensitivity Model]\label{lemma:identify}
Suppose Assumptions \ref{assump:source} and \ref{assump:pos.s} hold.
For a given $\Gamma_a \in \mathbb{R}^+$, the expected counterfactual outcome under treatment level $a\in\{0,1\}$ is

\begin{align}
  \E[\Ya_i\mid S_i=0]
   & = \E\left[
\dfrac{\Gamma_a\rho_a(\V_i)}
    {\Gamma_a\rho_a(\V_i) + 1 - \rho_a(\V_i)}
    \Bigg\vert S_i=0
    \right]
       = \theta_{a}(\Gamma_a). \label{eq:identify.rho}
\end{align}
\end{theorem}
To highlight the inclusion of the sensitivity analysis, we denote the mean counterfactual outcome under treatment level $a$ by $\theta_a(\Gamma_a)$ and the TATE by $\theta_1(\Gamma_1)-\theta_0(\Gamma_0)$. Despite the expanded notation, the interpretation of $\theta_a(\Gamma_a)$ as an average of the counterfactual outcome $\Ya_i$ in the target population remains the same.
For example, in our motivating application, $\theta_1(1)$ is the turnout that would have occurred  had all Georgia voters been assigned to \citet{aggarwal20232}'s anti-Trump digital ads and transportability held;
$\theta_1(1.01)$ is the turnout  that would have occurred had all Georgia voters been assigned to \citet{aggarwal20232}'s anti-Trump digital ads and transportability been violated by $\Gamma_1=1.01$.

We conclude this section with a couple of remarks on the sensitivity model \eqref{eq:tilt.binary}. First,
this model was first proposed by \cite{robins2000sensitivity} as a non-parametric (just) identified model for describing selection bias in missing data. The model was later called an  exponential tilting model \citep{rotnitzky2001methods,birmingham2003pattern} and an extrapolation-factorization model \citep{linero2018bayesian}.
The model was also used to conduct sensitivity analysis for unmeasured confounding in causal inference \citep{franks2019flexible,nabi2024semiparametric} and for violation of transportability in generalizability \citep{dahabreh2022global}. In particular, when $\calV = \calX$ (i.e., covariates are identical between the two populations), Theorem \ref{lemma:identify} recovers the identification of the TATE in \citet{dahabreh2022global}. Second, we choose this sensitivity model as it (a) posits no testable implications on the data, (b) makes statistical inference tractable (e.g., asymptotic normality), and (c) has a simple, odds ratio interpretation. Third, the sensitivity model can be extended in various ways. For example, it can be extended to handle a continuous, counterfactual outcome where the sensitivity model tilts the entire density of the counterfactual outcome; see Section A of the Supplementary Materials where we discuss identification, estimation, and interpretation of the TATE under a sensitivity model with a continuous, counterfactual outcome. Also, at the expense of more sensitivity parameters, model \eqref{eq:tilt.binary} can be extended to allow $\Gamma_a$ to depend on $\V_i$ and $\Ya_i$;  see \cite{franks2019flexible} and \citet{nabi2024semiparametric} for examples.
Fourth, model \eqref{eq:tilt.binary} can be reformulated under a selection model (Section A of the Supplementary Materials) or be reparametrized to be an $R^2$-based model \citep{franks2019flexible}.


\section{Estimation and inference}
\label{sec:estimation}

\subsection{Outcome regression and percentile bootstrap}\label{subsec:OR}

The analysis pipeline in this section  is appropriate  when the outcome regression model $\rho_a(\v)$ can be consistently estimated at a parametric rate. Even otherwise, we suggest investigators begin with this analysis since  it is not only  simple, but also the alternative analysis based on the efficient influence function (EIF) is not doubly robust; see Section \ref{subsec:eif}.

From the identification equation \eqref{eq:identify.rho}, a natural estimator of $\theta_a(\Gamma_a)$ is a plug-in estimator that takes a weighted average of an estimator of the outcome regression function $\rho_a(\v)$, denoted as $\wh\rho_a(\v)$, among the target sample. We call this estimator the outcome regression (OR) estimator:
\begin{align} \label{eq:or.est}
    \wh\theta\rega(\Gamma_a) =\meannt \dfrac{\Gamma_a\wh\rho_a(\V_i)}
    {\Gamma_a\wh\rho_a(\V_i) + 1 -\wh\rho_a(\V_i) }.
\end{align}
From the definition of $\rho_a(\v)$ in Section \ref{subsec:setup}, a simple estimator of $\wh\rho_a(\v)$ is to regress $\wh\mu_a(\x)$ on $\v$ using ordinary least squares (OLS) where $\wh\mu_a(\x)$ is an estimate of $\mu_a(\x)$. Other approaches to estimate  $\rho_a$ is discussed in Section \ref{subsec:est.nuisance}.

For inference, we recommend a nonparametric, percentile bootstrap \citep{efron1979bootstrap} where the source and the target data are resampled separately and we take the
$\alpha/2$ and $1-\alpha/2$ quantiles of the bootstrapped estimates of $ \wh\theta\rega(\Gamma_a)$, denoted $\wh L_a(\Gamma_a;1-\alpha)$ and $\wh U_a(\Gamma_a;1-\alpha)$ respectively. These quantiles are used to construct a $(1-\alpha)$ confidence interval (CI), denoted as  $\wh{\text{CI}}\rega(\Gamma_a;1-\alpha) = \left[\wh{L}_a(\Gamma_a;1-\alpha),\wh{U}_a(\Gamma_a;1-\alpha)\right]$.

Suppose $\rho_a(\v)$ is indexed by a finite-dimensional parameter $\bta_a$. Theorem \ref{thm:or.bootstrap.ci} shows that under regularity conditions, the plug-in, OR estimator $\wh\theta\rega(\Gamma_a)$ in equation (\ref{eq:or.est}) is consistent and the nonparametric, percentile bootstrap leads to a valid CI.
\begin{theorem}[Theoretical properties of the OR estimator and bootstrapped CI]\label{thm:or.bootstrap.ci}
     Suppose Assumptions \ref{assump:source} and \ref{assump:pos.s} hold and $\theta_a(\Gamma_a) \in \Theta$ where $\Theta$ is open and compact.  Also suppose $\rho_a(\v;\bta_a)$ is twice differentiable with respect to  $\bta_a$. If $\wh\bta_a$ is an asymptotically linear estimate of $\bta_a$ and $\ns\asymp\nt$, then $\wh\theta\rega(\Gamma_a) \to_p \theta_a(\Gamma_a)$. Furthermore, if regularity conditions (B1)-(B4) in Section B of the Supplementary Materials hold, the bootstrap interval $\wh{\textnormal{CI}}\rega(\Gamma_a;1-\alpha) $ for $\alpha \in (0,0.5)$ satisfies
$\P(\theta_a(\Gamma_a)\in\wh{\textnormal{CI}}\rega\left(\Gamma_a;1-\alpha)\right) \to 1-\alpha$.
\end{theorem}

\subsection{Efficient influence function}\label{subsec:eif}

The analysis pipeline in this section is based on the efficient influence function (EIF). While this estimator is optimal in the sense that it reaches the  semiparametric efficiency bound under some assumptions,
 the EIF-based estimator is complex and requires estimating multiple nuisance functions.

To motivate the estimator, we first derive the EIF of $\theta_a(\Gamma_a)$ below.
\begin{theorem}[Efficient Influence Function of $\theta_a(\Gamma_a)$] \label{thm:eif}
    Under Assumptions \ref{assump:source} and \ref{assump:pos.s},
    the EIF of $\theta_a(\Gamma_a)$ is
\begin{align*}
\EIF(\O_i,\theta_a(\Gamma_a))  =& \dfrac{S_iw(\V_i)}{\P(S_i=1)}
    \dfrac{\Gamma_a}{[\Gamma_a\rho_a(\V_i) + 1-\rho_a(\V_i)]^2}
        \bigg[\mu_a(\X_i)-\rho_a(\V_i)
          \\
& +\ind(A_i=a) \left\{\dfrac{A_i}{\pi(\X_i)} + \dfrac{(1-A_i)}{1-\pi(\X_i)}\right\}
        \{Y_i-\mu_a(\X_i)\}
\bigg] \\
&+ \dfrac{1-S_i}{\P(S_i=0)}
       \left[     \dfrac{\Gamma_a \rho_a(\V_i) }{\Gamma_a\rho_a(\V) + 1-\rho_a(\V_i)}
           - \theta_a(\Gamma_a) \right],
\end{align*}
 Also, if the propensity score $\pi(\X_i)$ is known, the EIF of $\theta_a(\Gamma_a)$ remains unchanged.
\end{theorem}
We remark that when transportability  holds, i.e., $\Gamma_a=1$, Theorem \ref{thm:eif} reduces to the EIF in \cite{zeng2025efficient}.

Following the current  trend in causal inference, we use cross-fitting
and the EIF (e.g., \citet{chernozhukov2017double,kennedy2022semiparametric}) to estimate $\theta_a(\Gamma_a)$. Specifically, we randomly partition the source and target sample indices $\Is$ and $\It$ into $K$ disjoint sets, $\calI_{s}^{(k)}$ and $\calI_{t}^{(k)}$, respectively, for $k=1,2,\cdots,K$, and let $\calI^{(k)}=\calI_{s}^{(k)}\cup\calI_{t}^{(k)}$. For each $k$, the nuisance functions are estimated with data in $\calI \setminus \calI^{(k)}$ and they are denoted as $\wh\pi^{(k)}(\x)$, $\wh\mu_a^{(k)}(\x)$, $\wh w^{(k)}(\v)$ and $\wh\rho_a^{(k)}(\v)$. We then plug them into the ``uncentered'' EIF and evaluate it with the data in $\calI^{(k)}$:
{\small \begin{align*}
    \wh\theta\eifa^{(k)}(\Gamma_a) =&
  \dfrac{1}{|\calI_{s}^{(k)}|}\sum_{i\in\calI_{s}^{(k)}}
    \dfrac{\Gamma_a\wh w^{(k)}(\V_i)}{[\Gamma_a\wh\rho_a^{(k)}(\V_i) + 1-\wh\rho_a^{(k)}(\V_i)]^2}
        \left[\left\{\dfrac{A_i}{\wh\pi^{(k)}(\X_i)} + \dfrac{1-A_i}{1-\wh\pi^{(k)}(\X_i)}\right\} \{Y_i-\wh\mu_a^{(k)}(\X_i)\} \right.\\
    &\left. \quad{} \quad{} \quad{} \quad{} +
       \wh \mu_a^{(k)}(\X_i)-\wh\rho_a^{(k)}(\V_i)\right]
+\dfrac{1}{|\calI_{t}^{(k)}|}\sum_{i\in\calI_{t}^{(k)}}  \dfrac{\Gamma_a \wh\rho_a^{(k)}(\V_i) }{\Gamma_a\wh\rho_a^{(k)}(\V_i) + 1-\wh\rho_a^{(k)}(\V_i)}.
\end{align*}}
Finally, we take an average of $ \wh\theta\eifa^{(k)}(\Gamma_a)$ to arrive at the EIF-based, cross-fitting estimator of $\theta_a(\Gamma_a)$, which we denote as $  \wh\theta\eifa(\Gamma_a) = K^{-1} \sum_{k=1}^K \wh\theta\eifa^{(k)}(\Gamma_a)$. A step-by-step algorithm can be found from Section C of the Supplementary Materials.
Theorem \ref{thm:eif.asymp} shows that  $\wh\theta\eifa(\Gamma_a)$ is consistent, asymptotically normal, and semiparametrically efficient.

\begin{theorem}[Theoretical properties of the EIF-based estimator] \label{thm:eif.asymp}
Suppose Assumptions \ref{assump:source} and \ref{assump:pos.s} hold and there exist $c,C>0$ such that $c < \wh\pi^{(k)}(\x)$, $\wh w^{(k)}(\v)<C$ and $\wh\rho_a^{(k)}(\v)\in[0,1]$ for  $\v\in\calV$ and $\x\in\calX$. Then, the following holds: \\
(i) [Consistency]. Suppose $\wh\rho^{(k)}_a$ is a consistent estimator of $\rho^{(k)}_a$ (i.e., $\lVert \wh\rho^{(k)}_a(\V_i) - \rho^{(k)}_a(\V_i)\rVert = o_p(1)$). Then, $\wh\theta\eifa(\Gamma_a) \to_p \theta_a(\Gamma_a)$ if
\begin{align}
  \lVert \wh\pi^{(k)}(\X_i)-\pi^{(k)}(\X_i)\rVert
    \cdot
    \lVert \wh\mu_a^{(k)}(\X_i) - \mu_a^{(k)}(\X_i)\rVert &= o_p(1) ,\label{eq:consistent.x}
    \end{align}
\noindent (ii) [Asymptotic normality and Semiparametric Efficiency] Suppose $\wh\rho^{(k)}_a$, $\wh \mu_a^{(k)}$, $\wh w\supk$, and $\wh\pi^{(k)}$ are consistent
with the following rates:
\begin{subequations}
\begin{alignat}{3}
   & \lVert \wh\pi^{(k)}(\X_i)-\pi^{(k)}(\X_i)\rVert
    \cdot
    \lVert \wh\mu_a^{(k)}(\X_i) - \mu_a^{(k)}(\X_i)\rVert  =o_p(n^{-1/2}) , \label{eq:DR.x}\\
   & \lVert \wh w^{(k)}(\V_i) - w^{(k)}(\V_i)\rVert\cdot
    \lVert\wh\rho^{(k)}_a(\V_i) - \rho^{(k)}_a(\V_i) \rVert =o_p(n^{-1/2}),\text{ and} \label{eq:DR.v}\\
&     \lVert \wh\rho^{(k)}_a(\V_i) - \rho^{(k)}_a(\V_i)\rVert^2 =o_p(n^{-1/2}). \label{eq:DR.rho}
\end{alignat}
\end{subequations}
Then, $ \sqrt n\left\{\wh\theta\eifa(\Gamma_a) - \theta_a(\Gamma_a)\right\}\to_d N\left(0,\sigma^2\eifa(\Gamma_a)\right)$ where $\sigma^2\eifa(\Gamma_a) = \mathbb{E}[\{\EIF(\O_i,\theta_a(\Gamma_a))\}^2]$.

\noindent (iii) [Consistent Estimator of Standard Error] Suppose the same assumptions in (ii) hold. Then, $\wh\sigma^2\eifa(\Gamma_a) \to_p \sigma^2\eifa(\Gamma_a)$, where
$$    \wh\sigma\eifa^2(\Gamma_a) = K\inv\sum_{k=1}^K\meanIk\left\{\wh\EIF^{(k)}(\O_i,\wh\theta\eifa(\Gamma_a))\right\}^2,$$
and $\wh\EIF^{(k)}(\O_i,\wh\theta\eifa(\Gamma_a))$ is the empirical counterpart of $\EIF^{(k)}(\O_i,\wh\theta\eifa(\Gamma_a))$ with plug-in estimates of the nuisance parameters $\wh\pi\supk$, $\wh\rho_a\supk$, $\wh w\supk$, and $\wh\mu_a\supk$.
\end{theorem}

Compared to \citet{zeng2025efficient}'s result, Theorem \ref{thm:eif.asymp} shows the EIF-based estimator is not doubly robust in terms of consistency when transportability is violated. Specifically, under transportability (i.e., $\Gamma_a=1$), $\wh\theta\eifa(1)$ remains consistent when we  correctly specify one nuisance function in each of the two pairs: $(\pi, \mu_a)$ and $(w, \rho_a)$. For example, if we can correctly specify the propensity score $\pi$ and the density ratio $w$, we do not need to correctly specify the two outcome regression functions $\rho_a$ or $\mu_a$ for consistency of $\wh\theta\eifa(1)$. However, when transportability is violated via the sensitivity model \eqref{eq:tilt.binary}, consistency of $\wh\theta\eifa(\Gamma_a)$ for
$\Gamma_a \neq 1$ requires a correct specification of $\rho_a$; correctly specifying the propensity score and the density ratio is insufficient for consistency of $\wh\theta\eifa(\Gamma_a)$.
Intuitively, this is because when transportability fails, the identification of TATE is based on the sensitivity model \eqref{eq:tilt.binary}. Since this model depends only on the outcome regression function $\rho_a$, knowing the other nuisance functions (e.g., propensity score or the density ratio) cannot provide information about the model.

Relatedly, the EIF-based estimator loses rate double robustness or the mixed bias property \citep{Rotnitzky2020} when transportability is violated. Specifically, when transportability holds ($\Gamma_a=1$), part (ii) of Theorem \ref{thm:eif.asymp} holds with only the two conditions \eqref{eq:DR.x} and \eqref{eq:DR.v}; these conditions are identical to those in \citet{zeng2025efficient}. Since both conditions involve products of convergence rates in two nuisance estimators, the estimator has rate double robustness whereby a slower convergence rate in one nuisance function can be compensated by a faster convergence rate in the other nuisance function. However, when transportability fails and $\Gamma_a\neq 1$, condition \eqref{eq:DR.rho} on $\rho_a$ is needed to guarantee asymptotic normality and semiparametric efficiency and we no longer have rate double robustness.

\begin{remark}[Double Robustness with Variationally Dependent Nuisance Functions]
    To be absolutely precise, we briefly remark that the double robustness we discuss here is with respect to variationally independent nuisance functions, in that constraints on one nuisance function do not place constraints on the other.
To illustrate, consider the pair of nuisance functions $(\mu_a,\pi)$ in our setup. Estimating $\mu_a$ at a slow rate does not imply that the estimator of $\pi$ has a slow rate, and vice versa. Therefore, double robustness with variationally independent functions offers some ``independent'' protection against model mis-specification if one nuisance function is particularly difficult to estimate.
In contrast, \citet{vansteelandt2007estimation} proposed non-efficient, doubly robust estimators with variationally dependent nuisance functions under model \eqref{eq:tilt.binary} in missing data problems. In Section C of the Supplementary Materials, we extend their construction to our setup with $\calX=\calV$ and construct a doubly robust estimator with variationally dependent nuisance parameters.
It remains an open question on whether doubly robust estimators with variationally independent nuisance functions for the TATE exist under model \eqref{eq:tilt.binary} and $\calV\subset\calX$.
\end{remark}

We conclude by discussing the trade-offs between the OR estimator and the EIF-based estimator. The OR estimator is simpler and easier to explain, but it requires a correct specification of the outcome regression function $\rho_a(\v)$.
The EIF-based estimator also requires a correct specification of the outcome regression function because the estimator is not doubly robust. Additionally, the EIF-based estimator requires estimating more nuisance functions. But, the  condition for the estimator of the outcome regression function is slightly weaker for the EIF-based estimator than for the OR estimator. In practice, we recommend starting with the OR estimator due to its simplicity. The EIF-based estimator may be preferred when the additional nuisance functions can be reliably estimated,
most notably when $w(\v)$ can be well-estimated; see  Section \ref{subsec:est.nuisance} for discussions on estimating nuisance functions.

\subsection{Estimation of nuisance functions}\label{subsec:est.nuisance}

    This section discusses estimation of the nuisance parameters, specifically $\rho_a(\v)$ and $\omega(\v)$. For the other, ``classical'' nuisance functions (i.e., propensity score $\pi(\x)$ and the outcome regression function $\mu_a(\x)$), we echo the modern recommendation of using the investigator's favorite classification and regression models. As a reminder, if the source data is from an RCT, investigators should use the design of the RCT to estimate $\pi(\x)$.

    The regression function $\rho_a(\v)$ can be estimated in a couple of different ways and we highlight each approach through the equalities below:
{\small\begin{align}
    \rho_a(\v) &= \E\{\mu_a(\X_i)\mid\V_i=\v,S_i=1\} \label{eq.rho.reg}\\
    &= \E \left[\left\{\dfrac{A_i\ind(A_i=a)}{\pi(\X_i)} + \dfrac{(1-A_i)\ind(A_i=1-a)}{1-\pi(\X_i)}\right\}Y_i\mid\V_i=\v,S_i=1\right]\label{eq.rho.ipw}\\
    &= \E \left[\left\{\dfrac{A_i\ind(A_i=a)}{\pi(\X_i)} + \dfrac{(1-A_i)\ind(A_i=1-a)}{1-\pi(\X_i)}\right\}\{Y_i-\mu_a(\X_i)\} + \mu_a(\X_i)\mid\V_i=\v,S_i=1\right].\label{eq.rho.aipw}
\end{align}}
The first equality \eqref{eq.rho.reg} suggests estimating $\rho_a$ by regressing the predicted outcome $\wh\mu_a(\X_i)$ on $\V_i$. The second equality \eqref{eq.rho.ipw} suggests
regressing an inverse-probability-weighted (IPW) outcome $\left[A_i\ind(A_i=a)/\wh\pi(\X_i) + (1-A_i)\ind(A_i=1-a)/\{1-\wh\pi(\X_i)\}\right]Y_i$ on $\V_i$. The third equality \eqref{eq.rho.aipw} suggests regressing an augmented IPW outcome,
\[\left[A_i\ind(A_i=a)/\wh\pi(\X_i) + (1-A_i)\ind(A_i=1-a)/\{1-\wh\pi(\X_i)\}\right]\{Y_i-\wh\mu_a(\X_i)\} + \wh\mu_a(\X_i)\] on $\V_i$. Under the first and the third approaches, the rate of convergence of $\wh\rho_a$ is dependent of the rate of convergence of  $\wh\mu_a$ \citep{kennedy2023towards}. Under the second approach, the rate of convergence of $\wh\rho_a$ is independent of the rate of convergence of $\wh\mu_a$. We remark that when all covariates are discrete and  $\pi$ and $\mu_a$ are estimated by taking means within subgroups defined by the covariates, the three approaches are equivalent.

For estimating the density ratio $w(\v)$, we recommend entropy balancing methods
\citep{hainmueller2012entropy,josey2022calibration,chen2023entropy},
 which obtains $\wh w(\V_i)$ as solutions to the
 following constrained optimization problem,
 \begin{align}\label{eq:EB}
 \begin{split}
     &\underset{w_i}{\rm argmin} \sum_{i\in\Is}w_i\log(w_i), \quad{}
    \textnormal{s.t. }\dfrac{1}{\ns}\sum_{i\in\Is}w_i\V_i = \dfrac{1}{\nt}\sum_{i\in\It}\V_i.
     \end{split}
 \end{align}
If the true $\P(S_i=1\mid\V_i)$ is a logistic regression model and the parameters of the logistic model are identifiable, the probability limit of the weights in \eqref{eq:EB} is equal to $\P(S_i=0\mid\V_i)\P(S_i=1)/\{\P(S_i=1\mid\V_i)\P(S_i=0)\}$.
Otherwise, the weights in \eqref{eq:EB} generally have favorable, finite-sample properties (e.g., \citet{chen2023entropy}). For more discussions on estimating $w(\v)$, see Section C of the Supplementary Materials.

\section{Calibrating sensitivity parameters}\label{sec:calibrate}

\subsection{Definition of a sensitive effect and motivation for calibration}\label{subsec:def.sensitive}

Traditionally in a sensitivity analysis, investigators assess how much their statistical conclusion about the TATE changes between $(\Gamma_0, \Gamma_1) =(1,1)$ and other $(\Gamma_0, \Gamma_1)$s of interest.  While Section \ref{sec:estimation} has provided two inference tools to draw statistical conclusions for any given $(\Gamma_0,\Gamma_1)$, this section focuses on finding a reasonable range of $(\Gamma_0,\Gamma_1)$s, henceforth denoted as $\calC$.

Formally, let $\calC \subset \mathbb{R}^+\times\mathbb{R}^+$  be the set of $(\Gamma_0,\Gamma_1)$s that the investigator considers for the sensitivity analysis where $\calC$ is not the singleton set $\{(1,1)\}$.
We say the TATE is \emph{sensitive} to transportability  if the decision to reject the null hypothesis of no effect at the significance level $\alpha$ changes between $(\Gamma_0,\Gamma_1) = (1,1)$ and another value of $(\Gamma_0, \Gamma_1) \in \calC$.

\begin{definition}[Sensitivity to Transportability]\label{def:sens}
   Consider the significance level $\alpha \in (0,1)$ and the set $\calC \subset \mathbb{R}^+\times\mathbb{R}^+$. For a given $(\Gamma_0, \Gamma_1)$, let $\wh{\textnormal{CI}}(\Gamma_0,\Gamma_1;1-\alpha)$ denote a $1-\alpha$ CI of the TATE from Section \ref{sec:estimation}. The TATE is \emph{sensitive} to transportability in $\calC$ if there exists $(\Gamma_0,\Gamma_1)\in\calC$ such that $(\Gamma_0,\Gamma_1)\neq(1,1)$ and either of the following holds:
   \begin{align*}
    \begin{split}
  \text{(i) from significant to insignificant:} \quad{} & 0\not\in   \wh{\textnormal{CI}}(1,1;1-\alpha) \text{ and } 0\in  \wh{\textnormal{CI}}(\Gamma_0,\Gamma_1;1-\alpha);\\
 \text{(ii) from insignificant to significant:} \quad{} &  0\in   \wh{\textnormal{CI}}(1,1;1-\alpha) \text{ and } 0\not\in  \wh{\textnormal{CI}}(\Gamma_0,\Gamma_1;1-\alpha).
     \end{split}
           \end{align*}
    
    Further, for interpretation purpose, we distinguish case (ii) by the sign of effect.
    We say that the TATE is sensitive to a positive (or negative) effect if for any $(\Gamma_0,\Gamma_1)$ satisfying (ii), the corresponding CI contains only positive (or negative) values.
    Otherwise, we say that the TATE is sensitive to an effect in either direction. \\
   Finally, if neither (i) nor (ii) holds, the TATE is \emph{insensitive} to transportability in $\calC$.
\end{definition}

Some investigators specify $\calC$ based on their belief about the unmeasured differences between the source and the target populations.
But in general, specifying a reasonable, ``reference'' set of sensitivity parameters has been a long-standing challenge in sensitivity analysis and this task has recently been referred to as calibration or benchmarking \citep{cinelli2020making,huang2024sensitivity}. One popular approach
is to omit an observed covariate and conduct a sensitivity analysis with the values of the sensitivity parameters that are comparable to the effects of the omitted covariate on the outcome or on the treatment (e.g., \citet{hsu2013calibrating,cinelli2020making,ek2023externally,huang2024sensitivity}). But, as discussed in Section 6.2 of \citet{cinelli2020making}, this can lead to a misleading understanding of the magnitude of unmeasured confounding, especially when the omitted variable is correlated with other confounders.
Furthermore, since the covariates used for calibration differ from those used for estimating the treatment effect, the resulting study design is not equivalent with the original design in terms of covariates.

To this end, we present a new, data-driven calibration procedure that generates a reference set of the sensitivity parameters. Our calibration procedure avoids the issue from the omitted variables approach stated above by using the same covariates for both sensitivity analysis and calibration. Instead, our procedure is inspired by an idea from design sensitivity  \citep{rosenbaum_designsens_2004,rosenbaum2020design}, which is used to benchmark designs of observational studies in terms of robustness against unmeasured confounding by measuring the limiting power to accept a particular type of alternative hypothesis, referred to as a  ``favorable situation'' (Chapter 15 of \cite{rosenbaum2020design}). The favorable situation is typically formulated as a hypothetical parametric model of the treatment effect.  But, in our proposed procedure,  we create the favorable situation by directly using the source data. We state the calibration procedure in Section \ref{subsec:cali.3steps} and describe the rationale in Section \ref{subsec:cali.explain}.

\subsection{Calibration procedure}\label{subsec:cali.3steps}
\begin{figure}
    \centering
    \includegraphics[width=1\linewidth]{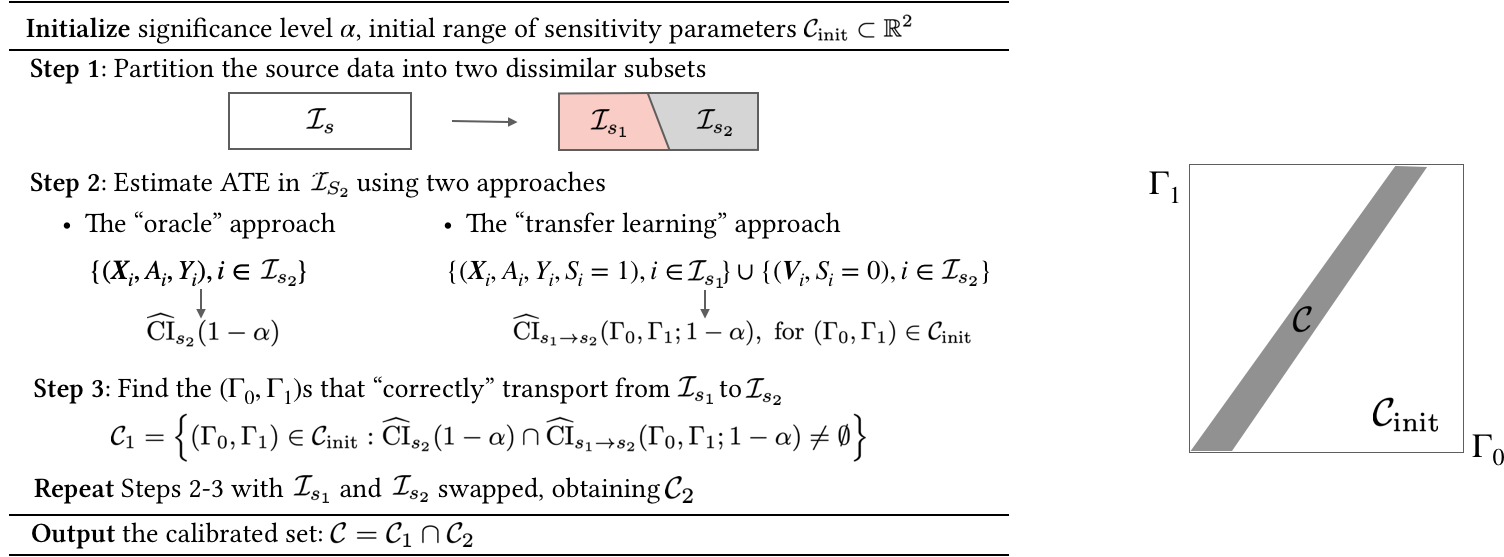}
    \caption{Left: Illustration of the calibration procedure. Right: Illustration of the initial range $\calC_{\text{init}}$ and the resulting calibrated range $\calC$.}
    \label{fig:calibraton}
\end{figure}

At a high level, the calibration procedure takes an initial range $\mathcal{C}_{\text{init}}\subset \mathbb{R}^+\times\mathbb{R}^+$ and produces a smaller calibrated subset $\calC$.
The procedure
can be divided into three steps; see Figure \ref{fig:calibraton} for a visualization.

The first step partitions the source data into two \emph{dissimilar} subsets, denoted as $\calI_{s_1}$ and $\calI_{s_2}$ such that $\calI_{s_1} \cup \calI_{s_2} =\calI_{s}$ and $\calI_{s_1} \cap \calI_{s_2} = \emptyset$.
The second step temporarily treats units in $\calI_{s_2}$ as the ``proxy'' target population and constructs two $1-\alpha$ CIs of the TATE:
\begin{itemize}
    \item (Our Transfer Learning Approach): We treat
    units in $\calI_{s_1}$ as the ``proxy'' source population and use the methods in
     Section \ref{sec:estimation} to
 infer the TATE in $\calI_{s_2}$,
 i.e., the proxy target population. We denote the resulting confidence interval as
 ${\wh{\text{CI}}}_{s_1 \to s_2}(\Gamma_0,\Gamma_1;1-\alpha)$.
    \item
    (The Oracle Approach): Using the data from $\calI_{s_2}$
    only (i.e., the proxy target population), we compute a valid $1-\alpha$ CI of the TATE, say
  the Wald confidence interval based on the difference-in-means estimator, and denote it as $\wh{{\text{CI}}}_{s_2}(1-\alpha)$.
\end{itemize}
The third step keeps the values of $(\Gamma_0,\Gamma_1)$ where the CIs from both approaches overlap, or formally, $\mathcal{C}_1 = \{(\Gamma_0,\Gamma_1)\in\calC_{\text{init}} \mid \wh{{\text{CI}}}_{s_1 \to s_2}(\Gamma_0,\Gamma_1;1-\alpha) \cap \wh{{\text{CI}}}_{s_2}(1-\alpha) \neq \emptyset\}$.
We repeat the three steps above, but with the roles of the proxy target and proxy source populations reversed, yielding another set of sensitivity parameters $\calC_2$, which is $\mathcal{C}_2 = \{(\Gamma_0,\Gamma_1)\in\calC_{\text{init}} \mid \wh{{\text{CI}}}_{s_2 \to s_1}(\Gamma_0,\Gamma_1;1-\alpha) \cap \wh{{\text{CI}}}_{s_1}(1-\alpha) \neq \emptyset\}$.
The intersection of the two sets, $\calC=\calC_1\cap\calC_2$,
is the calibrated set of sensitivity parameters. Further details are explained in Supplementary Materials, notably dealing with relative differences in sample sizes between $\mathcal{I}_{s_1}, \mathcal{I}_{s_2}$, and $\mathcal{I}_{t}$ when computing the confidence intervals.

\subsection{Rationale behind the calibration procedure}\label{subsec:cali.explain}

The first step partitions the source data into two subsets $\calI_{s_1}$ and $\calI_{s_2}$ such that there are scientifically meaningful, unobserved differences between them. In the motivating application, a natural choice is to let $\calI_{s_1}$ and $\calI_{s_2}$ be rust belt states and sun belt states, since there are meaningful differences in socioeconomic status, labor markets, and region-specific politics that are not measured by the shared covariates $\V_i$. Investigators can choose other partitions that are interpretable; see the last paragraph of this section for further discussion and Section \ref{sec:ga} for an illustration with real data.

After partitioning the data, the next two steps find $(\Gamma_0,\Gamma_1)$s that quantify the unmeasured differences between $\calI_{s_1}$ and $\calI_{s_2}$.
This is accomplished by finding a set of $(\Gamma_0,\Gamma_1)$s  such that the transported ATE (i.e., ``Our Transfer Learning Approach'' above)  is close to the true ATE of the proxy target population, up to sampling error. Note that the true TATE in the proxy target population can be inferred with only Assumption \ref{assump:source}, which has been  referred to as the oracle approach above. Then, the resulting set $\calC$ represents the magnitude of the unmeasured differences between the two subsets $\calI_{s_1}$ and $\calI_{s_2}$ where the transported estimate of the TATE matches the oracle ATE of the proxy target population.

It's important to recognize that the calibrated set $\calC$
is not the true unmeasurable differences between the original source population and the original target population.
As noted in Section \ref{sec:sens_model}, the $(\Gamma_0,\Gamma_1)$ that parametrizes the true unmeasured differences
cannot be identified or estimated, and this fact remains for any calibration procedure.

Instead, the calibrated set $\mathcal{C}$ is a data-driven approach to generate a qualitatively meaningful  set of sensitivity parameters.
For example,
if the partition is chosen to be the rust/sun belt states and the TATE in Georgia is sensitive with respect to the calibrated set $\mathcal{C}$, it suggests that the unmeasured differences that are as large as those between the sun belt states and the rust best states in the 2020 election can overturn the conclusion about the TATE. In contrast, if the TATE in Georgia is sensitive with respect to a $\mathcal{C}$ informed by investigator's prior belief, it suggests that the unmeasured differences that are as large as those conjectured by the investigator can overturn the conclusion about the TATE.

Finally, as mentioned in the beginning of this section, investigators can choose other partitions of the source data. But, some partitions are more useful than others. For example, a random partition of the source data such that
there are no unmeasurable differences between the two subsets is not meaningful. But, between two non-random partitions, some investigators may find one partition to be more interpretable than the other.
In Section \ref{sec:ga} we demonstrate two partitions of the source data: (i) partitioning voters from rust belt states and voter from sun belt states, as mentioned above and (ii) partitioning voters with a high Trump support score and a low Trump support score.
In fact, the investigators' unrestricted ability to choose a partition is a useful feature of our calibration procedure compared to the omitted variable approach where the investigators are restricted to the list of observed confounders or focus on the ``strongest'' confounder. More broadly, compared to existing approaches, we believe creating dissimilar partitions of the observed data is a promising way to study how unobservable differences may affect causal conclusions.

\section{Ad effects in Georgia for the 2020 election}\label{sec:ga}

\subsection{Setup}\label{sec:data_background}

\begin{figure}
    \centering
  \includegraphics[width=1\linewidth]{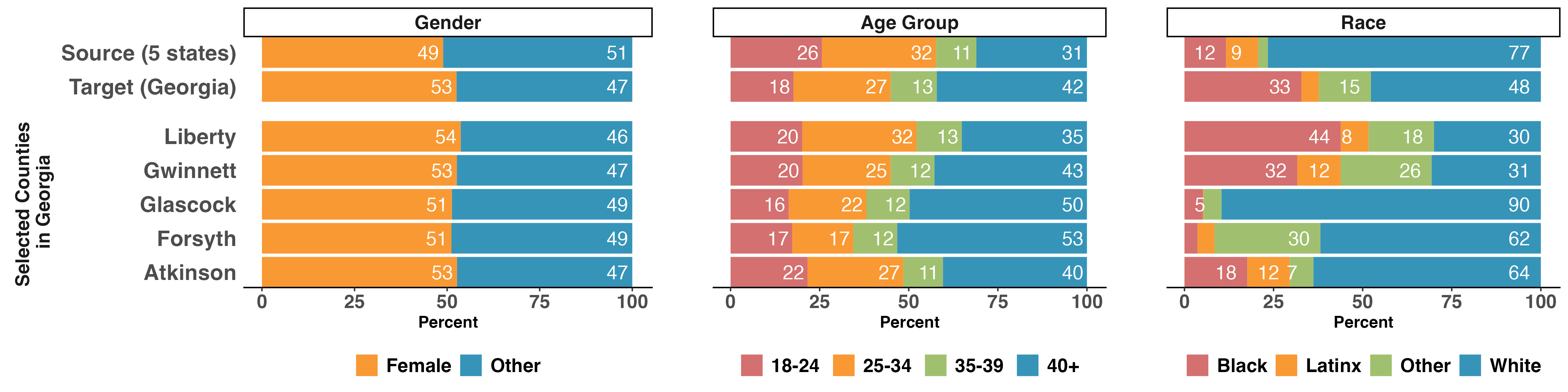}
    \caption{Distributions of voter demographics from the source data (Arizona, Michigan, North Carolina, Wisconsin, Pennsylvania) and the target data (Georgia).
    We also plot some some counties in Georgia.
    }
    \label{fig:demo.source.target}
\end{figure}

We apply our approach to study the main empirical question of the paper: how would running \citet{aggarwal20232}'s digital ad campaign against Trump affect turnout among registered voters in Georgia for the 2020 U.S.\ presidential election?
The target data are registered voters for the 2020 U.S.\ presidential election from Georgia's voter registration database; the database was accessed on August 13, 2025. To harmonize with
the source data by \citet{aggarwal20232}, we recoded age and race in Georgia's database to match the definitions in the source data. We also took a subset of voters in the database who are between 18 and 55 years old to guarantee Assumption \ref{assump:pos.s}; see Section \ref{subsubsec:plausible.assump} for more discussions. In the end, we had $n_{t} = 3,909,868$ registered voters in the target data and the shared covariates $\V_i$ included gender, age groups, and race; see Figure \ref{fig:demo.source.target} for the distribution of the shared covariates. The source covariates $\X_i$ included $\V_i$, party and voting history from \citet{aggarwal20232} and there were $\ns=  1,999,282$ registered voters from the source data.
For more details on the target and source data, see Section E of the Supplementary Materials and \citet{aggarwal20232}.

We estimate the ad effect for each of the 159 counties of Georgia.
Section \ref{subsec:ga.trans} presents the results under transportability. Section \ref{subsec:ga.sens} presents the results of the sensitivity analysis.
The calibration is performed on two different partitions of the source data:
\begin{itemize}
    \item RS partition: voters are from the rust belt states (i.e., PA, WI, MI) or the sun belt states (i.e., AZ, NC),
    \item TS partition: voters' Trump support scores are above 50 or below 50. For reference, Trump support score is an integer between 30 and 70 with 50 indicating a neutral attitude; see Supplementary Section D of \citet{aggarwal20232} for details.
\end{itemize}
Following \cite{aggarwal20232}, $\mu_{a}(\X_i)$ and
$\wh{\rm CI}_{s_1}, \wh{\rm CI}_{s_2}$ in the calibration procedure are based on weighted least squares that regresses the outcome on the treatment and pre-treatment covariates and the weights are the inverse of the propensity scores.
The regression function $\rho_a(\v)$ is estimated by regressing $\wh\mu_a(\X_i)$ on $\V_i$ in a linear model (without interactions). The density ratio $w\supk(\v)$ is estimated by entropy balancing from equation \eqref{eq:EB}.
Due to page limit, we present the results from the OR estimator and discuss the results from the EIF estimator in Section \ref{subsubsec:ga.eif}.

Throughout the analysis, the significance level is $\alpha=0.05$.
In Section E.4 of the Supplementary Materials, we apply Bonferroni correction to account for multiplicities in testing.
 In short, no county has a sensitive effect after the correction.
 We remark that the Bonferroni correction is likely sub-optimal due to potentially complex dependencies across the hypotheses.

Finally, as a reminder on the interpretation, a positive effect means that running the digital ad campaign  against Trump would increase voter turnout whereas a negative effect means that running  the digital ad campaign against Trump  would decrease voter turnout. Also, the sign of the ad effect  does not equate to more or fewer voters for Trump.

\subsection{Ad effect under transportability}\label{subsec:ga.trans}
When $\Gamma_0=\Gamma_1=1$ (i.e., transportability holds), the ad effect is insignificant in all counties; see the left panel of Figure \ref{fig:ga.given.gamma} for a visual illustration and Section E of the Supplementary Materials for the point estimates. In other words, if the difference in voter turnout between Georgia and the five battleground states can be completely adjusted with $\V_i$, then the negative ads will be ineffective in all counties across Georgia for the 2020 election.

\begin{figure}[!h]
    \centering
    \includegraphics[width=1\linewidth]{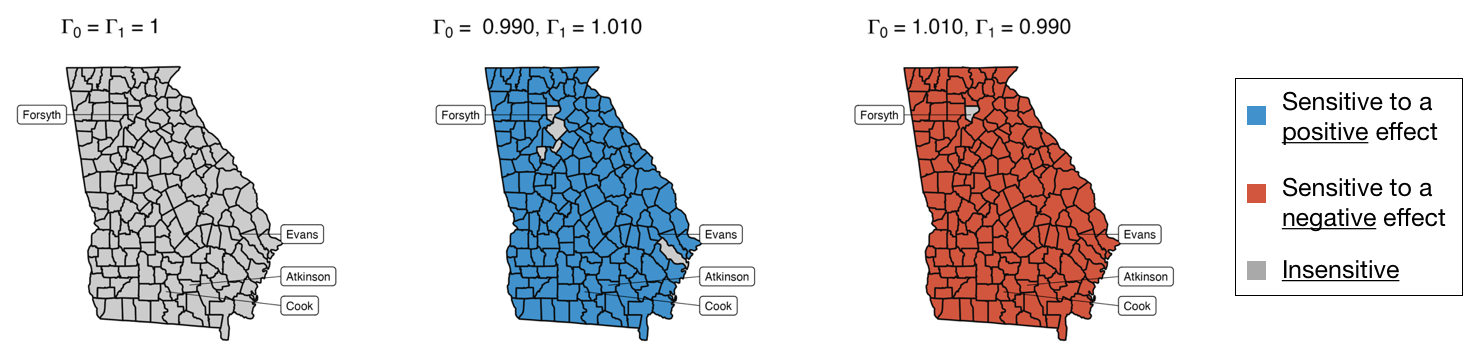}
    \caption{County-by-county ad effect in Georgia for the 2020 U.S.\ presidential election with pre-specified $(\Gamma_0,\Gamma_1)$s. Left: $(\Gamma_0,\Gamma_1)=(1,1)$; i.e., transportability holds. Middle: $(\Gamma_0,\Gamma_1)=(0.990,1.010)$. Right: $(\Gamma_0,\Gamma_1)=(1.010,0.990)$. }
    \label{fig:ga.given.gamma}
\end{figure}

\subsection{Ad effect under sensitivity analysis}\label{subsec:ga.sens}

\subsubsection{Ad effect with pre-specified $(\Gamma_0,\Gamma_1)$s}\label{sec:results_trad}

We study the ad effects for pre-specified $(\Gamma_0,\Gamma_1)\neq(1,1)$. This mirrors a ``traditional''
sensitivity analysis discussed in Section \ref{subsec:def.sensitive}
where the investigator pre-specifies a set $\calC$ based on prior beliefs about the unmeasured differences between Georgia and the five battleground states. For brevity, we consider two values of $(\Gamma_0, \Gamma_1)$, and present their results in terms of sensitive effects (Definition \ref{def:sens}); see the middle and right panels of Figure \ref{fig:ga.given.gamma}.
For the exact point estimates, see Section E of the Supplementary Materials.

Suppose $\Gamma_0=0.990$ and $\Gamma_1=1.010$, i.e., $\calC=\{(0.990,1.010)\}$. In words, under the control, the counterfactual odd of voting in Georgia is  $0.990$ times the counterfactual odd in the five battleground states and under treatment, the counterfactual odd of voting among Georgia voters is $1.010$ times the counterfactual odd of voting in the five battleground states. The ad effect is significant and positive in 154 counties and insignificant in 5 counties. Following Definition \ref{def:sens}, these 154 counties are sensitive to a positive effect and the rest are insensitive.
For the sensitive counties, the p-value is the smallest in Evans county ($p=0.010$), followed by Cook county ($p=0.014$) and Atkinson county ($p=0.019$).

Conversely, suppose $\Gamma_0=1.010$ and $\Gamma_1=0.990$, i.e., $\calC=\{(1.010,0.990)\}$.
The ad effect is significant and negative (i.e., sensitive to a negative effect) in 158 counties and insignificant (i.e., insensitive) in  Forsyth county.

\subsubsection{Ad effect with the calibrated set}\label{sec:calibrated_ga_analysis}

\begin{figure}[!h]
    \centering
    \includegraphics[width=1\linewidth]{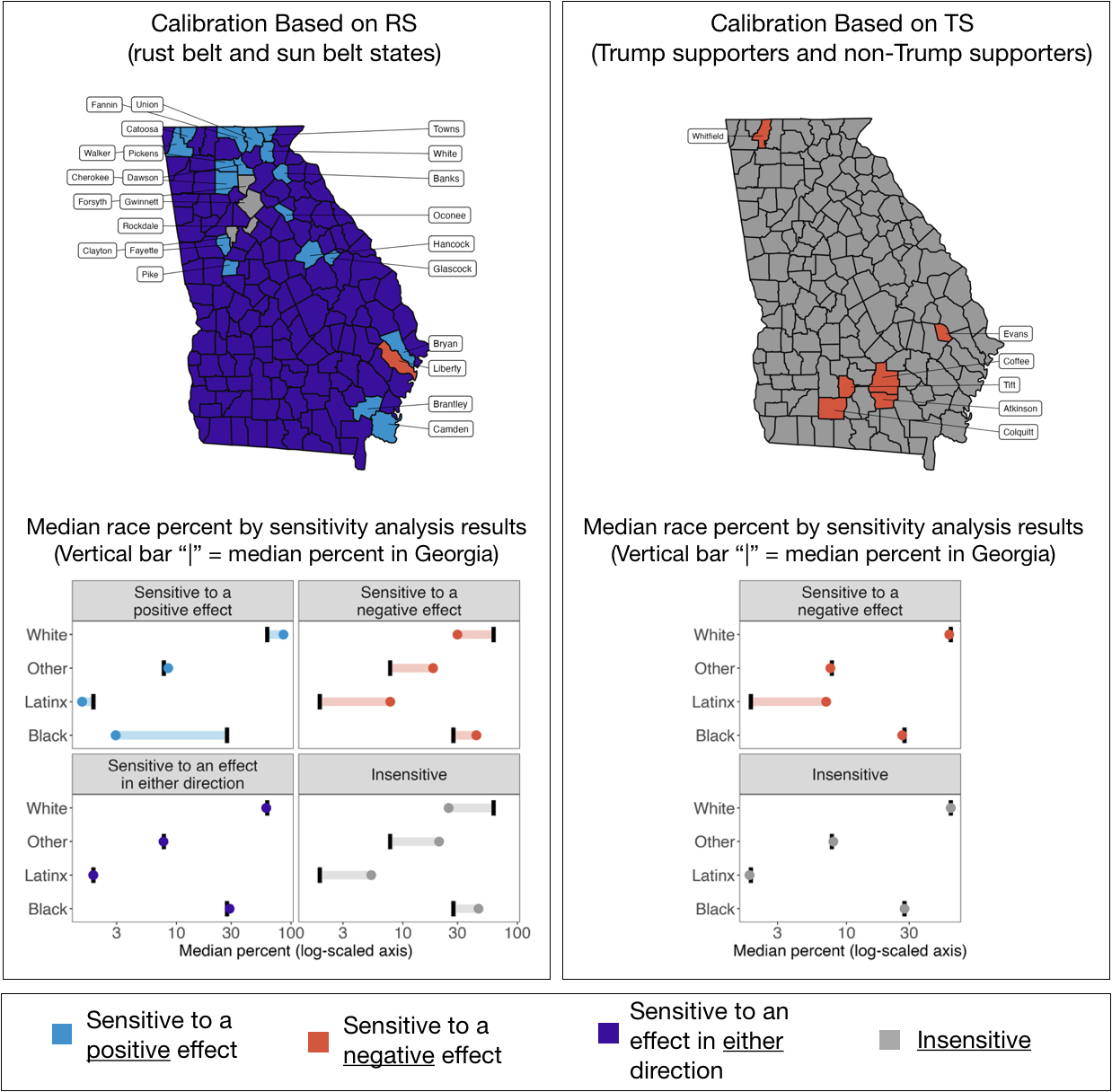}
    \caption{Calibrated sensitivity analysis results with two partitioning schemes: (left) sun belt states versus rust belt states and (right) Trump support score. Top panels show the calibrated result on the Georgia map.  Bottom panels compare the median of percents of a particular race among counties with a particular type of calibrated result (colored dots) with the median of percents of that race among all counties of Georgia (vertical bars). Counties are highlighted if they are sensitive to an effect in one direction (i.e., sensitive to a positive effect only, or sensitive to a negative effect only). For example, on the bottom right, the red dot for Latinx is the median percent of Latinx voters among six counties that were sensitive for a negative effect and the vertical bar is the median percent of Latinx voters among all 159 counties of Georgia.
    }
    \label{fig:ga.calibrated}
\end{figure}

We use the sensitivity parameters from the calibrated set to conduct the sensitivity analysis. Recall that we created two  partitions of the source data: The RS partition based on the rust belt states and the sun belt states and the TS partition based on the Trump support score.
Results are shown in Figure \ref{fig:ga.calibrated}.

We start by discussing the results under the RS partition; see the left part of Figure \ref{fig:ga.calibrated}. First, 18 counties (colored in blue) are sensitive to a positive effect only, meaning that \citet{aggarwal20232}'s negative ad campaign against Trump could increase voter turnout in these counties if the unmeasured difference between a target county and the five battleground states is comparable with the difference between the rust belt states and the sun belt states. Interestingly, these 18 counties have a high proportion of white voters and low proportion of black voters; see the blue dot in the left bottom panel in Figure \ref{fig:ga.calibrated} which indicates the median of these 18 counties against the median of all counties of Georgia. Second, Liberty county (colored in red) is sensitive to a negative effect only, meaning  that the ad campaign against Trump could significantly decrease voter turnout.
Note that Liberty county is a Democratic leaning county with a high proportion of Latinx voters (7.8\%, as opposed to 1.9\%, the median proportion of Latinx voters across all counties of Georgia).
Third, most counties (in particular, 136, colored in purple) are sensitive to either a positive or a negative effect.
This is because the calibrated set $\calC$ is fairly large, and within this set, some $(\Gamma_0,\Gamma_1)$s can lead to a statistically significant and positive effect while some others can lead to a statistically significant and negative effect.
Fourth, the other four counties (colored in gray) are insensitive and running the negative ad campaign wouldn't alter voter turnout in these counties.

Next, we discuss the results under the TS partition; see the right part of Figure \ref{fig:ga.calibrated}. First, 153 out of 159 counties are insensitive (colored in gray). In other words, if the unmeasured difference between a target county and the five experimental states are comparable with the unmeasured difference between Trump supporters and non-Trump supporters, then the ad campaign will not change voter turnout in most counties. This result stands in sharp contrast to the result based on RS partitioning. The difference can be attributed to the fact that the calibrated set under TS partitioning is much smaller than that under RS partitioning. In words, there are only few values of the sensitivity parameters that correctly transport the ad effect between Trump supporters and non-Trump supporters because the difference between these two subgroups of voters are, qualitatively speaking, large. In contrast, there are many values of the sensitivity parameters that correctly transport the ad effect between rust belt states and sun belt states because the difference between these two subgroups of voters are, qualitatively speaking, small to moderate.
Finally, we remark that the six leftover counties are all sensitive for a negative effect and these six counties have a high proportion of Latinx voters; see the vertical bar and the red dot in the bottom right panel of Figure \ref{fig:ga.calibrated} for the median proportion of Latinx voters in all of Georgia and the six counties, respectively.

\subsection{Key takeaway: small effects are more sensitive for generalization}

From the sensitivity analysis based on non-calibrated values in Section \ref{sec:results_trad}, a small, unmeasured difference between Georgia and the five battleground states leads to different conclusions about the ad effect across many counties.
For example, a small change in the odds of voting between Georgia and the five battleground states from $(\Gamma_0, \Gamma_1) = (1,1)$ to $(\Gamma_0, \Gamma_1) = (1.01, 0.99)$,
leads to 100 $\cdot$ 156/157 $\approx$ 99\% of counties having significant effects.
Similarly, the sensitivity analysis based on calibrated sets in Section \ref{sec:calibrated_ga_analysis}
 suggests that many counties will be sensitive if there exists small, unmeasured differences represented by the partitions of the source data.
 In both cases,
our paper supports the conjecture from \citet{aggarwal20232} that their null effect from the five battleground states will not generalize to most counties in GA.
Also, because the original effects by \citet{aggarwal20232} were close to null, our analysis empirically supports a simple, but under-appreciated point by \citet{rosenbaum2010design} that regardless of sample size, ``small effects are sensitive to small [unmeasured] biases'' and thus, they are harder to generalize. Or, modifying his quote in the context of transfer learning, irrespective of sample size, small effects are sensitive to small unmeasured differences between source and target populations.

\subsection{Discussion on the estimand and assumptions}\label{subsec:ga.remarks}

\subsubsection{Interpreting the exposure and the estimand}\label{subsubsec:interpretation.tate}

In the data analysis, the estimand (i.e., TATE) is the average effect of the anti-Trump digital ad campaign studied by \citet{aggarwal20232} on voters from Georgia during the 2020 U.S.\ presidential election. The estimand is not the average effect of \emph{all} anti-Trump digital advertising. Page 340 and Supplementary Section A of \citet{aggarwal20232} contain extensive discussions on how their advertising intensity dominated background online advertising during the 2020 election. Notably, in these discussions, the authors conjecture that voters in their treatment group
are likely seeing their online ads more than other online ads. Also, page 336 of \citet{aggarwal20232} argues that their negative ads are representative of online, negative advertisements during the 2020 election.

To generalize the results from \citet{aggarwal20232}, the exposures and the counterfactual outcomes are defined to be consistent with the original study. Specifically, $A_i=1$ denotes a voter who is randomized to the treatment group  and $A_i=0$ denotes a voter who is randomized to the control group where both groups are defined by  \citet{aggarwal20232}'s experiment. For the counterfactual outcomes, $Y_i(1)$ denotes the counterfactual outcome if, contrary to fact, a voter was randomized to the treatment group and $Y_i(0)$ denotes the counterfactual outcome if, contrary to fact, a voter was randomized to the control group. Then, the TATE measures the average intent-to-treat (ITT) effect on the target population or the average effect of being randomized to the treatment group versus the control group on voter turnout.
While we recognize the scientific importance of going beyond ITT effects (e.g., effect of actually seeing the ads; effect of different types of negative ads; dose-response relationship for negative ads), this requires an experiment that defined such interventions and studied such effects; see \citet[pp.\ 340]{aggarwal20232} for related discussions.

\subsubsection{Plausibility of  assumptions}\label{subsubsec:plausible.assump}
This section discusses the plausibility of Assumptions \ref{assump:source}-\ref{assump:pos.s} and the i.i.d.\ sampling assumption in the data analysis.
We start with the plausibility of Assumption \ref{assump:source}.
Assumption \ref{assump:source}b (strong ignorability) is satisfied by the randomization procedure used in \cite{aggarwal20232} to randomize voters into the treatment group and the control group.

Assumption \ref{assump:source}a (SUTVA) consists of two parts.
The first part, the no hidden variation in treatment, is satisfied because our exposure definition matches those from \cite{aggarwal20232}. Also, we are measuring the ITT effect and not other types of effects that can be defined with different exposures or counterfactual outcomes;
see Section \ref{subsubsec:interpretation.tate} for related discussions.

The second part of SUTVA, the no-interference assumption, is arguably the most questionable.
\citet{aggarwal20232} implicitly assumes no interference. Also, to their credit,  \citet{aggarwal20232} designed their randomized experiment to minimize interference by matching voters with similar social media profiles and tweaking the ad delivery mechanism to target a single online user. But, after randomization, voters may interact with each other through social media or in real life, which can lead to violation of the no-interference assumption.

 For some context, \citet{aggarwal20232} follows a long line of literature on large-scale political advertising  experiments where the no-interference assumption has been adopted and various justifications have been provided to support the assumption (e.g., \citet{kendall2015voters,kalla2018minimal,spenkuch2018political}). If these justifications are satisfactory, then the TATE measures the ITT effect of \citet{aggarwal20232}'s ad campaign on the target population. If the justifications   are unsatisfactory, we conjecture that the TATE is a ``super-population'' counterpart of the overall ITT effect of \citet{aggarwal20232}'s ad campaign on the target population \citep{kang2016peer,savje2021average,li2022random,keele2022introduction}.

Next we discuss Assumption \ref{assump:pos.s} (positivity). Broadly, this assumption is satisfied if the target population consists of voters who are demographically represented in the source population. Accordingly, we have restricted the target population to be voters under 55 since  \citet{aggarwal20232}'s experiment excluded voters over 55; see \citet[pp.\ 339]{aggarwal20232} for an explanation. Figure \ref{fig:demo.source.target} presents a visual comparison of the demographic distributions in the source and target samples, providing empirical support for Assumption \ref{assump:pos.s}.
We also note that had \citet{aggarwal20232}  not excluded older voters, we would be able to study the effect among a wider range of voters in the target population.

Lastly, we discuss the plausibility of independent target and source data and i.i.d.\ sampling within each population.
First, the data from Georgia voters and the data from the five battleground states are likely independent given that by U.S.\ law, a voter cannot vote twice in the same presidential election across two different states. Second, the i.i.d.\ sampling within each population is a common, working assumption in the analysis of large-scale campaign experiments  (e.g., \citet{kendall2015voters,kalla2018minimal,spenkuch2018political,aggarwal20232}) or in generalization of causal effects (e.g., \citet{tipton2013improving,dahabreh2019generalizing,degtiar2023review,zeng2025efficient}). But, we acknowledge that this may not be realistic in practice. In our setup, deviation from i.i.d.\ sampling assumption can lead to inconsistent estimates of the standard error.
Nevertheless, the estimate of the TATE will still be consistent under a version of the law of large numbers that allows for departure from the i.i.d.\ assumption.

\subsection{Additional results}
\label{subsubsec:ga.eif}

\begin{figure}[!h]
    \centering
    \includegraphics[width=.8\linewidth]{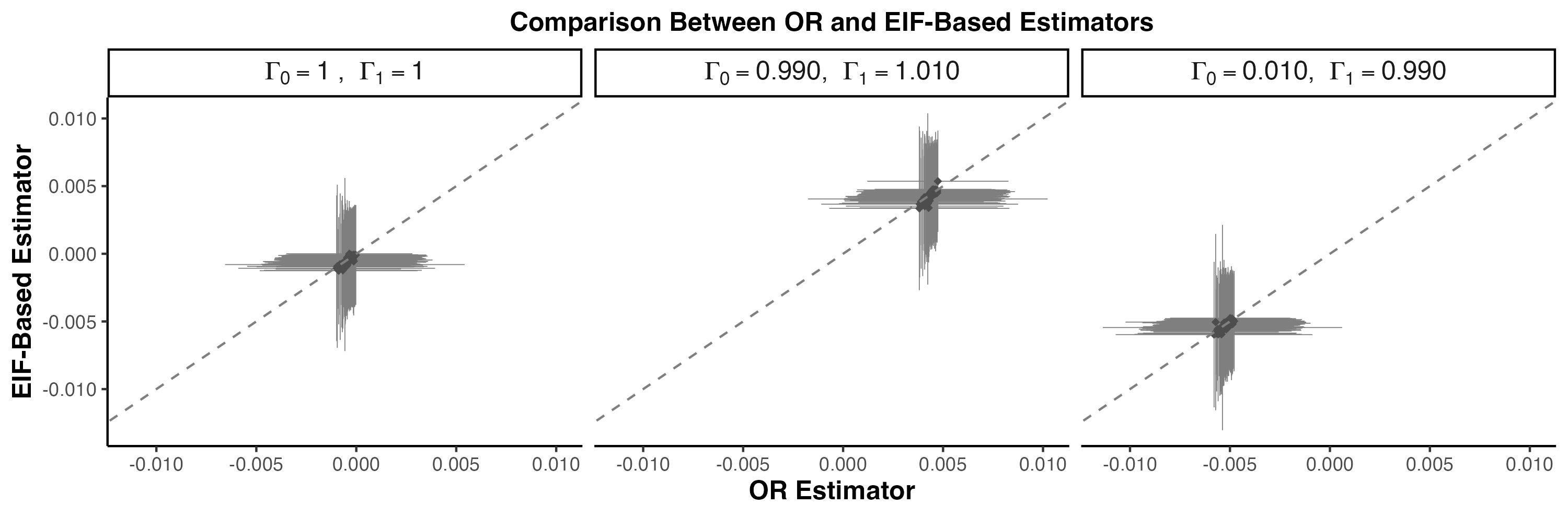}
    \caption{Comparison between the OR and EIF-based estimators for estimating ad effects in every county of Georgia.
    In each panel, x- and y- axes represent results from the OR and the EIF-based estimator, respectively. The points represent point estimates and the gray bars represent 95\% CIs. The dashed line represents the 45 degree line through origin (i.e., $y=x$).}
   \label{fig:compare.estimators}
\end{figure}

We briefly summarize the results from the EIF-based estimator and the subgroup analysis. First, the analyses based on the OR estimator and the EIF-based estimator were very similar, but not identical. For example, in Figure \ref{fig:compare.estimators}, we see that for all 159 counties, the point estimates between the OR estimator and the EIF-based estimator fall closely to the 45 degree line and all the 95\% confidence intervals generated from the two estimators overlap;  note that the widths of the CIs from the two estimators did not uniformly dominate one another.
A simulation study that compares both estimators with semi-synthetic data is given in Section G of the Supplementary Materials.

Second, for the subgroup analysis, we defined eight subgroups by a three-way interaction between gender (female versus not female), race (Black or Latinx versus Others), and recent voting history (new voters versus returning voters). Under transportability, all subgroup effects are small and statistically insignificant. When transportability is violated, the ad effect is sensitive in six out of eight subgroups, most notably female, returning voters that are not Black or Latinx; see details in Section E of the Supplementary Materials for more discussions.

\section{Discussion and future work}\label{sec:disc}
This paper proposes a framework to study the TATE based on transfer learning with sensitivity analysis. We present two estimation procedures for the TATE, one based on OR with bootstrapped CIs (i.e., the recommended procedure) and another based on the EIF. For each procedure, we show that it leads to consistent estimates of the TATE and asymptotically valid  CIs.
Finally, inspired by ideas from design sensitivity, we present a calibration procedure based on partitioning the source population and use it to generate a set of reference magnitudes of the sensitivity parameters for the sensitivity analysis. We use the framework to address whether running \citet{aggarwal20232}'s digital ad campaign against Trump is effective in changing voter turnout in Georgia for the 2020 U.S.\ presidential election.

\begin{figure}[!h]
    \centering
    \includegraphics[width=1\linewidth]{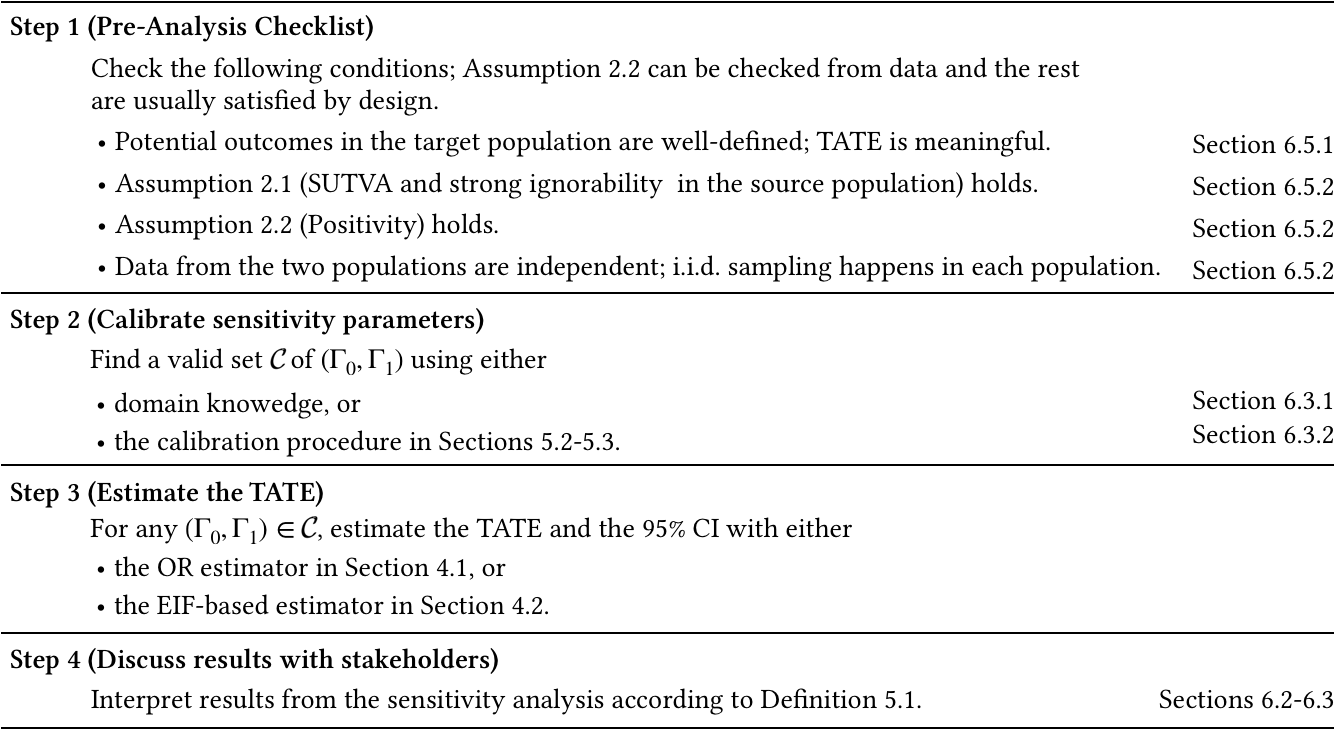}
    \caption{The proposed analysis pipeline for transfer learning with sensitivity analysis. Corresponding sections for the real data application are shown on the right.}
    \label{fig:pipeline}
\end{figure}

 Beyond elections, our framework provides statistically valid solutions to important, practical issues that arise in transportability and generalizability, such as (1) dealing with mis-matched covariates between the source and the target populations (i.e., $\mathcal{X} \neq \mathcal{V}$), (2) addressing violation of transportability, (3) providing a theoretical justification for a commonly used bootstrap procedure in transfer learning to quantify statistical uncertainty of the TATE, and (4) proposing a new calibration procedure for sensitivity analysis without omitting measured confounders. For convenience, we provide a summary of the analysis pipeline in Figure \ref{fig:pipeline} for researchers who wish to use our framework.

\section*{Acknowledgments}
 The authors would like to thank the area editor, the associate editor and the anonymous reviewer for their constructive comments and valuable feedback. The authors would also like to thank  Sameer Deshpande, Melody Huang, Adeline Lo, Ying Jin, Chan Park, Eric J.\ Tchetgen Tchetgen, Ang Yu, Xiaobin Zhou, statistics student seminar participants at University of Wisconsin-Madison on April 29, 2024,
 participants in the Online Causal Inference Seminar on April 30, 2024,
  participants of the Models, Experiments, and Data Workshop in the Department of Political Science at the University of Wisconsin-Madison on February 14, 2025, participants of New England Statistics Symposium on June 2, 2025, participants of Joint Statistics Meetings on August 5, 2025, and participants of the Isaac Newton Institute on April 9, 2026.

\bibliographystyle{apalike}
\bibliography{ref}

\clearpage

\setcounter{section}{0}
\renewcommand\thesection{\Alph{section}}
\numberwithin{equation}{section}
\numberwithin{figure}{section}
\numberwithin{table}{section}
\renewcommand{\theHsection}{\Alph{section}}
\renewcommand{\theHsubsection}{\Alph{section}.\arabic{subsection}}
\renewcommand{\theHsubsubsection}{\Alph{section}.\arabic{subsection}.\arabic{subsubsection}}
\renewcommand{\theHequation}{\Alph{section}.\arabic{equation}}
\renewcommand{\theHfigure}{\Alph{section}.\arabic{figure}}
\renewcommand{\theHtable}{\Alph{section}.\arabic{table}}
\renewcommand{\theHtheorem}{\Alph{section}.\arabic{theorem}}

\begin{center}
{\Large\bf Supplementary Materials}\\[6pt]
{\large Transporting causal effects from a randomized trial without ``transportability:'' a case study of political advertising during U.S.\ elections}
\end{center}

This document provides supplementary materials for the manuscript titled ``Transporting causal effects from a randomized trial without `transportability:' a case study of political advertising during U.S.\ elections''.
Section \ref{sec:alt.form.sens} details the alternative formulations and extensions of the sensitivity model (2). Sections \ref{supp.sec:or} and \ref{supp.sec:eif} provide additional details and proofs for inference procedures based on the OR estimator and the EIF-based estimator introduced in Sections 4.1 and 4.2 of the main text, respectively. Section \ref{supp.sec:calibration} provides details of the calibration procedure introduced in Section 5. Section \ref{supp.sec:pa} provides additional results for analyzing the ad effect in GA.  Section \ref{supp.sec:sim} presents a simulation study where the simulation data were generated to mimic the data from \cite{aggarwal20232}.
Throughout this supplement, references to the main text are not hyperlinked and references to this supplement are hyperlinked.

\tableofcontents

\section{Extensions and Interpretations of the Sensitivity Model}\label{sec:alt.form.sens}
\subsection{Exponential Tilting for Continuous Outcomes}\label{subsec:sensitivity.continuous}
The proposed sensitivity model (2) is not limited to binary outcomes. It can be equivalently expressed for a general, possibly continuous outcome with support $\calY$. To ease communication, we let $\gamma_a=\log(\Gamma_a)\in\mathbb{R}$.
Suppose the conditional density of the potential outcome on the target population is shifted from that of the source by an exponential tilting shift,
\begin{align}\label{eq:tilt.general}
    p_{\Ya\mid\V,S=0}(y_a\mid\v,S_i=0)\propto \exp(\gamma_a y_a)\cdot p_{\Ya\mid\V,S=1}(y_a\mid\v,S_i=1),\quad\forall\v\in\calV,
\end{align}
where $\propto$ represents ``proportional to'' and $p_{\Ya\mid\V,S=s}$ represents the conditional probability density function of $\Ya_i\mid\V_i,S_i=s$ for $s=0,1$. When $\gamma_a=0$ (i.e., $\Gamma_a=1$), (\ref{eq:tilt.general}) reduces to $ p_{\Ya\mid\V,S=0}(y_a\mid\v,S_i=0)=p_{\Ya\mid\V,S=0}(y_a\mid\V_i=\v,S_i=1)$
and thereby  transportability  (Assumption 2.3) holds.
When $\gamma_a\neq 0$, $\gamma_a$ measures the violation to the transportability assumption by the degree in shifts of the conditional densities.

Under (\ref{eq:tilt.general}) and for a given $\gamma_a$, the expected potential outcome under treatment level $a$ can be identified as follows.
\begin{lemma}[Identification of TATE for A General Outcome Under Sensitivity Model]
\label{lemma:identifiy.tate.general}
    Suppose Assumptions 2.1, 2.2 and the sensitivity model in equation \eqref{eq:tilt.general} hold.
    For a given $\gamma_a \in \mathbb{R}$, the expected potential outcome under treatment level $a\in\{0,1\}$ is
\begin{align}
  \E[\Ya_i\mid S_i=0]
    &=    \E\left(
        \dfrac
        {\E[\E\{\exp(\gamma_a Y_i)Y_i\mid\X_i,A_i=a,S_i=1\}\mid\V_i,S_i=1]}
        {\E[\E\{\exp(\gamma_a Y_i)\mid\X_i,A_i=a,S_i=1\}\mid\V_i,S_i=1]}
       \mid S_i=0 \right), \label{eq:identify.continuous}\\
       &=\theta_a(\gamma_a)\nonumber.
\end{align}
\end{lemma}

For a binary outcome, Lemma \ref{lemma:identifiy.tate.general} reduces to Lemma 3.1.
When $\calX=\calV$, Lemma \ref{lemma:identifiy.tate.general} recovers the identification result in \cite{dahabreh2022global}. When $\gamma_a=0$, i.e.,  transportability holds, Lemma \ref{lemma:identifiy.tate.general} recovers the identification result in \cite{zeng2023efficient}.

From (\ref{eq:tilt.general}), the difference between the two conditional densities at $y_a\in\calY$ is quantified by $\exp(\gamma_ay_a)$ up to some normalizing constant.
An extension is to replace $\exp(\gamma_ay_a)$ with $\exp\{\gamma_a\delta(y_a,\v)\}$ where $\delta(y_a,\v)$\footnote{If $\delta(y_a,\v,\gamma_a)$ can be factorized to $\delta_1(y_a,\gamma_a)\delta_2(\v,\gamma_a)$ then it can be replaced with $\delta_1(y_a,\gamma_a)$.} is a statistic including $y_a$ and $\v$. One may also further generalize $\gamma_a$ to  a vector or generalize the exponential function to other forms based on experts' knowledge. We note that the choice should ensure the density $p_{\Ya\mid\V,S=0}$ is well-defined and we refer readers to \cite{franks2019flexible,scharfstein2021semiparametric} for practical choices.

\subsection{Selection Model}
 An alternative view to the sensitivity model is via the selection to the source, in particular, via the probability of $S_i=1$. From this perspective, sensitivity model (\ref{eq:tilt.general}) implies a partially linear logistic regression model \citep{carroll1997generalized} on the selection of $S_i$:
\begin{align}\label{eq:selection}
    &\P(S_i=1\mid \Ya_i=y_a,\V_i=\v)
    = \text{expit}\left(-\gamma_ay_a-\eta(\v,\gamma_a)\right),\quad\forall y_a\in\calY,\v\in\calV,\\
    &\eta(\v,\gamma_a)= \log\left(\frac{\P(S_i=0)}{\P(S_i=1)}\frac{w(\v)}{\E\left\{\exp(\gamma_a\Ya_i)\mid\V_i=\v,S_i=1\right\}}\right)\n,
\end{align}
where $\text{expit}(t)=1/\{1+\exp(-t)\}$ for any $t\in\mathbb R$ is known as the logistic function.
The selection model  (\ref{eq:selection}) indicates that the participation $S_i$ is determined by both the potential outcome and the covariate $\V_i$. After the logistic transformation, the selection probability is associated with $\Ya_i$ linearly with coefficient $\gamma_a$. If $\gamma_a=0$, then the selection will depend on $\V_i$ only, which reduces to the case where the difference between the target and the source is fully characterized by $\V_i$, i.e., when the transportability holds.

\subsection{Estimation for a Continuous Outcome}
The identification condition (\ref{eq:identify.continuous}) directs an OR estimator through
\begin{align*}
    \wh\theta\rega\cont(\gamma_a) &=
    \meannt \dfrac{\wh\E\left\{\exp(\gamma_a\Ya_i)\Ya_i\mid\V_i,S_i=1\right\}}{\wh\E\left\{\exp(\gamma_a\Ya_i)\mid\V_i,S_i=1\right\}}.
\end{align*}

To motivate an EIF-based estimator, we  present the EIF in Theorem \ref{supp.thm:eif}, which is a generalization of Theorem 4.2
to continuous outcomes.
\begin{theorem}\label{supp.thm:eif}
    Under Assumptions 2.1 and 2.2 and sensitivity model (\ref{eq:tilt.general}),
    the EIF for $\theta_a(\gamma_a)$ is
{\footnotesize{
 \begin{align*}
        &\EIF\cont(\O_i,\theta_a(\gamma_a)) \\
        =& \dfrac{S_iw(\V_i)}{\P(S_i=1) }\left\{\dfrac{A_i}{\pi(\X_i)} + \dfrac{1-A_i}{1-\pi(\X_i)}\right\}\left[
        \dfrac{\exp(\gamma_a Y_i)Y_i}{\E\{\exp(\gamma_a \Ya_i)\mid\V_i,S_i=1\}}
                - \dfrac{\E\{\exp(\gamma_aY_i)Y_i\mid\X_i,A=A_i,S_i=1\}}{\E\{\exp(\gamma_a\Ya_i)\mid\V_i,S_i=1\}} \right.\\
     &\left.
        - \dfrac{\exp(\gamma_aY_i)\E\{\exp(\gamma_a\Ya_i)\Ya_i\mid\V_i,S_i=1\}}{[\E\{\exp(\gamma_a\Ya_i)\mid\V_i,S_i=1\}]^2}
         + \dfrac{\E\{\exp(\gamma_aY_i)\mid\X_i,A=A_i,S_i=1\}\E\{\exp(\gamma_a\Ya_i)\Ya_i\mid\V_i,S_i=1\}}{[\E\{\exp(\gamma_a\Ya_i)\mid\V_i,S_i=1\}]^2}
        \right]\\
    & +\dfrac{S_iw(\V_i)}{\P(S_i=1)}
   \left( \dfrac{\E\{e^{\gamma_aY_i}Y_i\mid\X_i,A=A_i,S_i=1\}}{\E\left\{\exp(\gamma_a\Ya_i)\mid\V_i,S_i=1\right\}}-
    \dfrac{\E\{e^{\gamma_a\Ya_i}\Ya_i\mid\V_i,S_i=1\}\E\{e^{\gamma_aY_i}\mid\X_i,A=A_i,S_i=1\}}{[\E\{\exp(\gamma_a\Ya_i)\mid\V_i,S_i=1\}]^2}
    \right)\\
    & + \dfrac{1-S_i}{\P(S_i=0)}
    \left[\dfrac{\E\left\{\exp(\gamma_a\Ya_i)\Ya_i\mid\V_i,S_i=1\right\}}
    {\E\left\{\exp(\gamma_a\Ya_i)\mid\V_i,S_i=1\right\}} - \theta_a(\gamma_a)\right].
\end{align*}}
}
\end{theorem}
$\EIF\cont(\O_i,\theta_a(\gamma_a))$ reduces to $\EIF(\O_i,\theta_a(\gamma_a))$ in Theorem 4.2
for a binary outcome.
It motivates the following EIF-based cross-fitting estimator:
\begin{align*}
    \wh\theta\eifa\cont(\gamma_a) = \meanK\wh\theta\eifa^{\textnormal{cont},(k)}(\gamma_a),
\end{align*}
where $\wh\theta\eifa^{\textnormal{cont},(k)}(\gamma_a)$ is the estimate at $k$-th partition of the cross-fitting procedure as described in Section 4.2,
{\footnotesize{
\begin{align*}
   & \wh\theta\eifa^{\textnormal{cont},(k)}(\gamma_a) \\
    =& \dfrac{1}{|\calI_{s}^{(k)}|}\sum_{i\in\calI_{s}^{(k)}}
    \wh w\supk(\V_i)\left( \left\{\dfrac{A_i}{\wh \pi\supk(\X_i)} + \dfrac{1-A_i}{1-\wh \pi\supk(\X_i)}\right\}\left[
        \dfrac{\exp(\gamma_a Y_i)Y_i}{\wh\E\supk\{e^{\gamma_a \Ya_i}\mid\V_i,S_i=1\}}
                - \dfrac{\wh\E\supk\{e^{\gamma_aY_i}Y_i\mid\X_i,A_i,S_i=1\}}{\wh\E\supk\{e^{\gamma_1\Ya_i}\mid\V_i,S_i=1\}} \right.\right.\\
     &\left.
        - \dfrac{e^{\gamma_aY_i}\wh\E\supk\{e^{\gamma_a\Ya_i}\Ya_i\mid\V_i,S_i=1\}}{[\wh\E\supk\{e^{\gamma_a\Ya_i}\mid\V_i,S_i=1\}]^2}
         + \dfrac{\wh\E\supk\{e^{\gamma_aY_i}\mid\X_i,A_i,S_i=1\}\wh\E\supk\{e^{\gamma_a\Ya_i}\Ya_i\mid\V_i,S_i=1\}}
         {\left[\wh\E\{e^{\gamma_a\Ya_i}\mid\V_i,S_i=1\}\right]^2}
        \right]\\
    & \left. +\dfrac{
    \wh\E\supk\{e^{\gamma_aY_i}Y_i\mid\X_i,A_i,S_i=1\}\wh \E\supk\{e^{\gamma_a\Ya_i}\mid\V_i,S_i=1\} -
    \wh \E\supk\{e^{\gamma_a\Ya_i}\Ya_i\mid\V_i,S_i=1\}\wh\E\supk\{e^{\gamma_aY_i}\mid\X_i,A_i,S_i=1\}
    }{\left[\wh \E\supk\{\exp(\gamma_1\Ya_i)\mid\V_i,S_i=1\}\right]^2}\right)\\
    & +  \dfrac{1}{|\calI_{t}^{(k)}|}\sum_{i\in\calI_{t}^{(k)}}
    \dfrac{\wh \E\supk\left\{\exp(\gamma_1\Ya_i)\Ya_i\mid\V_i,S_i=1\right\}}
    {\wh \E\supk\left\{\exp(\gamma_1\Ya_i)\mid\V_i,S_i=1\right\}} .
\end{align*}
}}

\input{supp_body}

\end{document}

%% file: macros.tex
\usepackage[dvipsnames]{xcolor}
\usepackage{url}
\usepackage{amssymb, amsfonts,mathtools,mathrsfs}
\allowdisplaybreaks 
\usepackage{mathabx}

\usepackage{algorithm}
\usepackage{algpseudocode}

\usepackage{longtable,array,calc}
\usepackage{tabularx} 
\usepackage{makecell} 
\usepackage{multirow}
\usepackage{longtable} 
\usepackage{physics}
\usepackage{comment}
\usepackage{soul} 
\usepackage{subcaption} 

\newtheorem{theorem}{Theorem}[section]

\newtheorem{lemma}[theorem]{Lemma}

\newtheorem{definition}[theorem]{Definition}
\newtheorem{assumption}[theorem]{Assumption}
\newtheorem{remark}[theorem]{Remark}

\def\b{{\bf b}}

\def\g{{\bf g}}

\def\O{{\bf O}}

\def\S{{\bf S}}

\def\V{{\bf V}}
\def\v{{\bf v}}

\def\X{{\bf X}}
\def\x{{\bf x}}

\def\calC{{\cal C}}

\def\calI{{\cal I}}

\def\calS{{\cal S}}

\def\calV{{\cal V}}
\def\calX{{\cal X}}
\def\calY{{\cal Y}}

\def\ba{{\boldsymbol\alpha}}
\def\bb{{\boldsymbol\beta}}

\def\bphi{\boldsymbol\phi}

\def\bta{\boldsymbol\eta}

\def\bS{{\boldsymbol\Sigma}}

\newcommand{\indep}{\perp \!\!\! \perp}

\usepackage{dsfont}
\def\ind{\mathds{1}}

\def\0{{\bf 0}}
\def\trans{^{\rm T}}
\def\star{^{*}}
\def\pr{\textnormal{pr}}
\def\P{\mathbb{P}}
\def\wh{\widehat}
\def\wt{\widetilde}

\def\E{\mathbb{E}}
\def\var{\textnormal{Var}}
\def\cov{\textnormal{Cov}}
\def\n{\nonumber}

\def\log{{\textnormal{log}}}

\def\bnum{\begin{enumerate}}
\def\enum{\end{enumerate}}
\def\bit{\begin{itemize}}
\def\eit{\end{itemize}}

\newcommand{\pf}[1]{\noindent \textit{Proof of #1}. }

\def\boxit#1{\vbox{\hrule\hbox{\vrule\kern6pt\vbox{\kern6pt#1\kern6pt}\kern6pt\vrule}\hrule}}

\def\Is{\calI_{s}}
\def\It{\calI_{t}}

\def\ns{n_s}
\def\nt{n_t}

\def\sumn{\sum_{i=1}^n}

\def\meanK{\dfrac{1}{K}\sum_{k=1}^K}
\def\meanIk{\dfrac{1}{|\calI^{(k)}|}\sum_{i\in\calI^{(k)}}}
\def\meann{\dfrac{1}{n}\sum_{i=1}^n}
\def\meanns{\dfrac{1}{\ns}\sum_{i\in\Is}}
\def\meannt{\dfrac{1}{\nt}\sum_{i\in\It}}

\def\yone{y^{(1)}}
\def\Yone{Y^{(1)}}

\def\Yzero{Y^{(0)}}

\def\Ya{Y^{(a)}}

\def\sqrtn{\dfrac{1}{\sqrt n}\sum_{i=1}^n}

\def\opone{o_p(1)}

\def\inv{^{-1}}
\def\supk{^{(k)}}

\def\eifone{_{\textnormal{EIF,1}}}
\def\eifzero{_{\textnormal{EIF,0}}}
\def\eifa{_{\textnormal{EIF,a}}}
\def\eif{_{\textnormal{EIF}}}

\def\EIF{\textnormal{EIF}}

\def\reg{_{\textnormal{OR}}}
\def\rega{_{\textnormal{OR,a}}}

\def\cont{^{\textnormal{cont}}}

\usepackage[most]{tcolorbox}
\usepackage{enumitem}
\usepackage{tabularray}

%% file: supp_body.tex
\section{Details and Proofs for the Outcome Regression Based Estimation}\label{supp.sec:or}
This section provides details and proofs for the inference procedure with the OR estimator proposed in Section 4.2.
We detail the bootstrap procedure in Section \ref{supp.subsec:details.boot}, state regularity conditions for the bootstrap consistency in Section \ref{supp.subsec:regular.bootstrap},
and prove Theorem 4.1 in Section \ref{supp.subsec:proof.or.boot}.

\subsection{Estimation of $\rho_a$}
We verify that when $\calX$ and $\calV$ are discrete and $\pi$ and $\mu_a$ are estimated by group averages, then estimators motivated from (8), (9) and (10) are equivalent. To be explicit, for a given $\x\in\calX$, the estimates of $\pi$ and $\mu_a$ are
\begin{align*}
    \wh\pi(\x) &= \dfrac{\sum_{i\in\Is}A_i\ind(\X_i=\x)}{\sum_{i\in\Is}\ind(\X_i=\x)},\\
    \wh\mu_a(\x) &= \dfrac{\sum_{i\in\Is}\ind(A_i=a,\X_i=\x)Y_i}{\sum_{j\in\Is}\ind(A_i=a,\X_i=\x)},
\end{align*}
respectively.
The equalities in (8), (9), (10) suggest an outcome regression typed estimator $\wh\rho_a^{\text{OR}}$, an inverse probability weighting estimator $\wh\rho_a^{\text{IPW}}$, and
The equality in (8) suggests an outcome regression typed estimator that we denote as  an augmented inverse probability weighting estimator $\wh\rho_a^{\text{AIPW}}$, respectively, where for $\v\in\calV$,
\begin{align*}
    \wh\rho_a^{\textnormal{OR}}(\v) &= \dfrac{
    \sum_{i\in\Is}\ind(\V_i=\v)\wh\mu_a(\X_i)}
    {\sum_{i\in\Is}\ind(\V_i=\v)},\\
    \wh\rho_a^{\textnormal{IPW}}(\v) & =
    \dfrac{
    \sum_{i\in\Is} \left\{\dfrac{A_iY_i}{\wh\pi(\X_i)} +
    \dfrac{(1-A_i)Y_i}{1-\wh\pi(\X_i)}
    \right\}\ind(\V_i=\v)
    }
    {\sum_{i\in\Is}\ind(\V_i=\v)},\\
     \wh\rho_a^{\textnormal{AIPW}}(\v) &=
      \dfrac{
     \sum_{i\in\Is} \left[
     \left\{\dfrac{A_i}{\wh\pi(\X_i)} +
    \dfrac{(1-A_i)}{1-\wh\pi(\X_i)}\right\}\{Y_i-\wh\mu_a(\X_i)\} + \wh\mu_a(\X_i)
    \right]\ind(\V_i=\v)
    }
    {\sum_{i\in\Is}\ind(\V_i=\v)}.
\end{align*}

\begin{lemma}\label{lemma:rho.discrete}
    When $\calX$ and $\calV$ are discrete, $ \wh\rho_a^{\textnormal{OR}}(\v) = \wh\rho_a^{\textnormal{IPW}}(\v) =  \wh\rho_a^{\textnormal{AIPW}}(\v) $ for any $\v\in\calV$.
\end{lemma}
\pf{Lemma \ref{lemma:rho.discrete}}
Without loss of generality, we prove for the case when $a=1$.

\noindent First, we show that $\wh\rho_1^{\textnormal{OR}}(\v) = \wh\rho_1^{\textnormal{IPW}}(\v)$.
We can simplify  $\wh\rho_1^{\textnormal{OR}}(\v)$ as
\begin{align}
    \wh\rho_1(\v) &= \dfrac{\sum_{i\in\Is}\ind(\V_i=\v)\wh\mu_a(\X_i)}
    {\sum_{i\in\Is}\ind(\V_i=\v)}\n\\
    &= \dfrac{
    \sum_{i\in\Is}\ind(\V_i=\v)\cdot
   \dfrac{\sum_{j\in\Is}\ind(A_j=1,\X_j=\X_i)Y_j}{\sum_{k\in\Is}\ind(A_k=1,\X_k=\X_i)}
    }{  \sum_{i\in\Is}\ind(\V_i=\v)} \n\\
    \label{eq:rho.discrete.or}
    &= \dfrac{1}{  \sum_{i\in\Is}\ind(\V_i=\v)}
    \sum_{i\in\Is}\left\{
     \dfrac{\sum_{j\in\Is}\ind(A_j=1,\X_j=\X_i)Y_j}{\sum_{k\in\Is}\ind(A_k=1,\X_k=\X_i)}\right\}.
\end{align}
We can simplify $\wh\rho_1^{\textnormal{IPW}}(\v)$ as
\begin{align}
    \wh\rho_1^{\textnormal{IPW}}(\v) &= \dfrac{
    \sum_{i\in\Is} \left\{\dfrac{\ind(A_i=1)Y_i}{\wh\pi(\X_i)}
    \right\}\ind(\V_i=\v)
    }
    {\sum_{i\in\Is}\ind(\V_i=\v)} \n\\
&=\dfrac{1}{\sum_{i\in\Is}\ind(\V_i=\v)} \cdot\sum_{i\in\Is}
        \left[
            \dfrac{\ind(A_i=1)Y_i}{
            \wh\pi(\X_i)
            }
            \right]\n\\
&= \dfrac{1}{\sum_{i\in\Is}\ind(\V_i=\v)} \cdot\sum_{i\in\Is}
        \left\{
        \dfrac{\ind(A_i=1)Y_i}{
        \sum_{j\in\Is}\ind(A_j=1)\ind(\X_j=\X_i) / \left\{\sum_{k\in\Is}\ind(\X_k=\X_i)\right\}
        }
            \right\}\n\\
    \label{eq:rho.discrete.ipw}
&= \dfrac{1}{\sum_{i\in\Is}\ind(\V_i=\v)} \cdot\sum_{i\in\Is}
        \left\{
        \dfrac{\sum_{k\in\Is}\ind(A_i=1)Y_i\ind(\X_k=\X_i)}{
        \sum_{j\in\Is}\ind(A_j=1)\ind(\X_j=\X_i)
        }
            \right\}
\end{align}
Since $\eqref{eq:rho.discrete.ipw}=\eqref{eq:rho.discrete.or}$, we have that $\wh\rho_1^{\textnormal{OR}}(\v)=\wh\rho_1^{\textnormal{IPW}}(\v)$.

\noindent Next, we show that $\wh\rho_1^{\textnormal{AIPW}}(\v)=\wh\rho_1^{\textnormal{IPW}}(\v)$.
\begin{align*}
  \wh\rho_1^{\textnormal{AIPW}}(\v)  -\wh\rho_1^{\textnormal{IPW}}(\v)
  =&
  \dfrac{
  \sum_{i\in\Is} \left\{-\dfrac{A_i}{\wh\pi(\X_i)}  + 1\right\}\wh\mu_a(\X_i)\ind(\V_i=\v)
  }
  {\sum_{i\in\Is}\ind(\V_i=\v)},
\end{align*}
where the numerator is
\begin{align*}
  & \sum_{i\in\Is} \left\{-\dfrac{\ind(A_i=1)}{\wh\pi(\X_i)}  + 1\right\}\wh\mu_a(\X_i)\ind(\V_i=\v) \\
  =& \sum_{i\in\Is}\ind(\V_i=\v)\wh\mu_a(\X_i) \left\{
  \dfrac{-\ind(A_i=1) + \dfrac{\sum_{j\in\Is}\ind(A_j=1,\X_j=\X_i)}{\sum_{k\in\Is}}\ind(\X_k=\X_i)}
  {
  \dfrac{\sum_{j\in\Is}\ind(A_j=1,\X_j=\X_i)}{\sum_{k\in\Is}\ind(\X_k=\X_i)}
  }
  \right\}\\
  =&\sum_{i\in\Is}\ind(\V_i=\v)\wh\mu_a(\X_i)
  \dfrac{-\ind(A_i=1)\sum_{k\in\Is}\ind(\X_k=\X_i) +
  \sum_{j\in\Is}\ind(A_j=1,\X_j=\X_i)
  }{\sum_{j\in\Is}\ind(A_j=1,\X_j=\X_i)}\\
=&  \sum_{i\in\Is}\ind(\V_i=\v)\wh\mu_a(\X_i)
  \dfrac{-\sum_{k\in\Is}\ind(A_i=1)\ind(\X_k=\X_i) +
  \sum_{j\in\Is}\ind(A_j=1,\X_j=\X_i)
  }{\sum_{j\in\Is}\ind(A_j=1,\X_j=\X_i)}\\
=& 0.
\end{align*}

Therefore, $\wh\rho_1^{\textnormal{AIPW}}(\v)=\wh\rho_1^{\textnormal{IPW}}(\v)$.

$\square$

\subsection{Details for the Bootstrap}\label{supp.subsec:details.boot}
We detail the nonparametric, percentile bootstrap for the inference with the OR estimator. In each bootstrap iteration, we resample with replacement the source and target samples, respectively,  to have sizes $\ns$ and $\nt$, and construct an OR estimator with the resampled data.
After repeating the bootstrap iterations for a large number of times, say $B$ times, we calculate the $\alpha/2$ and $1-\alpha/2$ quantiles of the resulting bootstrap estimates, denoted as $\wh L_a(\Gamma_a;1-\alpha)$ and $\wh U_a(\Gamma_a;1-\alpha)$. By Theorem 4.1,
the interval $\wh{\text{CI}}\rega(\Gamma_a) = [\wh{L}_a(\Gamma_a;1-\alpha),\wh{U}_a(\Gamma_a;1-\alpha)]$ is a consistent confidence interval for $\theta_a(\Gamma_a)$. A step-by-step procedure is provided in Algorithm \ref{alg:or}.

We note that underlying true quantiles of the bootstrap estimates are estimated by their empirical counterparts ($\wh L_a(\Gamma_a;1-\alpha)$ and $\wh U_a(\Gamma_a;1-\alpha)$). This estimation step introduces an additional random error. Since this error can be made arbitrarily small by resampling the data for sufficiently many times, our proof supposes that $\wh L_a(\Gamma_a;1-\alpha)$ and $\wh U_a(\Gamma_a;1-\alpha)$ are the exact quantiles of bootstrap estimates. This argument follows the approach in Chapter 23 of \cite{vandervaart1998}. For numerical results throughout the paper, the bootstrap iterations are repeated for $B=1000$ times.

 \begin{algorithm}[!h]
 \caption{Outcome regression estimator with nonparametric, percentile bootstrap}\label{alg:or}
     \begin{algorithmic}[1]
 \Require Sensitivity parameters $\Gamma_a$, confidence level $1-\alpha$, bootstrap iteration $B$.
 \State \textbf{Step 1}: Estimate $\wh\rho_a(\v)$ using the source data.
 \State \textbf{Step 2}: Estimate $\wh\theta\rega(\Gamma_a)$ as in (4). 
 \State\textbf{Step 3}: Nonparametric, percentile bootstrap
     \For{$b$ in $1,\cdots B$}
     \State  Resample source and target data with replacement at sizes $\ns$ and $\nt$, respectively.
     \State With the resampled data, obtain $\wh\theta\rega^{*,b}(\Gamma_a)$.
     \EndFor
     \State Calculate the $\alpha/2$ and $1-\alpha/2$ quantiles of $\left\{\wh\theta\rega^{*,b}(\Gamma_a)\right\}_{b=1}^B$, denoted as $\wh L_a(\Gamma_a;1-\alpha)$ and $\wh U_a(\Gamma_a;1-\alpha)$ where
     \begin{align*}
         \wh L_a(\Gamma_a;1-\alpha) = \wh Q\star(\alpha/2),\quad\wh U_a(\Gamma_a;1-\alpha) = \wh Q\star(1-\alpha/2),\\
          \wh Q\star(\tau) = \inf_{t}\left\{
          \frac{1}{B}\sum_{b=1}^B\ind(\wh\theta\reg^{*,b}(\Gamma_a) \leq t) \geq \tau
          \right\},\ \forall\tau\in(0,1).
     \end{align*}
 \Ensure The OR estimator $\wh\theta\reg(\Gamma_a)$ with a $(1-\alpha)$ confidence interval $\wh{\text{CI}}\rega(\Gamma_a;1-\alpha) = [\wh L_a(\Gamma_a;1-\alpha), \wh U_a(\Gamma_a;1-\alpha)]$.
     \end{algorithmic}
 \end{algorithm}

\subsection{Regularity Conditions for the Bootstrap}
\label{supp.subsec:regular.bootstrap}
Recall that we suppose the $\rho_a(\v)$ is indexed by a finite-dimensional parameter $\bta_a$. Specifically, suppose the parameter $\bta_a$ is estimated through an estimating equation,
\begin{align*}
    \meanns \S(\O_i,\wh\bta_a) = \0
\end{align*}
with a known $\S(\O_i,\bta_a)$.
Let $\bb_a(\Gamma_a)= [\bta_a\trans,\theta_a(\Gamma_a)]\trans$ and
\begin{align*}
\bphi_a(\O_i,\bb_a(\Gamma_a)) &=
       \left [\dfrac{S_i}{\P(S_i=1)}\S(\O_i,\bta_a)\trans,
        \dfrac{1-S_i}{\P(S_i=0)}\phi_a(\V_i,\theta_a(\Gamma_a),\bta_a)\right]\trans, \text{ where}\\
\phi_a(\V_i,\theta_a(\Gamma_a),\bta_a) &= \dfrac{ \Gamma_a \rho_a(\V_i,\bta_a)}{ \Gamma_a \rho_a(\V_i,\bta_a) + 1-\rho_a(\V_i,\bta_a)} - \theta_a(\Gamma_a).
\end{align*}
Then $\wh\bb_a(\Gamma_a)= [\wh\bta_a\trans,\wh\theta_a(\Gamma_a)]\trans$ can be alternatively expressed as the solution to the estimating equation
\begin{align*}
    \meann\bphi_a(\O_i,\wh\bb(\Gamma_a)) = \0.
\end{align*}
We define the bootstrap estimator  $\wh\bb\star_a(\Gamma_a)$ as the solution to
\begin{align*}
    \meann W_{n,i}\bphi(\O_i,\wh\bb\star_a(\Gamma_a)) = \0,
\end{align*}
where $(W_{n,1},\cdots,W_{n,\ns})\sim\text{Multinomial}(\ns;1/\ns,\cdots,1/\ns)$ and \\ $(W_{n,\ns+1},\cdots,W_{n,n})\sim\text{Multinomial}(\nt;1/\nt,\cdots,1/\nt)$.

We assume the following regularity conditions.
\bit
\item[(B1)] $\E\{\bphi(\O_i,\bb_a(\Gamma_a))\} = \0$ with a unique solution $\bb(\Gamma_a)$.
\item[(B2)] Parameter $\bb_a(\Gamma_a)$ is contained in a compact parameter space $\Xi$ and $\E\sup_{\bb_a(\Gamma_a)\in\Xi}\lVert \bphi\rVert_1 < \infty$.
\item[(B3)] $\E\left(\sup_{\bb_a(\Gamma_a)\in\Xi}\lVert\partial\bphi^2_a/\partial\bb_a(\Gamma_a)^2\rVert\right)<\infty$.
\item[(B4)] The function class $\{\bphi_a(\O_i,\bb_a(\Gamma_a)), \bb_a(\Gamma_a)\in\Xi\}$ is $\P$-Donsker and $\E\lVert\bphi_a(\O_i,\wt\bb(\Gamma_a)) - \bphi_a(\O_i,\bb_a(\Gamma_a))\rVert^2\to 0$ as long as $\lVert\wt\bb_a(\Gamma_a)-\bb_a(\Gamma_a)\rVert\to 0$.
\eit
Condition (B1) is essentially assuming $\E\{\S(\O,\bta_a)\}=\0$ with the unique solution being the true parameter $\bta_a$. Condition (B2) guarantees that $\bphi_a$ is $\P$-Gilvenko-Cantelli by \citet[Lemma 6.1]{wellner2005empirical}. Condition (B3) and (B4) are standard regularity conditions for the complexity of the function class and the smoothness of the estimating equation.

\subsection{Proof of Theorem 4.1}\label{supp.subsec:proof.or.boot}
Before proving Theorem 4.1, 
we state the asymptotic Normality of the OR estimator in Theorem \ref{thm:or.an}. Next in Theorem \ref{thm:bootstrap.an}, we show that the bootstrap estimator is also asymptotically Normal with the same asymptotic variance. Finally we prove the bootstrap CI consistency in Theorem 4.1.

\begin{theorem}[OR estimator]\label{thm:or.an}
    Suppose Assumptions 2 and 3 
    hold and $\ns\asymp\nt$.
    Also suppose
$\rho_a(\v,\bta_a)$ is twice differentiable with respect to $\bta_a$ and $\wh\bta_a$ is an asymptotically linear estimate of $\bta_a$ with some influence function $\g_a(\O,\bta_a)$; i.e., $\sqrt n(\wh\bta_a - \bta_a) = \sqrtn\g(\O_i,\bta_a) + o_p(1)$.
If $\theta_a(\Gamma_a) \in \Theta$ where $\Theta$ is open and compact,
then $\wh\theta\rega(\Gamma_a)\to_p\theta_a(\Gamma_a)$ and $\wh\theta\rega(\Gamma_a)$ is asymptotically linear with influence function
    \begin{align*}
       \psi_a(\O_i,\theta_a(\Gamma_a),\bta_a) &=  \dfrac{1-S_i}{\P(S_i=0)} \phi_a(\V_i,\theta_a(\Gamma_a),\bta_a) + \E\left(\partial\phi_a/\partial\bta_a\trans\mid S_i=0\right) \g_a(\O_i,\bta_a).
       \end{align*}
Consequently,
   \begin{align*}
  \sqrt n(\wh\theta\rega - \theta_a) &= \sqrtn\psi_a(\O_i,\theta_a(\Gamma_a),\bta_a) + \opone\to_dN(0,\sigma\rega^2(\Gamma_a)),\text{ where}\\
  \sigma\rega^2(\Gamma_a)&=\E\{\psi^2_a(\O_i,\theta_a(\Gamma_a),\bta_a)\}.
\end{align*}
\end{theorem}
\pf{Theorem \ref{thm:or.an}}
Without loss of generality, we prove the results for $\theta_1(\Gamma_1)$. We suppress the dependence of $\theta_1$ on $\Gamma_1$ for notation simplicity.

Since $\Theta$ is compact and $\rho_1(\v)$ is between zero and one, by \citet[Lemma 2.4]{newey1994large}, we have that
\begin{align*}
    \sup_{\theta_1\in\Theta} \left\lVert \meannt\dfrac{ \Gamma_1 \rho_1(\V_i)}{ \Gamma_1 \rho_1(\V_i) + 1 - \rho_1(\V_i)} - \theta_1\right\rVert =o_p(1).
\end{align*}

In addition, we note that by the asymptotic linearity of $\wh\bta$,
\begin{align}\label{eq:rho.opone}
    \left\lVert \wh\rho_1(\V_i)-\rho_1(\V_i)\right\rVert =
        \partial\rho_1/\partial\bta_1\trans(\bta_1 - \wh\bta_1) + o_p(1)
        =
    o_p(1).
\end{align}

Now we establish consistency by \cite[Theorem 5.9]{vandervaart1998}.
Note that
\begin{align*}
  &\sup_{\theta_1\in\Theta} \left\lVert \meannt\dfrac{ \Gamma_1 \wh\rho_1(\V_i)}{ \Gamma_1 \wh\rho_1(\V_i) + 1 - \wh\rho_1(\V_i)} - \theta_1\right\rVert\\
  \leq & \left\lVert \meannt\dfrac{ \Gamma_1 \wh\rho_1(\V_i)}{ \Gamma_1 \wh\rho_1(\V_i) + 1 - \wh\rho_1(\V_i)} - \meannt\dfrac{ \Gamma_1 \rho_1(\V_i)}{ \Gamma_1 \rho_1(\V_i) + 1 - \rho_1(\V_i)}\right\rVert \\
  &+
    \sup_{\theta_1\in\Theta}\left\lVert\meannt\dfrac{ \Gamma_1 \rho_1(\V_i)}{ \Gamma_1 \rho_1(\V_i) + 1 - \rho_1(\V_i)} - \theta_1\right\rVert\\
  \leq &  \dfrac{ \Gamma_1 }{\min\{1, \Gamma_1 \}}
            \left\lVert\meannt\wh\rho_1(\V_i) - \meannt\rho_1(\V_i)\right\rVert
  + o_p(1),\\
  =& o_p(1),
\end{align*}
where the first inequality follows from triangle inequality, the second inequality follows from the boundedness of $\rho_1(\v)$ and the compactness of the parameter space, and the last inequality follows from (\ref{eq:rho.opone}). By \citet[Theorem 5.9]{vandervaart1998}, $\wh\theta_1$ is consistent for $\theta_1$.

Finally we prove the asymptotic Normality. With Taylor expansion, we have
\begin{align*}
    0 =& \meannt \phi_1(\V_i,\wh\theta_1,\wh\bta_1)\\
    =& \meannt \phi_1(\V_i,\theta_1, \bta_1) + \meannt \dfrac{\partial\phi_1}{\partial\theta_1}(\wh\theta_1 - \theta_1)\\
     &+ \meannt \dfrac{\partial\phi_1}{\partial\bta_1\trans}(\wh\bta_1 - \bta_1) +
        \meannt (\wt\bta_1 - \bta_1)\trans \dfrac{\partial^2\phi_1}{\partial\bta_1\partial\bta_1\trans}(\wt\bta_1 - \bta_1)/2,
\end{align*}
where $\wt\bta_1$ is between $\bta_1$ and $\wh\bta_1$.
Multiplying both sides with $\sqrt n$ and rearranging terms, we have
\begin{align*}
    \sqrt n(\wh\theta_1 - \theta_1) =& \sqrt n \dfrac{1-S_i}{\wh\P(S_i=0)}\phi_1(\V_i,\theta_1,\bta_1) +
    \sqrtn\dfrac{\P(S_i=0)}{\wh\P(S_i=0)} \E\left(\partial\phi_1/\partial\bta_1\trans\mid S_i=0\right) \g_1(\O_i,\bta_1) + \opone.
\end{align*}
Since $\wh\P(S_i=0)=\nt/n$ converges to $\P(S_i=0)$ almost surely, we have
\begin{align*}
    \sqrt n(\wh\theta_1 - \theta_1) =& \sqrt n \dfrac{S_i}{\P(S_i=1)}\phi_1(\V_i,\theta_1,\bta_1) +
    \sqrtn \E\left(\partial\phi_1/\partial\bta_1\trans\mid S_i=0\right) \g_1(\O_i,\bta_1) + \opone.
\end{align*}
The proof is completed.
$\square$


Next we consider the asymptotic properties for the bootstrap estimator.
The resampling procedure during each bootstrap iteration can be viewed as using a weighted sample, where the weights are determined by Multinomial distributions.
Therefore, for a bootstrap quantity, for example $\wh\theta\rega\star(\Gamma_a)$, there are two sources of randomness: the randomness from the observed data and the randomness from the bootstrap weights. To distinguish between them, until the end of this subsection we denote by $\P_{\O}$ the probability measure for the observed data  and $\P_{W}$ the  probability measure for bootstrap weights, and $\P_{\O W}$ the probability measure on the product space (recall that the bootstrap weights are independent of data). Similar rules apply to the notation of expectations: $\E_{\O}$, $\E_{W}$ and $\E_{\O W}$, respectively. A formal treatment of these notations can be found from \cite{cheng2010bootstrap}.

\begin{theorem}[Bootstrap Consistency]\label{thm:bootstrap.an}
    Suppose conditions in Theorem \ref{thm:or.an} as well as conditions (B1) and (B2) hold, then $\wh\theta_a\star(\Gamma_a)\to\theta_a(\Gamma_a)$ in $\P_{\O W}$-probability.
    Suppose additionally conditions (B3) and (B4), then conditional on observations, the bootstrap estimate $\wh\theta\rega\star(\Gamma_a)$ satisfies
    \begin{align*}
     \sqrt n( \wh\theta\rega\star(\Gamma_a) - \wh\theta\rega(\Gamma_a))\mid \{\O_i\}_{i=1}^n
     \to_d  N(0,\E\{ \psi_a^2(\O_i,\theta_a(\Gamma_a),\bta_a)\})\text{ in } \P_{\O}\text{-probability}.
    \end{align*}
\end{theorem}

\pf{Theorem \ref{thm:bootstrap.an}}
We start by proving the consistency, i.e., $\wh\theta_a\star(\Gamma_a)\to\theta_a(\Gamma_a)$ in $\P_{\O W}$-probability. By Lemma 6.1 of \cite{wellner2005empirical}, condition (B2) guarantees that $\bphi_a$ is $\P$-Gilvenko-Cantelli. Together with condition (B1), by the multiplier Gilvenko-Cantelli theorem \citep[3.6.16]{vaart1996weak},
\begin{align*}
    \sup_{\bb_a(\Gamma_a)\in\Xi}\left|\meann W_{n,i} \bphi_a(\theta_a(\Gamma_a),\bta_a)-\P_{\O}\bphi_a(\theta_a(\Gamma_a),\bta_a)\right|\to0\text{ in }\E_{\O W}\text{ probability}.
\end{align*}
Then the consistency for $\wh\theta\star\rega(\Gamma_a)$ follows from Corollary 3.2.3 of \cite{vaart1996weak}.

Next, to prove the asymptotic Normality, it's sufficient to show
\begin{align*}\label{eq:pf.boot.eq1}
    \sqrt n(\wh\bb\star_a(\Gamma_a)-\wh\bb_a(\Gamma_a))\mid \{\O_i\}_{i=1}^n \to_d N\left(0,\bS_a(\Gamma_a)\right),
\end{align*}
in $\P_{\O}$-probability,
where
\begin{align*}
\bS_a(\Gamma_a) = \E_{\O}\left\{\dfrac{\partial\bphi_a(\O,\bb_a(\Gamma_a))}{\partial\bb_a(\Gamma_a)}\right\}\inv
\E_{\O}\{\bphi_a(\O,\bb_a(\Gamma_a))\bphi_a(\O,\bb_a(\Gamma_a))\trans\}
\left[\E_{\O} \left\{\dfrac{\partial\bphi_a(\O,\bb_a(\Gamma_a))}{\partial\bb_a(\Gamma_a)}\right\}\inv\right]\trans.
\end{align*}
From there, the asymptotic Normality of $\wh\theta\rega\star(\Gamma_a)$ follows from Delta Method.

To show \eqref{eq:pf.boot.eq1}, we follow \citet{wellner1996bootstrapping} or \citet{cheng2010bootstrap}. In particular, the asymptotic Normality in \eqref{eq:pf.boot.eq1} holds under regularity conditions (B1) to (B4) and additional conditions (W1) to (W3) on the bootstrap weights:
\begin{itemize}
    \item[(W1)] $\int_{0}^\infty \{\P_{W}(|W_{ni}|>t)\}^{1/2}\dd t\leq C <\infty$ for some constant $C$.
    \item[(W2)] $\lim_{\lambda\to\infty}\lim\sup_{n\to\infty}\sup_{t\geq\lambda}t^2\P_{W}(W_{ni}\geq t)=0$.
    \item[(W3)] $\sumn(W_{ni}-1)^2/n\to c$ for some constant $c$.
\end{itemize}
We are left to verify (W1)-(W3),
which can be implied from conditions (W1')-(W3') by Lemma 3.1 of \cite{praestgaard1993exchangeably}.
\begin{itemize}
    \item[(W1')] $\lim\sup_{n\to\infty}\E_{W}(W_{n,i}^4)<\infty$.
    \item[(W2')] There exists a constant $c$ such that  $\E_{W}(W_{ni}^2)\to 1+ c^2$.
    \item[(W3')] $\cov_{W}(W_{n,i}^2,W_{n,j}^2)\leq 0$, $i\neq j$.
\end{itemize}
Finally we verity (W1')-(W3').
Let $n^{(k)}=n(n-1)\cdots (n-k+1)$ for integer $k$. Without loss of generality suppose $i,j\in\Is$.
\begin{align*}
    \E_{W}(W_{n,i}^2) =& 2-1/\ns \to 2,\\
    \E_{W}(W_{n,i}^4) =& 1 + 7\ns^{(2)}/\ns^2  +
                    6\ns^{(3)}/\ns^3 +\ns^{(4)}/\ns^4  \leq 15,\\
    \cov_{W}(W_{ni}^2,W_{nj}^2) =&  \dfrac{1}{\ns^4}\left[
                    \left\{\ns^{(4)} - \left(\ns^{(2)}\right)^2\right\} +
                   2\ns\left\{\ns^{(3)}-\ns\cdot\ns^{(2)}\right\} +
                   \ns^2\left\{\ns^{(2)}-\ns^2\right\}
                    \right]\\
                    \leq&0.
\end{align*}
Hence, (W1')-(W3') are satisfied.
$\square$

Now we are ready to prove the confidence interval consistency result in  Theorem 4.1.
This proof resembles the classic proofs for bootstrap CI consistency \citep{shao1995jackknife,vandervaart1998}.

\pf{Theorem 4.1}
The consistency of $\wh\theta\rega(\Gamma_a)$ has been proven in Theorem \ref{thm:or.an}. Here we prove the bootstrap confidence interval consistency.

Let $\Psi_a$ be the cumulative distribution function (c.d.f.) of $N(0, \sigma\rega^2(\Gamma_a))$. Let $\wh\Psi_a$ and $\wh\Psi\star_a$ be the empirical  distribution functions of $\sqrt n(\wh\theta\rega(\Gamma_a)-\theta_a(\Gamma_a))$ and $\sqrt n(\wh\theta\rega\star(\Gamma_a) - \wh\theta\rega(\Gamma_a))$, respectively. Then $\wh\Psi_a\to_d \Psi_a$ by Theorem \ref{thm:or.an} and $\wh\Psi\star_a\mid\{\O_i\}_{i=1}^n\to_d\Psi_a$ in $\P_{\O}$-probability by Theorem \ref{thm:bootstrap.an}. For the latter, there exists a subsequence that converges almost surely. For simplicity we assume the whole sequence converges almost surely; similar arguments have been made in Lemma 23.3 of \cite{vandervaart1998} and \cite{cheng2010bootstrap}. Applying the quantile convergence theorem \citep[Lemma 21.2]{vandervaart1998} onto the random distribution functions $\wh\Psi\star_a$, we have $(\wh\Psi\star)\inv _a(\tau)$ converges to $\Psi_a\inv(\tau)$ almost surely for any $\tau\in(0,1)$. By Slutsky's theorem,
\begin{align*}
    \sqrt n(\wh\theta\rega(\Gamma_a) - \theta_a(\Gamma_a)) - (\wh\Psi\star)\inv(\alpha/2) \to_d N(0, \sigma^2\rega(\Gamma_a)) - \Psi\inv(\alpha/2).
\end{align*}

Further noting $\sqrt n\left(\wh L_a(\Gamma_a)-\wh\theta_a(\Gamma_a)\right) = (\wh\Psi\star)\inv(\alpha/2)$, we have
\begin{align}
    \P\left(\wh L_a(\Gamma_a) \leq \theta_a(\Gamma_a)\right)
    =& \P\left(
    \sqrt n \{\wh L_a(\Gamma_a)-\wh\theta\rega(\Gamma_a)\}
    \leq \sqrt n (\theta_a(\Gamma_a) - \wh\theta\rega(\Gamma_a))
    \right)\\
    =& \P\left((\wh\Psi\star)\inv(\alpha/2) \leq \sqrt n\{\theta_a(\Gamma_a)-\wh\theta\rega(\Gamma_a)\}\right)\\
    =& \P \left(
    \sqrt n\{\theta_a-\wh\theta\rega(\Gamma_a)\} \leq -(\wh\Psi\star)\inv(\alpha/2)
    \right)\\
    \to& 1-\alpha/2\text{ as }n\to\infty.
\end{align}
The proof of $\P\left(\wh U_a(\Gamma_a)\geq \theta_a(\Gamma_a)\right)\to 1-\alpha/2$ follows similarly and is therefore omitted. The confidence interval consistency follows. $\square$

\section{Details and Proofs for the EIF-Based Estimation}\label{supp.sec:eif}
In this section we provide details and proofs for the EIF-based estimator $\wh\theta\eifa(\Gamma_a)$ proposed in Section 4.2 of the main text.
\subsection{Implementation Details}
For the EIF-based estimator $\wh\theta\eifa\supk(\Gamma_a)$, we remark that the second sum may be replaced by a counterpart that does not use sample splitting, i.e., $\dfrac{1}{|\calI_{t}|}\sum_{i\in\calI_{t}}  \dfrac{ \Gamma_a  \wh\rho_a(\V_i) }{ \Gamma_a \wh\rho_a(\V_i) + 1-\wh\rho_a(\V_i)}$, and the resulting estimator will have the same asymptotic distribution as $\wh\theta\supk\eifa$. For simplicity of presentation and implementation, we focus our attention on $\wh\theta\supk\eifa$ throughout as the first sum in the EIF estimator requires sample splitting.

A step-by-step implementation of the EIF-based estimation is provided in Algorithm \ref{alg:eif}.
\begin{algorithm}[!h]
\caption{EIF-Based Estimation with Cross-Fitting}\label{alg:eif}
    \begin{algorithmic}[1]
\Require $\Is$, $\It$; integer $K\geq 2$; sensitivity parameters $\Gamma_1,\Gamma_0$; confidence level $(1-\alpha)$.
\State \textbf{Step 1 (Partitioning)}: Randomly split $\Is$ and $\It$ to $\calI_{s,k}$ and  $\calI_{t,k}$, respectively, $1\leq k\leq K$.
\State \textbf{Step 2 (Cross Fitting)}:
     \For{$k$ in $1,2,\cdots K$}
    \State With $\{\Is\backslash\calI_{s,k}\}\cup\{\It\backslash\calI_{t,k}\}$, obtain $\wh\pi^{(k)}(\X_i)$, $\wh w^{(k)}(\V_i)$, $\wh\mu^{(k)}_a(\X_i)$ and  $\wh\rho^{(k)}(\V_i)$. 
    \State With $\calI_{s,k}\cup\calI_{t,k}$, calculate
$\wh\theta\eif^{(k)}(\Gamma_0,\Gamma_1) = \wh\theta\eifone^{(k)}(\Gamma_1) - \wh\theta\eifzero^{(k)}(\Gamma_0)$
        with $\wh\theta\eifa^{(k)}(\Gamma_a)$ in Section 4.2.
    \EndFor
\State \textbf{Step 3 (Building Estimator)}: Construct the estimator
        $\wh\theta\eif(\Gamma_0,\Gamma_1) = \meanK\wh\theta\eif^{(k)}(\Gamma_0,\Gamma_1)$.
\State\textbf{Step 4 (Variance Estimation)}: Construct the variance estimator
    \begin{align*}
       \wh \sigma^2\eif(\Gamma_0,\Gamma_1) = \meanK
       \left[\dfrac{1}{|\calI_k|}\sum_{i\in\calI_k}
       \left\{\wh\EIF^{(k)}(\O_i,\wh\theta\eifone(\Gamma_1))-\wh\EIF^{(k)}(\O_i,\wh\theta\eifzero(\Gamma_0))\right\}^2
       \right].
    \end{align*}
\Ensure The EIF-based estimator $\wh\theta\eif(\Gamma_0,\Gamma_1)$ with a $(1-\alpha)$ confidence interval
\begin{align*}
\left(
\wh\theta\eif(\Gamma_0,\Gamma_1)-z_{\alpha/2}\wh\sigma\eif(\Gamma_0,\Gamma_1)/\sqrt{n},\
\wh\theta\eif(\Gamma_0,\Gamma_1)+z_{1-\alpha/2}\wh\sigma\eif(\Gamma_0,\Gamma_1)/\sqrt{n}
\right),\end{align*}
where $z_{\beta}$ is the $\beta$ quantile of the standard Normal distribution for any $\beta\in(0,1)$ and $\wh\sigma\eif(\Gamma_0,\Gamma_1)=\sqrt{\wh\sigma^2\eif(\Gamma_0,\Gamma_1)}$.
    \end{algorithmic}
\end{algorithm}

\subsection{Estimating the Density Ratio}\label{supp.subsec:wv}

The estimation of the density ratio can proceed in two methods falling into two categories. The first category is to recognize the relationship between $w(\v)$ and $\P(S_i=1\mid\V_i=\v)$ via the Bayes rule, i.e.,
\begin{align}\label{eq:w.selection}
    w(\v) &= \dfrac{\P(S_i=1)}{\P(S_i=0)}\dfrac{\P(S_i=0\mid\V_i=\v)}{\P(S_i=1\mid\V_i=\v)},
\end{align}
and estimate $w(\v)$ by estimating  $\P(S_i=1\mid\V_i=\v)$ with a binary classifier and estimating $\P(S_i=1)$ as $\ns/n$; see \cite{kallus2020role} and \citet{zeng2023efficient}.
For example, when $\calV$ is discrete, one may estimate this probability for any $\v\in\calV$ by calculating the proportion of source samples among all samples with the same covariate:
 \begin{align}\label{eq:wv.est.discrete}
     \wh\P(S_i=1\mid\V_i=\v) = \dfrac{
 \sumn\ind(S_i=1,\V_i=\v)}{\sum_{i=1}^n\ind(\V_i=\v)}, \ \ \wh w(\v)  = \dfrac{\ns}{\nt}\dfrac{\wh\P(S_i=0\mid\V_i=\v)}{\wh\P(S_i=1\mid\V_i=\v)}.
 \end{align}

Equation \eqref{eq:w.selection} also reveals the necessity of having a sufficiently large target sample (i.e., the second part of Assumption 2.2). Intuitively, a substantially small target sample will make  estimation of $\P(S_i=1\mid\V_i=\v)$ challenging due to class imbalance. Also, when $\P(S_i=0)$ is close to zero, $w(\v)$ can be large in magnitude, which will generally increase the bias and variance of the estimated TATE.

The second category is to use principles behind covariate balance to estimate $w(\v)$. Specifically,
$w$ serves as a balancing score between the source and the target population, i.e.,
\begin{align*}
    \E \{f(\V_i)w(\V_i)\mid S_i=1\} = \E\{f(\V_i)\mid S_i=0\}, \text{ any measurable }f.
\end{align*}
\citet{han2021federated} considered this connection to construct an exponential tilting estimator of $w(\V_i)$. Relatedly,
\citet{josey2022calibration,chen2023entropy} used entropy balancing of \citet{hainmueller2012entropy}
 to estimate $w(\V_i)$.

To account for the possible imbalance between the source and target samples (i.e., $\ns$ and $\nt$ may differ a lot) and to enable covariate balancing, we proceed with the entropy balancing method in (12) that falls into the second category. 
The solutions $\wh w_i$ of entropy balancing are characterized in Lemma \ref{lemma:EB}.
\begin{lemma}\label{lemma:EB}
    The solution of (12) 
    is
    $\wh w_i =\exp(\wh\alpha + \wh\bb\trans\V_i)$,
    where $(\wh\alpha,\wh\bb)$ is solution to
    \begin{align}\label{eq:EB.dual}
        \min_{\alpha,\bb}\dfrac{1}{\ns}\sum_{i\in\Is}\exp(\alpha+\bb\trans\V_i)-\alpha-\dfrac{1}{\nt}\sum_{i\in\It}\bb\trans \V_i.
    \end{align}
\end{lemma}

Lemma \ref{lemma:EB} is a special case of Proposition 1 of \cite{chen2023entropy}. The dual problem \eqref{eq:EB.dual} is an unconstrained convex optimization problem and numeric solutions can be  efficiently solved by  algorithms like the Newton-Raphson method.  The implementation is performed using the \texttt{optim} function in R.

\subsection{Extended Remarks on Doubly Robust Estimation}
    We note that  \citet{vansteelandt2007estimation} have proposed a doubly robust estimator in missing data settings under sensitivity model (2). However, their construction involves nuisance functions are \emph{variation dependent}. To illustrate, suppose $\calX=\calV$, and we construct an estimator based on  nuisance functions $\pi(\v)$, $m_a(\v) =\E\{Y_i(a)\mid \V_i=\v,S_i=0 \}$, $h_a(\v) =  \dfrac{\P(S_i=0)}{\P(S_i=1)}\omega(\v)/\E\left[\exp\{\gamma_aY_i(a)\}\mid \V_i=\v,S_i=0\right].$ Suppose these nuisance functions are estimated from an independent sample and their estimates are $\wh\pi(\v)$, $\wh m_a(\v)$, and $\wh h_a(\v)$. Then the following estimator $\wh\theta_{\textnormal{DR},a}$ is doubly robust with respect to (i) the pair $\pi(\v)$ and $m_a(\x)$, and (ii) the pair $h_a(\x)$ and $m_a(\x)$,
    \begin{align*}
 \wh\theta_{\textnormal{DR},a}
    =&
   \dfrac{S_i}{\P(S_i=0)}
\exp\{\gamma_a Y_i\}\wh h_a(\X_i)
 \ind(A_i=a)  \left[\dfrac{\ind(A_i=1)}{\wh \pi(\X_i)} + \dfrac{\ind(A_i=0)}{1-\wh\pi(\X_i)} \right]
    \left\{Y_i-\wh m_a(\X_i)\right\} \\
    &+
    \dfrac{1-S_i}{\P(S_i=0)} \wh m_a(\X_i),
\end{align*}
where we let $\gamma_a=\log(\Gamma_a)$.
    However, part (ii) of the double robustness involves variation dependent nuisance functions, $h_a(\X_i)$ and $m_a(\X_i)$, in the sense that constraints on one function can place constraints on the other. To see that, notice that for the binary outcome $Y_i(a)$, these two nuisance functions can be expressed as
    \begin{align*}
        m_a(\X_i) = \dfrac{\Gamma_a\mu_a(\X_i)}{\Gamma_a\mu_a(\X_i) + 1-\mu_a(\X_i)}, \quad
        h_a(\X_i) =  \dfrac{\pr(S_i=0)}{\pr(S_i=1)}\dfrac{w(\X_i)}{\Gamma_a\mu_a(\X_i) + 1-\mu_a(\X_i)}.
    \end{align*}
    Since both functions depend on the outcome regression functional $\mu_a(\x)$ (which equals to $\rho_a(\v)$), they are not variation independent. In other words, a slow convergence rate for one nuisance function may imply a slow convergence rate for the other, thereby limiting the degree of robustness in practice. In general, to the best of our knowledge, there is no doubly robust estimator of the average treatment effect  with variation independent nuisance functions under sensitivity model (2).

\subsection{Proof of Theorem 4.2 and Theorem \ref{supp.thm:eif}}
\label{subsec:eif.derivation}

We prove Theorem \ref{supp.thm:eif}, the EIF for a general outcome under sensitivity model (\ref{eq:tilt.general}). It includes Theorem 4.2
as a special case for a binary outcome.
To simplify notation, we suppress the dependence of the TATE on $\Gamma_a$ and denote the expected potential outcome on the target population at treatment level $a$ as $\theta_a$ for $a=0,1$. We also drop the subscript $i$ and denote by $\O$ a generic random variable, which consists of $(\X,Y,S=1)$ for the source and $(\V,S=0)$ for the target.
 We recall that we have defined $\gamma_a=\log(\Gamma_a)$.

We start with the case where $\pi(\X)$ is unknown and therefore considered as a nuisance parameter. For clarify we denote its true value as $\pi_0(\X)$. Denote by $p_{\V\mid S=1}$, $p_{\X\mid\V,S=1}$, $p_{Y\mid\X,A,S=1}$  the density functions of the conditional distributions of $\V\mid S$, $\X\mid\V,S=1$ and $Y\mid\X,A,S=1$, respectively.
For a generic observation $\O$, the log-likelihood can be written as
\begin{align*}
    l(\O) =& (1-S)\log\left( p_{\V\mid S=0}(\V\mid S=0)\right) +
    S\log\left( p_{\V\mid S=1}(\V\mid S=1) \right)\\
    &+S\log \left(p_{\X\mid\V,S=1} (\X\mid\V,S=1) \right)+
    AS\log\left(\pi(\X)\right) + S(1-A)\log(1-\pi(\X)) \\
    &+
    SA\log \left(p_{Y\mid\X,A=1,S=1}(Y\mid\X,A=1,S=1)\right) +
    S(1-A)\log \left(p_{Y\mid\X,A=0,S=1}(Y\mid\X,A=0,S=1)\right).
\end{align*}
Consider the Hilbert space $\Lambda$ that contains all one-dimensional zero-mean measurable functions of the observed data with finite variance. Consider $p_{Y\mid\X,A=0,S=1}$, $p_{Y\mid\X,A=1,S=1}$, $\pi(\X)$, $p_{\X\mid\V,S=1}$, $p_{\V\mid S=0}$ and $p_{\V\mid S=1}$ as nuisance functions and denote their nuisance tangent spaces as $\Lambda_{Y\mid\X,A=1,S=1}$, $\Lambda_{Y\mid\X,A=0,S=1}$, $\Lambda_{\pi}$, $\Lambda_{\X\mid S=1}$, $\Lambda_{\V\mid S=1}$ and $\Lambda_{\V\mid S=0}$, respectively. Then
\begin{align*}
    \Lambda = \Lambda_{Y\mid\X,A=1,S=1}\oplus \Lambda_{Y\mid\X,A=0,S=1} \oplus \Lambda_{\pi}\oplus \Lambda_{\X\mid S=1} \oplus \Lambda_{\V\mid S=1} \oplus \Lambda_{\V\mid S=0},
\end{align*}
where $\oplus$ is the direct sum between orthogonal spaces, and
\begin{align*}
    \Lambda_{Y\mid\X,A=1,S=1} &= \left\{SA\b_1(Y,\X): \E\left[\b_1(Y,\X)\mid\X,A=1,S=1\right]=\0\right\},\\
    \Lambda_{Y\mid\X,A=0,S=1} &= \left\{S (1-A)\b_2(Y,\X): \E\left[\b_2(Y,\X)\mid\X,A=0,S=1\right]=\0\right\},\\
    \Lambda_{\pi} &= \left\{S[A-\pi_0(\x)]\b_3(\X): 0<\pi_0(\X)<1\right\},\\
    \Lambda_{\X\mid S=1} &= \left\{S\b_4(\X):\E\left[\b_4(\X)\mid\V,S=1\right]=\0\right\},\\
    \Lambda_{\V\mid S=1} &= \left\{S\b_5(\V):\E\left[\b_5(\V)\mid S=1\right]=\0\right\},\\
     \Lambda_{\V\mid S=0} &= \left\{(1-S)\b_6(\V):\E\left[\b_6(\V)\mid S=0\right]=\0\right\}.\\
\end{align*}
Without loss of generality, we derive the EIF for $\theta_1$. The EIF for $\theta_0$ is analogous and thus omitted for brevity. Consider parametric submodels indexed by parameter $\ba$ where $\ba=\0$ represents the true data generating process.
We re-express the log-likelihood under the parametric submodel,
\begin{align*}
    l(\O,\ba) =& (1-S)\log p_{\V\mid S=0}(\V\mid S=0;\ba) + S\log p_{\V\mid S=1}(\V\mid S=1;\ba) \\
    &+S\log p_{\X\mid\V,S=1} (\X\mid\V,S=1;\ba) + AS\log\pi(\x;\ba) + S(1-A)\log(1-\pi(\X;\ba)) \\
    &+
    SA\log p_{Y\mid\X,A=1,S=1}(Y\mid\X,A=1,S=1;\ba)\\
    &+
    S(1-A)\log p_{Y\mid\X,A=0,S=1}(Y\mid\X,A=0,S=1;\ba).
\end{align*}
Define the score function
\begin{align*}
    \S(\O) =& \dfrac{\partial l(\O,\ba)}{\partial\ba}\Bigg\vert_{\ba=\0}\\
    =&SA\S_1(Y,\X) + S(1-A)\S_2(Y,\X)+
    S\dfrac{\partial\left[\{ A\log(\pi(\X;\ba)) + (1-A)\log(1-\pi(\X;\ba))\right] }{\partial\ba}\Bigg\vert_{\ba=\0}\\
   & +S\S_4(\X) + S\S_5(\V)+(1-S)\S_6(\V),\text{ where}\\
    \S_1(Y,\X)=& \dfrac{\partial\log p_{Y\mid\X,A=1,S=1}(Y\mid\X,A=1,S=1;\ba)}{\partial\ba}\Bigg\vert_{\ba=\0},\\
    \S_2(Y,\X) =& \dfrac{\partial\log p_{Y\mid\X,A=0,S=1}(Y\mid\X,A=0,S=1;\ba)}{\partial\ba}\Bigg\vert_{\ba=\0},\\
    \S_4(\X) =& \dfrac{\partial\log p_{\X\mid\V,S=1}(\X\mid\V,S=1;\ba)}{\partial\ba}\Bigg\vert_{\ba=\0},\\
    \S_5(\V) =& \dfrac{\partial\log p_{\V\mid S=1}(\V\mid S=1;\ba)}{\partial\ba}\Bigg\vert_{\ba=\0},\\
    \S_6(\V) =&  \dfrac{\partial\log p_{\V\mid S=0}(\V\mid S=0;\ba)}{\partial\ba}\Bigg\vert_{\ba=\0},
\end{align*}
and $SA\S_1(Y,\X) \in\Lambda_{Y\mid\X,A=1,S=1}$, $S(1-A)\S_2(Y,\X)\in\Lambda_{Y\mid\X,A=0,S=1}$, $S\S_4(\X)\in\Lambda_{\X\mid S=1}$, $S\S_5(\V)\in\Lambda_{\V\mid S=1}$, $(1-S)\S_6(\V)\in\Lambda_{\V\mid S=0}$.

Next, we show that
\begin{align}\label{eq:proof.eif.unknownT.wts}
    \E\left[\phi\cont_1(\O,\theta_1)\S(\O)\right] &= \dfrac{\partial\theta_1}{\partial\ba}\Bigg\vert_{\ba=\0},
\end{align}
where
{\footnotesize{
\begin{align*}
    &\phi\cont_1(\O,\theta_1(\gamma_1)) \\
        =& \dfrac{Sw(\V)}{\P(S=1)\pi(\X) }\left[
        \dfrac{\exp(\gamma_1 Y)Y}{\E\{\exp(\gamma_1 \Yone)\mid\V,S=1\}}
                - \dfrac{\E\{\exp(\gamma_1Y)Y\mid\X,A=1,S=1\}}{\E\{\exp(\gamma_1\Yone)\mid\V,S=1\}} \right.\\
     &\left.
        - \dfrac{\exp(\gamma_1Y)\E\{\exp(\gamma_1\Yone)\Yone\mid\V,S=1\}}{[\E\{\exp(\gamma_1\Yone)\mid\V,S=1\}]^2}
         + \dfrac{\E\{\exp(\gamma_1Y)\mid\X,A=1,S=1\}\E\{\exp(\gamma_1\Yone)\Yone\mid\V,S=1\}}{[\E\{\exp(\gamma_1\Yone)\mid\V,S=1\}]^2}
        \right]\\
    & +\dfrac{Sw(\V)}{\P(S=1)}\dfrac{
    \E\{e^{\gamma_1Y}Y\mid\X,A=1,S=1\}\E\{e^{\gamma_1\Yone}\mid\V,S=1\} -
    \E\{e^{\gamma_1\Yone}\Yone\mid\V,S=1\}\E\{e^{\gamma_1Y}\mid\X,A=1,S=1\}
    }{[\E\{\exp(\gamma_1\Yone)\mid\V,S=1\}]^2}\\
    & + \dfrac{1-S}{\P(S=0)}
    \left[\dfrac{\E\left\{\exp(\gamma_1\Yone)\Yone\mid\V,S=1\right\}}
    {\E\left\{\exp(\gamma_1\Yone)\mid\V,S=1\right\}} - \theta_1\right].
\end{align*}
}}

To show (\ref{eq:proof.eif.unknownT.wts}), we calculate its right-hand side:
\begin{align}
    \dfrac{\partial\theta_1}{\partial\ba}\Bigg\vert_{\ba=\0}
    =& \E \left(w(\V)\E \left[ \E\{B_1(\Yone,\X)\S_1(Y,\X)\mid\X,A=1,S=1\}\mid\V,S=1\right]\mid S=1\right) \label{eq:proof.eif.unknownT.rhs1} \\
    &+ \E\left[\E\left\{
    w(\V)B_4(\X)\S_4(\X)
    \mid\V,S=1\right\}\mid S=1\right] \label{eq:proof.eif.unknownT.rhs2} \\
    &+\E\left\{
    \E(\Yone\S_6(\V)\mid S=0)
    \right\} ,  \label{eq:proof.eif.unknownT.rhs3}
\end{align}
where
\begin{align*}
     B_1(\Yone,\X) =& \dfrac{e^{\gamma_1\Yone}\Yone}{\E\{e^{\gamma_1\Yone}\mid\V,S=1\}}
     - \dfrac{e^{\gamma_1\Yone}\E\{e^{\gamma_1\Yone}\Yone\mid\V,S=1\}}{[\E\{e^{\gamma_1\Yone}\mid\V,S=1\}]^2},\\
      B_4(\X) =& \dfrac{
    \E\{e^{\gamma_1 Y}Y\mid\X,A=1,S=1\}\E\{e^{\gamma_1\Yone}\mid\V,S=1\}}{[\E\{e^{\gamma_1\Yone}\mid\V,S=1\}]^2}\\
    &-\dfrac{\E\{e^{\gamma_1\Yone}\Yone\mid\V,S=1\}\E\{e^{\gamma_1 Y}\mid\X,A=1,S=1\}}{[\E\{e^{\gamma_1\Yone}\mid\V,S=1\}]^2}.
\end{align*}

Further, note that
\begin{align*}
    (\ref{eq:proof.eif.unknownT.rhs1}) &=\E \left(w(\V)\E \left[ \E\{B_1(\Yone,\X)\S_1(Y,\X)\mid\X,A=1,S=1\}\mid\V,S=1\right]\mid S=1\right)\\
    &= \E \left(\dfrac{SAw(\V)}{\P(S=1) \pi(\X)}
    \left[B_1(\Yone,\X)-\E\{B_1(\Yone,\X)\mid\X,A=1,S=1\}\right]
    \S(\O)
    \right),\\
    (\ref{eq:proof.eif.unknownT.rhs2}) &=
    \E\left(
    \dfrac{S}{\P(S=1)} \left\{B_4(\X) - \E[B_4(\X)\mid\V,S=1]\right\}
    \S(\O)
    \right),\\
    (\ref{eq:proof.eif.unknownT.rhs3}) &=  \E\left\{ \dfrac{1-S}{\P(S=0)} \left[
    \E(\Yone\mid\V,S=0)-\theta_1
    \right]
    \S(\O)\right\}\\
    &=
    \E\left\{ \dfrac{1-S}{\P(S=0)} \left[
    \dfrac{\E\left\{\exp(\gamma_1\Yone)\Yone\mid\v,S=1\right\}}
    {\E\left\{\exp(\gamma_1\Yone)\mid\v,S=1\right\}} - \theta_1
    \right]
    \S(\O)\right\},
\end{align*}
we have
\begin{align*}
   \dfrac{\partial\theta_1}{\partial\ba}\Bigg\vert_{\ba=\0}  =   (\ref{eq:proof.eif.unknownT.rhs1}) + (\ref{eq:proof.eif.unknownT.rhs2}) + (\ref{eq:proof.eif.unknownT.rhs3})
   = \E\left[\phi_1\cont(\O,\theta_1)\S(\O)\right].
\end{align*}
Finally, we verify that $\phi\cont_1(\O,\theta_1)\in\Lambda$ since
\begin{align*}
   & \dfrac{SAw(\V)}{\P(S=1) \pi(\X)}
    \left[B_1(\yone,\x)-\E\{B_1(\Yone,\X)\mid\x,A=1,S=1\}\right] \in \Lambda_{Y\mid\X,A=1,S=1},\\
& \dfrac{S}{\P(S=1)} \left\{B_4(\X) - \E[B_4(\X)\mid\v,S=1]\right\} \in\Lambda_{\X\mid S=1},\text{ and}\\
& \dfrac{1-S}{\P(S=0)} \left[
    \dfrac{\E\left\{\exp(\gamma_1\Yone)\Yone\mid\V,S=1\right\}}
    {\E\left\{\exp(\gamma_1\Yone)\mid\V,S=1\right\}} - \theta_1
    \right] \in\Lambda_{\V\mid S=0}.
\end{align*}
Therefore, $\phi\cont_1(\O,\theta_1)$ is the EIF  in Theorem \ref{supp.thm:eif}, i.e., $\EIF\cont_1(\O,\theta_1)$.
Moreover, if the outcome is binary,  we can re-express the followings:
\begin{align*}
    \E\{\exp(\gamma_1Y)Y\mid\X,A=1,S=1\} =&  \Gamma_1 \mu_1(\X),\\
    \E\{\exp(\gamma_1Y)\mid\X,A=1,S=1\} =&  \Gamma_1 \mu_1(\X) + 1-\mu_1(\X),\\
    \E\{\exp(\gamma_1\Yone)\Yone\mid\V,S=1\} =&  \Gamma_1 \rho_1(\V),\\
    \E\{\exp(\gamma_1\Yone)\mid\V,S=1\} =&  \Gamma_1 \rho_1(\V) + 1-\rho_1(\V).
\end{align*}
Plugging in them to $\EIF\cont(\O,\theta_1)$  yields $\EIF(\O,\theta_1)$ as the expression of the EIF for a binary outcome in Theorem 4.2. 

Next, we suppose $\pi(\X)$ is known  as its true value $\pi_0(\X)$. Then $\pi(\X)$ is no longer considered as a nuisance function and the Hilbert space $\Lambda$ can now be decomposed as
\begin{align*}
    \Lambda = \Lambda_{Y\mid\X,A=1,S=1}\oplus\Lambda_{Y\mid\X,A=0,S=1}\oplus\Lambda_{\X\mid S=0}\oplus\Lambda_{\V\mid S=1} \oplus\Lambda_{\V\mid S=0}.
\end{align*}
Under the parametric submodel, the log-likelihood becomes
\begin{align*}
    l(\O,\ba) =& (1-S)\log p_{\V\mid S=0}(\V\mid S=0;\ba) + S\log p_{\V\mid S=1}(\V\mid S=1;\ba) \\
    &+S\log p_{\X\mid\V,S=1} (\X\mid\V,S=1;\ba) + AS\log\pi_0(\X) + S(1-A)\log(1-\pi_0(\X)) \\
    &+
    SA\log p_{Y\mid\X,A=1,S=1}(Y\mid\X,A=1,S=1;\ba)\\
    &+
    S(1-A)\log p_{Y\mid\X,A=0,S=1}(Y\mid\X,A=0,S=1;\ba).
\end{align*}

Then the score function becomes
\begin{align*}
    \S(\O) =& \dfrac{\partial l(\O,\ba)}{\partial\ba}\Bigg\vert_{\ba=\0}\\
    =&SA\S_1(Y,\X) + S(1-A)\S_2(Y,\X)+S\S_4(\X) + S\S_5(\V)+(1-S)\S_6(\V),
\end{align*}
where we still have $SA\S_1(Y,\X) \in\Lambda_{Y\mid\X,A=1,S=1}$, $S(1-A)\S_2(Y,\X)\in\Lambda_{Y\mid\X,A=0,S=1}$, $S\S_4(\X)\in\Lambda_{\X\mid S=1}$, $S\S_5(\V)\in\Lambda_{\V\mid S=1}$, $(1-S)\S_6(\V)\in\Lambda_{\V\mid S=0}$.
Therefore, $\E\left[\phi\cont_1(\O,\theta_1)\S(\O)\right] = \dfrac{\partial\theta_1}{\partial\ba}\Bigg\vert_{\ba=\0}$  holds following the same argument as we've shown.

\subsection{Lemma \ref{lemma:eif.plugin}}
In this section we characterize the plug-in bias for the EIF-based  estimator $\wh\theta\eifa(\Gamma_a)$. For the generality of the conclusion and to avoid overloading the notation, we assume the nuisance functions are estimated from an independent sample.
We introduce the general notation for the uncentered EIF,
 \begin{align*}
 &\varphi_a(\O_i) \\
 =& \dfrac{S_iw(\V_i)}{\P(S_i=1)}
    \dfrac{ \Gamma_a }{[\Gamma_a\rho_a(\V_i) + 1-\rho_a(\V_i)]^2}
        \left[\left\{\dfrac{A_i}{\pi(\X_i)} + \dfrac{1-A_i}{1-\pi(\X_i)}\right\} \{Y_i-\mu_a(\X_i)\} +
        \mu_a(\X_i)-\rho_a(\V_i)\right]\\
&+ \dfrac{1-S_i}{\P(S_i=0)}
        \dfrac{ \Gamma_a  \rho_a(\V_i) }{ \Gamma_a \rho_a(\V_i) + 1-\rho_a(\V_i)},
\end{align*}
  and its estimate
\begin{align*}
& \wh\varphi_a(\O_i) \\
=& \dfrac{S_i\wh w(\V_i)}{\wh\P(S_i=1)}
    \dfrac{ \Gamma_a }{[ \Gamma_a \wh\rho_a(\V_i) + 1-\wh\rho_a(\V_i)]^2}
        \left[\left\{\dfrac{A_i}{\wh\pi(\X_i)} + \dfrac{1-A_i}{1-\wh\pi(\X_i)}\right\} \{Y_i-\wh\mu_a(\X_i)\} +
       \wh \mu_a(\X_i)-\wh\rho_a(\V_i)\right]\\
&+ \dfrac{1-S_i}{\wh\P(S_i=0)}
        \dfrac{ \Gamma_a  \wh\rho_a(\V_i) }{ \Gamma_a \wh\rho_a(\V_i) + 1-\wh\rho_a(\V_i)},
\end{align*}
where $\wh \P(S_i=1)=n_s/n$, $\wh\mu_a(\X_i)$, $\wh\rho_a(\V_i)$, $\wh\pi(\X_i)$ and $\wh w(\V_i)$ are estimated from an independent sample.

\begin{lemma}\label{lemma:eif.plugin}
There exists a  constant $C$ such that
\begin{align*}
    |\E\{\wh\varphi_a(\O_i)-\varphi_a(\O_i)\}|
    \leq&
    C\left(\lVert\wh\mu_a(\X_i)-\mu_a(\X_i)\rVert\cdot
    \lVert\wh\pi(\X_i)-\pi(\X_i)\rVert +
    \lVert\wh \rho_a(\V_i)-\rho_a(\V_i)\rVert\cdot
    \lVert\wh w(\V_i)-w(\V_i)\rVert\right. \\
    &+\left.
    \lVert\wh \rho_a(\V_i)-\rho_a(\V_i)\rVert^2\right).
\end{align*}
In particular, if $\Gamma_1=\gamma_0=1$, there exists a constant $C$ such that
\begin{align*}
    |\E\{\wh\varphi_a(\O_i)-\varphi_a(\O_i)\}|
    \leq& C\left(\lVert\wh\mu_a(\X_i)-\mu_a(\X_i)\rVert\cdot
    \lVert\wh\pi(\X_i)-\pi(\X_i)\rVert +
    \lVert\wh \rho_a(\V_i)-\rho_a(\V_i)\rVert\cdot
    \lVert\wh w(\V_i)-w(\V_i)\rVert\right).
\end{align*}
\end{lemma}

\noindent\textit{Proof of Lemma \ref{lemma:eif.plugin}:}
Without loss of generality, we prove the case for $a=1$.
\begin{align*}
&\E\{\wh\varphi_1(\O_i)-\varphi_1(\O_i)\} \\
=& \E\{\wh\varphi_1(\O_i)\}-\theta_1\\
=&\E\{\wh\varphi_1(\O_i)\} -\E\left[\dfrac{1-S_i}{\P(S_i=0)}
        \dfrac{ \Gamma_1  \rho_1(\V_i) }{ \Gamma_1 \rho_1(\V_i) + 1-\rho_1(\V_i)}
        \right]\\
=& \E  \left[\dfrac{S_i\wh w(\V_i)}{\wh\P(S_i=1)}
    \dfrac{ \Gamma_1 }{[ \Gamma_1 \wh\rho_1(\V_i) + 1-\wh\rho_1(\V_i)]^2}
        \dfrac{A}{\wh\pi(\X_i)}\{\mu_1(\X_i)-\wh\mu_1(\X_i)\}\right] \\
&+ \E  \left[\dfrac{S_i\wh w(\V_i)}{\wh\P(S_i=1)}
    \dfrac{ \Gamma_1 }{[ \Gamma_1 \wh\rho_1(\V_i) + 1-\wh\rho_a(\V_i)]^2}\{\wh\mu_1(\X_i)-\mu_1(\X_i)\}\right]\\
&- \E  \left[\dfrac{S_i\wh w(\V_i)}{\wh\P(S_i=1)}
    \dfrac{ \Gamma_1 }{[ \Gamma_1 \wh\rho_1(\V_i) + 1-\wh\rho_1(\V_i)]^2}\{\wh\rho_1(\V_i)-\rho_1(\V_i)\}\right]\\
&+\E\left[ \dfrac{1-S_i}{\wh\P(S_i=0)}
        \dfrac{ \Gamma_1  \wh\rho_1(\V_i) }{ \Gamma_1 \wh\rho_1(\V_i) + 1-\wh\rho_1(\V_i)}-
        \dfrac{1-S_i}{\P(S_i=0)}
        \dfrac{ \Gamma_1  \rho_1(\V_i) }{ \Gamma_1 \rho_1(\V_i) + 1-\rho_1(\V_i)}
        \right]\\
=&
\E  \left[\dfrac{S_i\wh w(\V_i)}{\wh\P(S_i=1)\wh\pi(\X_i)}
    \dfrac{ \Gamma_1 }{[ \Gamma_1 \wh\rho_1(\V_i) + 1-\wh\rho_1(\V_i)]^2}
       \{\wh\pi(\X_i)-\pi(\X_i)\}\{\wh\mu_1(\X_i)-\mu_1(\X_i)\}\right]\\
&       - \E  \left[
    \left\{\dfrac{\P(S_i=1)-\wh\P(S_i=1)}{\wh\P(S_i=1)\P(S_i=1)} +
    \dfrac{1}{\P(S_i=1)}\right\}
    S_i\wh w(\V_i)
    \dfrac{ \Gamma_1 }{[ \Gamma_1 \wh\rho_1(\V_i) + 1-\wh\rho_1(\V_i)]^2}\{\wh\rho_1(\V_i)-\rho_1(\V_i)\}\right]\\
&+
\E\left(
\dfrac{
(1-S_i) \Gamma_1
\left[\wh\rho_1(\V_i) \{ \Gamma_1 \rho_1(\V_i) + 1-\rho_1(\V_i)\}
- \rho_1(\V_i) \{ \Gamma_1 \wh\rho_1(\V_i) + 1-\wh\rho_1(\V_i)\}
\right]
}{\P(S_i=0)\wh\P(S_i=0)
\{ \Gamma_1 \wh\rho_1(\V_i) + 1-\wh\rho_1(\V_i)\}
\{ \Gamma_1 \rho_1(\V_i) + 1-\rho_1(\V_i)\}
}
\right)\\
\leq & O(1)\cdot\E  \left[\{\wh\pi(\X_i)-\pi(\X_i)\}\{\wh\mu_1(\X_i)-\mu_1(\X_i)\}\right] \\
&       - \E  \left[
   \dfrac{S_i\wh w(\V_i)}{\P(S_i=1)}
    \dfrac{ \Gamma_1 }{[ \Gamma_1 \wh\rho_1(\V_i) + 1-\wh\rho_1(\V_i)]^2}\{\wh\rho_1(\V_i)-\rho_1(\V_i)\}\right]\\
&+
\E\left[
\dfrac{
(1-S_i)}{\P(S_i=0)}
\dfrac{ \Gamma_1 }{
\{ \Gamma_1 \wh\rho_1(\V_i) + 1-\wh\rho_1(\V_i)\}
\{ \Gamma_1 \rho_1(\V_i) + 1-\rho_1(\V_i)\}
}\left\{\wh\rho_1(\V_i)
- \rho_1(\V_i) \right\}
\right]
\\
\leq & O(1)\cdot\E  \left[\{\wh\pi(\X_i)-\pi(\X_i)\}\{\wh\mu_1(\X_i)-\mu_1(\X_i)\}\right] \\
&+
\E\left[
\dfrac{
S_i}{\P(S_i=1)}
\dfrac{ \Gamma_1 \{1- \Gamma_1 \}\left\{\wh\rho_1(\V_i)
- \rho_1(\V_i) \right\}
\left\{\wh w(\V_i)
- w(\V_i) \right\}
}{
\{ \Gamma_1 \wh\rho_1(\V_i) + 1-\wh\rho_1(\V_i)\}^2
\{ \Gamma_1 \rho_1(\V_i) + 1-\rho_1(\V_i)\}
}
\right]\\
&+
\E\left[
\dfrac{
Sw(\V_i)}{\P(S_i=1)}
\dfrac{ \Gamma_1 \{1- \Gamma_1 \}\left\{\wh\rho_1(\V_i)
- \rho_1(\V_i) \right\}^2
}{
\{ \Gamma_1 \wh\rho_1(\V_i) + 1-\wh\rho_1(\V_i)\}^2
\{ \Gamma_1 \rho_1(\V_i) + 1-\rho_1(\V_i)\}
}
\right]\\
\leq & O(1)\cdot\E  \left[\{\wh\pi(\X_i)-\pi(\X_i)\}\{\wh\mu_1(\X_i)-\mu_1(\X_i)\}\right]  +
 O(1)\cdot\E  \left[\{\wh w(\V_i)-w(\V_i)\}\{\wh\rho_1(\V_i)-\rho_1(\V_i)\}\right] \\
 &+
 O(1)\cdot\E  \left[\{\wh\rho_1(\V_i)-\rho_1(\V_i)\}^2\right]\\
\leq& O(1)\left\{\lVert\wh\mu_a(\X_i)-\mu_a(\X_i)\rVert\cdot
    \lVert\wh\pi(\X_i)-\pi(\X_i)\rVert +
    \lVert\wh \rho_a(\V_i)-\rho_a(\V_i)\rVert\cdot
    \lVert\wh w(\V_i)-w(\V_i)\rVert\right.\\
    &\left.+
    \lVert\wh \rho_a(\V_i)-\rho_a(\V_i)\rVert^2\right\}
\end{align*}

When $\Gamma_1=0$, following the same procedure and using the fact that  $ \Gamma_1 =1$, we have
\begin{align*}
  & \E\{\wh\varphi_1(\O_i)-\varphi_1(\O_i)\}  \\
 \leq& O(1)\left\{\lVert\wh\mu_a(\X_i)-\mu_a(\X_i)\rVert\cdot
    \lVert\wh\pi(\X_i)-\pi(\X_i)\rVert +
    \lVert\wh \rho_a(\V_i)-\rho_a(\V_i)\rVert\cdot
    \lVert\wh w(\V_i)-w(\V_i)\rVert\right\}.
\end{align*}

\subsection{Proof of Theorem 4.3}\label{pf:prop.dr.asymp}
The EIF-based estimator is $\wh\theta\eif(\Gamma_a)=\wh\theta\eifone-\wh\theta\eifzero$ with $\wh\theta\eifa(\Gamma_a) = \meanK\meanIk\wh\varphi(\O_i)$. Without loss of generality, we consider the proof for $\wh\theta\eifa(\Gamma_a)$ and drop $\Gamma_a$ in notation for simplicity.
We have
\begin{align}
\wh\theta\eifa - \theta_a
    &=\left\{\meanK\meanIk\wh\varphi_a\supk(\O_i) - \theta_a\right\}\\
    &=\left\{\meanK\meanIk\wh\EIF\supk(\O_i,\theta_a)\right\}\\
    \label{eq:proof.eif.est.decomp}
    &=\meann\EIF(\O_i,\theta_a)+
    \left\{
    \meanK\meanIk\wh\EIF\supk(\O_i) -\meann\EIF(\O_i)
    \right\}\\
    &= \meann\EIF(\O_i,\theta_a) +
    \left\{
    \meanK\meanIk \left[\wh\EIF\supk(\O_i,\theta_a) -\EIF(\O_i,\theta_a)\right]
    \right\}.\n
\end{align}
We define
\begin{align*}
    R_k= \meanIk\left\{\wh\EIF\supk(\O_i,\theta_a) -\EIF(\O_i,\theta_a)\right\},\text{ for }k=1,\cdots K.
\end{align*}

\subsubsection{Part (i)}
Since $K$ is independent of data, to show that $\wh\theta\eifa$ is consistent, it suffices to show
\begin{align*}
    R_1=o_p(1).
\end{align*}
From Lemma \ref{lemma:eif.plugin},
\begin{align*}
    \E(R_1)\leq O(1)&\cdot\left\{
    \lVert \wh w^{(k)}(\V_i) - w^{(k)}(\V_i)\rVert\cdot
    \lVert\wh\rho^{(k)}_a(\V_i) - \rho^{(k)}_a(\V_i) \rVert  + \lVert \wh\rho^{(k)}_a(\V_i) - \rho^{(k)}_a(\V_i)\rVert^2\right\}\\
    &+O(1)\cdot \lVert \wh\pi^{(k)}(\X_i)-\pi^{(k)}(\X_i)\rVert
    \cdot
    \lVert \wh\mu_a^{(k)}(\X_i) - \mu_a^{(k)}(\X_i)\rVert\\
    \leq &o_p(1),
\end{align*}
where the second inequality follows from the conditions that  $\lVert \wh\rho^{(k)}_a(\V_i) - \rho^{(k)}_a(\V_i)\rVert = o_p(1)$ and (6). 
Next, we show $R_1-\E(R_1)=o_p(1)$.
Conditioning on $\calI_k^c=\calI\backslash\calI_k$, we calculate the mean and variance for $R_1-\E(R_1)$:
\begin{align*}
    \E\{R_1-\E(R_1)\mid\calI_k^c\} &= \E\left[\wh\EIF\supk(\O_i,\theta_a)-\E\{\wh\phi\supk\eifa(\O_i,\theta_a)\}\mid\calI_k^c\right]-\E\left[\EIF(\O_i,\theta_a)-\E\{\EIF(\O_i,\theta_a)\}\right]\\
    &=0,\\
    \var(R_1-\E(R_1)\mid\calI_k^c) &= \var(R_1\mid\calI_k^c)\leq K\lVert\wh\EIF\supk(\O_i,\theta_a)-\EIF(\O_i,\theta_a)\rVert^2/n.
\end{align*}
Then for any $\varepsilon>0$, by Chebyshev's inequality,
\begin{align*}
    \P\left(
    \dfrac{R_1-\E(R_1)}{\lVert\wh\EIF\supk(\O_i,\theta_a)-\EIF(\O_i,\theta_a)\rVert/\sqrt n}
    \geq\varepsilon   \right)
    &=
    \E\left\{\P\left(
    \dfrac{R_1-\E(R_1)}{K\lVert\wh\EIF\supk(\O_i,\theta_a)-\EIF(\O_i,\theta_a)\rVert/\sqrt n}
    \geq\varepsilon\mid\calI_k^c\right)\right\}\\
    &\leq 1/\varepsilon^2.
\end{align*}
Therefore,
\begin{align*}
    R_1-\E(R_1) &=K O_p(\lVert\wh\EIF\supk(\O_i,\theta_a)-\EIF(\O_i,\theta_a)\rVert)/\sqrt{n}\leq O_p(1/\sqrt n)=o_p(1).
\end{align*}

\subsubsection{Part (ii)}

The decomposition at the beginning of the proof suggests
\begin{align*}
    \sqrt n(\wh\theta\eifa - \theta_a)
    &= \sqrtn\EIF(\O_i,\theta_a) +
    \sqrt n \left\{
    \meanK\meanIk \left[\wh\EIF\supk(\O_i,\theta_a) -\EIF(\O_i,\theta_a)\right]
    \right\}
\end{align*}
Since $K$ is independent of the data,  it suffices to show
\begin{align*}
    R_1=o_p(n^{-1/2}).
\end{align*}
From Lemma \ref{lemma:eif.plugin} and the rate conditions (7a), (7b) and (7c) 
in Theorem 4.3, 
we have \begin{align*}
    \E(R_1)=o_p(n^{-1/2}).
\end{align*}
In what follows we show $R_1-\E(R_1)=o_p(n^{-1/2})$.
Conditioning on $\calI_k^c=\calI\backslash\calI_k$, we calculate the mean and variance for $R_1-\E(R_1)$:
\begin{align*}
    \E\{R_1-\E(R_1)\mid\calI_k^c\} &= \E\left[\wh\EIF\supk(\O_i,\theta_a)-\E\{\wh\phi\supk\eifa(\O_i,\theta_a)\}\mid\calI_k^c\right]-\E\left[\EIF(\O_i,\theta_a)-\E\{\EIF(\O_i,\theta_a)\}\right]\\
    &=0,\\
    \var(R_1-\E(R_1)\mid\calI_k^c) &= \var(R_1\mid\calI_k^c)\leq K\lVert\wh\EIF\supk(\O_i,\theta_a)-\EIF(\O_i,\theta_a)\rVert^2/n.
\end{align*}
Then for any $\varepsilon>0$, by Chebyshev's inequality,
\begin{align*}
    \P\left(
    \dfrac{R_1-\E(R_1)}{\lVert\wh\EIF\supk(\O_i,\theta_a)-\EIF(\O_i,\theta_a)\rVert/\sqrt n}
    \geq\varepsilon   \right)
    &=
    \E\left\{\P\left(
    \dfrac{R_1-\E(R_1)}{K\lVert\wh\EIF\supk(\O_i,\theta_a)-\EIF(\O_i,\theta_a)\rVert/\sqrt n}
    \geq\varepsilon\mid\calI_k^c\right)\right\}\\
    &\leq 1/\varepsilon^2.
\end{align*}
Since all nuisance parameters are consistently estimated by assumption (i.e., $\lVert \wh\rho^{(k)}_a(\V_i) - \rho^{(k)}_a(\V_i)\rVert = o_p(1)$, $\lVert \wh\mu_a^{(k)}(\X_i) - \mu_a^{(k)}(\X_i)\rVert = o_p(1)$, $\lVert \wh w\supk(\V_i) - w\supk(\V_i)\rVert = o_p(1)$, $\lVert \wh\pi^{(k)}(\X_i)-\pi^{(k)}(\X_i)\rVert =o_p(1)$),
Lemma \ref{lemma:eif.plugin} suggests that $\lVert\wh\EIF\supk(\O_i,\theta_a)-\EIF(\O_i,\theta_a)\rVert=o_p(1)$.
Therefore,
\begin{align*}
    R_1-\E(R_1) &=K O_p(\lVert\wh\EIF\supk(\O_i,\theta_a)-\EIF(\O_i,\theta_a)\rVert)/\sqrt{n}=o_p(1/\sqrt n).
\end{align*}


\subsubsection{Part (iii)}
In order to show
\begin{align*}
\wh\sigma^2\eifa(\Gamma_a)-\sigma^2\eifa(\Gamma_a)
    = \meanK\meanIk\wh\EIF^2(\O_i,\wh\theta\eifa(\Gamma_a))-\E\{\EIF^2(\O_i,\theta_a(\Gamma_a))\}=o_p(1),
\end{align*}
it's sufficient to show

\begin{align}
 R_{k,1}-R_{k,2}=  \meanIk\wh\EIF^2(\O_i,\wh\theta\eifa(\Gamma_a))-\E\{\EIF^2(\O_i,\theta_a(\Gamma_a))\}=o_p(1),\label{eq:pf.var.est.R3R3}
\end{align}

where
\begin{align*}
    R_{k,1} &= \meanIk \left\{\wh\EIF^2(\O_i,\wh\theta\eifa(\Gamma_a))-\EIF^2(\O_i,\theta_a(\Gamma_a))\right\},\\
    R_{k,2} &= \meanIk \left[\EIF^2(\O_i,\theta_a(\Gamma_a))-\E\{\EIF^2(\O_i,\theta(\Gamma_a))\}\right].
\end{align*}
(\ref{eq:pf.var.est.R3R3}) can be concluded since
$R_{k,2}=O_p(n^{-1/2})$ by $\E\{\EIF^4(\O_i,\theta(\Gamma_a))\}<\infty$, and $R_{k,2}=O_p(n^{-1/2})$ by the following argument.
Note that
{\footnotesize{
\begin{align*}
    |R_{k,1}|
    \leq& \meanIk\left|\wh\EIF^2(\O_i,\wh\theta\eifa(\Gamma_a))-\EIF^2(\O_i,\theta_a(\Gamma_a))\right|\\
    =& \meanIk\left|\wh\EIF(\O_i,\wh\theta\eifa(\Gamma_a))-\EIF(\O_i,\theta_a(\Gamma_a))\right|\cdot
\left|\wh\EIF(\O_i,\wh\theta\eifa(\Gamma_a))+\EIF(\O_i,\theta_a(\Gamma_a))\right|\\
    \leq&\sqrt{\meanIk\left|\wh\EIF(\O_i,\wh\theta\eifa(\Gamma_a))-\EIF(\O_i,\theta(\Gamma_a))\right|^2}
\sqrt{\meanIk\left|\wh\EIF(\O_i,\wh\theta\eifa(\Gamma_a))+\EIF(\O_i,\theta(\Gamma_a))\right|^2}\\
    \leq & \sqrt{\meanIk\left|\wh\EIF(\O_i,\wh\theta\eifa(\Gamma_a))-\EIF(\O_i,\theta_a(\Gamma_a))\right|^2}
   \left( \sqrt{\meanIk\left|\wh\EIF(\O_i,\wh\theta\eifa(\Gamma_a))-\EIF(\O_i,\theta(\Gamma_a))\right|^2}\right.\\
  & \left.+ \sqrt{\frac{4}{|\calI_k|}\sum_{i\in\calI_k}\phi^2\eifa(\O_i,\theta(\Gamma_a))}
   \right),
\end{align*}
}}
we have
\begin{align*}
   R_{k,1}^2 \lesssim R_n\left\{
    \frac{4}{|\calI\supk|}\sum_{i\in\calI_k}\EIF^2(\O_i,\theta_a(\Gamma_a)) +
    R_n\right\}
\end{align*}
where $R_n =\meanIk\left|\wh\EIF(\O_i,\wh\theta\eifa(\Gamma_a))-\EIF(\O_i,\theta_a(\Gamma_a))\right|^2$. Since $ \meanIk\EIF^2(\O_i,\theta_a(\Gamma_a))=O_p(1)$, it's sufficient to show $R_n = O_p(n^{-1/2})$, which holds by the proof of Theorem 4.3. 

\section{Details and Examples of the Calibration Procedure}\label{supp.sec:calibration}
This section provides details and illustrations for the calibration procedure introduced in Section 5. 
\subsection{Analysis Pipeline}
\begin{figure}
    \centering
    \includegraphics[width=1\linewidth]{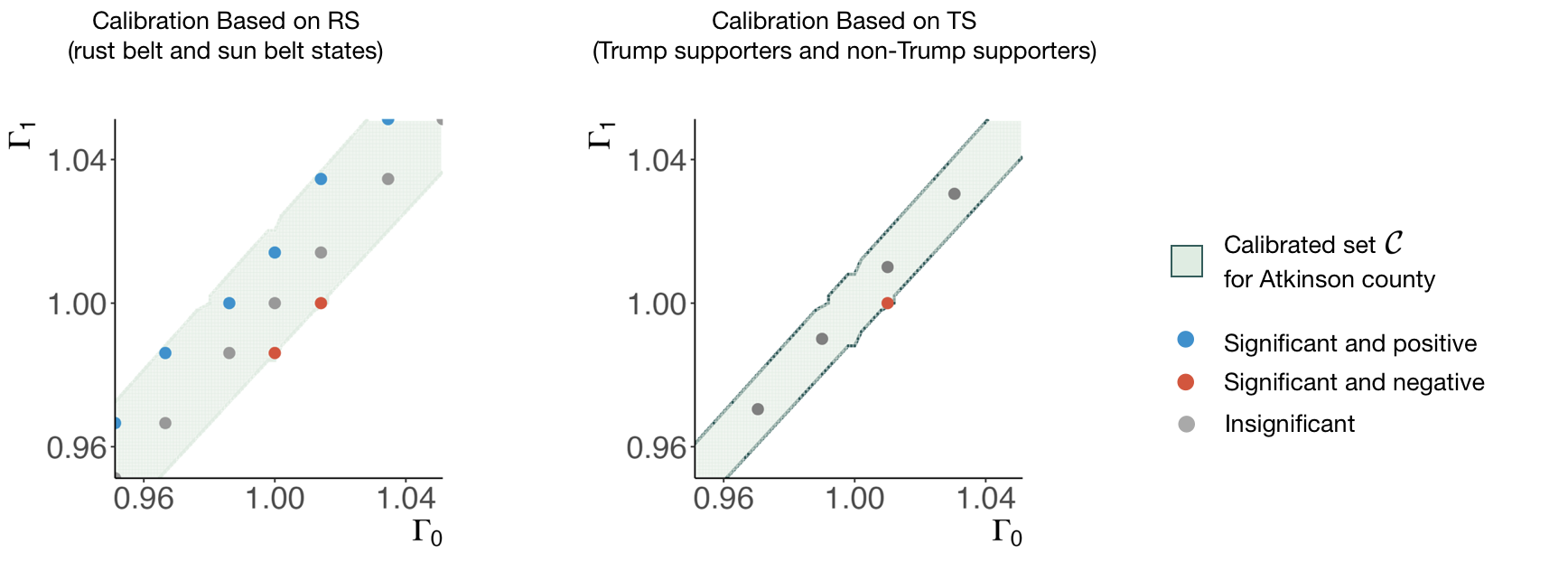}
    \caption{Calibrated region $\calC$ in Atkinson under (left) the  RS calibration and  (right) the TS calibration.}
    \label{fig:cali.ex.atkinson}
\end{figure}

We start with some remarks about the implementation of our calibration procedure. First, it's important to have the ratio of the sample sizes between the proxy source and target data be equal to that of the original source and the target data. This can be accomplished by downsampling one of the two proxy data. Relatedly, to make the comparisons fairer, it's useful to rescale the standard error estimate in the transported CI from the calibration procedure by multiplying it with $\sqrt{|\calI_{s_2}|/\nt}$ in order to mimic the length of the CI for the original TATE.
This was mentioned in Algorithm 1 under Step 1. Third, one should make sure the shared covariates in constructing  $\wh{\text{CI}}_{s\to t}(\Gamma_0,\Gamma_1;1-\alpha)$ should match the shared covariates $\V_i$ in the actual target sample.  See Algorithm \ref{alg:cali.practice} for the implementation and Section D of the Supplementary Materials for more discussions.
Algorithm \ref{alg:cali.practice} provides a step-by-step procedure for calibrating the sensitivity parameters.
As an example, Figure \ref{fig:cali.ex.atkinson} illustrates  the final calibration region $\calC$ for estimating the ad effect in Atkinson in Georgia under the two ways of partitioning  (RS, TS) introduced in Section 6 of the main text.

\begin{algorithm}[!h]
\caption{Calibrating Sensitivity Parameters}\label{alg:cali.practice}
\begin{algorithmic}[1]
 \Require Source data, confidence level $1-\alpha$, set $\calC_{\text{init}}\in\mathbb{R}\times\mathbb{R}$.
 \State \textbf{Step 1 (Partition source data)}: Partition the source data into two parts and denote their corresponding indices as $\calI_{s_1}$ and $ \calI_{s_2}$ where $\calI_{s_1} \cup \calI_{s_2} = \calI_s$ and $\calI_{s_1} \cap \calI_{s_2} = \emptyset$.
 \If{$\calI_{s_1}/\calI_{s_2}>\ns/\nt$}
 \State Randomly subset $\calI_{s_1}$ of size $|\calI_{s_2}|\cdot\ns.\nt$ and denote the resulting set of indices as $\calI_{s_1}$.
 \Else
 \State Randomly subset $\calI_{s_2}$ of size $|\calI_{s_1}|\cdot\nt/\ns$ and denote the resulting set of indices as $\calI_{s_2}$.
 \EndIf
 \State \textbf{Step 2.1 (Construct CI via the standard approach)}: With data in $\calI_{s_2}$, estimate the ATE and its $(1-\alpha)$ confidence interval, denoted as  $\wh{\text{CI}}_{s_2}(1-\alpha)$.
 \State \textbf{Step 2.2 (Construct CI via our transfer learning approach) }:
 \State With $\{(\X_i,A_i,Y_i,S_i=1):i\in\calI_{s_1}\}\cup\{(\V_i,S_i=0):i\in\calI_{s_2}\}$, estimate the ATE on $\calS_2$ and its standard error with any $(\Gamma_0,\Gamma_1)\in\calC_{\text{init}}$, denoted as $\wh\theta_{s_1\to s_2}(\Gamma_0,\Gamma_1)$ and $\wh{\text{SE}}_{s_1\to s_2}$. Denote the re-scaled confidence interval as
         \begin{align}
             \wh{\text{CI}}_{s_1\to s_2}(\Gamma_0,\Gamma_1;1-\alpha)=
             \left[\wh\theta_{s_1\to s_2}(\Gamma_0,\Gamma_1) \mp z_{1-\alpha/2}\cdot\wh{\text{SE}}_{s_1\to s_2}(\Gamma_0,\Gamma_1)\cdot \sqrt{|\calI_{s_2|}/\nt}
             \right].
         \end{align}
 \State \textbf{Step 3 (Find the plausible range) }: Find the plausible range of sensitivity parameters when transporting from $\calS_1$ to $\calS_2$:
 \begin{align}
     \calC_1 = \left\{
     (\Gamma_0,\Gamma_1)\in\calC_{\text{init}}:
       \wh{\text{CI}}_{s_2}(1-\alpha)\cap \wh{\text{CI}}_{s_1\to s_2}(\Gamma_0,\Gamma_1;1-\alpha) \neq\emptyset
     \right\}.
 \end{align}
 \State \textbf{Calibration in the other direction} Exchange $\calS_1$ and $\calS_2$ and repeat Steps 1-3, resulting in the plausible range $\calC_2$.
 \Ensure Intersect two plausible regions to construct the final region: $\calC = \calC_1\cap\calC_2$.
\end{algorithmic}
\end{algorithm}

\subsection{Interpretations}

The sensitivity parameters $\Gamma_0$ and $\Gamma_1$ quantify the change in turnout between the five states in \citet{aggarwal20232}'s experiment and Georgia in the control arm and the treatment arm, respectively, and different values of $\Gamma_0$ and $\Gamma_1$ will generally correspond to different effect sizes and direction. Some examples are listed below and Table \ref{tab:explain.sign} enumerates more examples.
\begin{enumerate}
\item Suppose $\Gamma_0=1$ and $\Gamma_1>1$ (i.e., the $y>1$ part in Figure \ref{fig:cali.ex.atkinson}). Then the turnout in Georgia if voters are not exposed to anti-Trump ads will be the same as that in the five states of \citet{aggarwal20232}'s experiment, but the turnout in Georiga if the voters are exposed to anti-Trump ads will be larger than that in the five states of \citet{aggarwal20232}'s experiment. Also, the ad effect in Georgia will be higher than that in the five states of \citet{aggarwal20232}'s experiment and if $\Gamma_1$ is sufficiently large, the effect will be positive and statistically significant.
\item Suppose $\Gamma_0>1$ and $\Gamma_1=1$ (i.e., the $x>1$ part in Figure \ref{fig:cali.ex.atkinson}). Then the turnout in Georgia if voters are not exposed to anti-Trump ads will be higher than that in the five states of \citet{aggarwal20232}'s experiment, but the turnout in Georgia if the voters are exposed to anti-Trump ads will be the same as that in the five states of \citet{aggarwal20232}'s experiment. Also, the ad effect in Georgia is likely smaller than that in the five states of \citet{aggarwal20232}'s experiment, and if $\Gamma_0$ is large enough, the effect may be negative and significant.
\item Suppose $\Gamma_0 > 1$ and $\Gamma_1 > 1$ (i.e., top right region of Figure \ref{fig:cali.ex.atkinson}). Then, the odds of turnout in both treatment and controls will be higher in Georgia than those in the five states of \citet{aggarwal20232}'s experiment. In this case, the ad effect in Georgia may be similar to that in the five states of \citet{aggarwal20232}'s experiment, especially if the shift in the turnouts between states are comparable between the control and treatment arms. A similar phenomena would occur if $\Gamma_0 < 1$ and $\Gamma_1 < 1$ (i.e., bottom left region of Figure \ref{fig:cali.ex.atkinson}).
\item Suppose $\Gamma_0 < 1$ and $\Gamma_1 > 1$ (i.e., top left region of Figure \ref{fig:cali.ex.atkinson}). Then, the odd of turnout if voters are not exposed to anti-Trump ads will be lower in Georgia than that in the five states of \citet{aggarwal20232}'s experiment, but the odd of turnout if voter are exposed to anti-Trump ads will be higher than Georgia than that in the five states of \citet{aggarwal20232}'s experiment. Then, the combined effect of the changes in the odds would be a large and positive value of the ad effect in Georgia.
\item Suppose $\Gamma_0 > 1$ and $\Gamma_1 < 1$ (i.e., bottom right region of Figure \ref{fig:cali.ex.atkinson}). Then, the odd of turnout if voters are not exposed to anti-Trump ads will be higher in Georgia than that in the five states of \citet{aggarwal20232}'s experiment, but the odd of turnout if voter are exposed to anti-Trump ads will be lower than Georgia than that in the five states of \citet{aggarwal20232}'s experiment. Then, the combined effect of the changes in the odds would be a negative ad effect in Georgia that is large in magnitude.
\end{enumerate}
\begin{table}[h!]
\caption{Examples on the signs of $\Gamma_0$, $\Gamma_1$ and the ad effect in Georgia (i.e., TATE) compared with the ad effect in the five states of \citet{aggarwal20232}'s experiment.}\label{tab:explain.sign}
    \begin{tabular}{l l p{4.2cm} p{4.2cm} p{5cm} }
    \hline
    \hline
    $\Gamma_0$ & $\Gamma_1$ & Odd of turnout in Georgia if unexposed to negative ads (i.e., $\Yzero$) & Odd of turnout in Georgia if exposed to negative ads (i.e., $\Yone$) & ATE in Georgia (i.e., TATE) \\ \hline \hline
$=1$ & $>1$ & same as the corresponding odd in the five states & higher than the corresponding odd in the five states &
 higher than ATE in the five states \\
 \hline
 $>1$ & $=1$ & the corresponding odd in the five states & same as the corresponding odd in the five states & lower than ATE in the five states\\
 \hline
 $>1$ & $>1$ & higher than the corresponding odd in the five states & higher than the corresponding odd in the five states & may be similar with ATE in the five states\\
 \hline
 $<1$ & $>1$ & lower than the corresponding odd in the five states & higher than the corresponding odd in the five states & higher than ATE in the five states\\
 \hline
 $>1$ & $<1$ & higher than the corresponding odd in the five states & lower than the corresponding odd in the five states & lower than ATE in the five states\\
 \hline
 \hline
    \end{tabular}
\end{table}
As discussed in Section 5, 
not all values of $\Gamma_0, \Gamma_1$ are meaningful and the calibration procedure, which produces the set $\mathcal{C}$ (i.e., the green area in Figure \ref{fig:cali.ex.atkinson}) allows us to focus on values of $\Gamma_0$ and $\Gamma_1$ that are more interpretable.
\clearpage

\section{Supplementary Materials for the Ad Effect in Georgia}\label{supp.sec:pa}

\subsection{Additional Data Description}\label{supp.subsec:data.describe}
Our analysis consists of two datasets, the source data derived from the  RCT data from \cite{aggarwal20232} and the target data derived from  GA voter database. Prior to analysis, we recoded the shared covariates $\V_i$ from these two datasets for them to match. A description is provided as follows.

The age was coded as four groups (18-24, 25-24, 35-39, and 40+) in the  RCT data and as date of birth in the GA voter database. For the target data, we calculated their age by the year of 2020 and excluded voters above 55 years' old to match the range of age in \cite{aggarwal20232}, and then constructed a variable of age groups according to the source data. The resulting age group variable for analysis is a discrete variable with four levels.

The race information is categorical variable coded in  four levels (White, Black, Latinx, and Other) in the RCT data and seven levels (White, Black, Hispanic/Latino, American Indian, Alaskan native, Other, Unknown) in the GA voter database. We recoded the GA voter database in which we matched the first three levels with the RCT data combined the last four levels as ``Other'' to match the RCT data.

The gender was coded in two levels (female and other) in the  RCT data and  four levels (female, male, other, X) in the GA voter database. Our gender variable for analysis has two levels: female and non-female where the non-female level includes voters whose gender weren't coded as female.

In addition to the common covariates, the 2020 RCT data also contains the party information and voting history.
The party information was coded as one of the four levels: Democratic, Republican, Unknown and Other.  Note that the party information was inaccurate with 72\% being unknown and we refer readers to \cite{aggarwal20232} for details.
The voting history was coded as ten binary variables. Each variable indicated whether a voter has voted in a specific year for every other year between 2000 and 2018 (i.e., voted in 2000, voted in 2002, voted in 2004, voted in 2006, voted in 2008, voted in 2010, voted in 2012, voted in 2014, voted in 2016, voted in 2018).
Table \ref{tab:data.description} summarizes the covariates (which are all discrete) and their levels.
\begin{table}[h]
    \centering
    \begin{tabular}{lll}
\hline
Covariate & Levels & Available from target?\\
\hline
Age group & 18-24, 25-34, 35-39, 40+ & Yes\\
Gender & Female, non-female & Yes \\
Race& White, Black, Latinx, other & Yes\\
Party& Democratic, Republican, Other, Unknown & No\\
Voted in 2000 & 0, 1 & No\\
Voted in 2002 & 0, 1 & No\\
Voted in 2004 & 0, 1 & No\\
Voted in 2006 & 0, 1 & No\\
Voted in 2008 & 0, 1 & No\\
Voted in 2010 & 0, 1 & No\\
Voted in 2012 & 0, 1 & No\\
Voted in 2014 & 0, 1 & No\\
Voted in 2016 & 0, 1 & No\\
Voted in 2018 & 0, 1 & No\\
\hline
    \end{tabular}
    \caption{Descriptions on covariates in pooled data.}
    \label{tab:data.description}
\end{table}

\clearpage
\subsection{Details for County-Level Ad Effects in Section 6.2}
We provide a comprehensive result (i.e., the specific numbers of confidence intervals) for the results presented in Section 6.2.
Specifically,
Table \ref{tab:ga.county.ate} lists the ad effect estimated by the OR estimator under the three choices of sensitivity parameters in Figure 4
for each county in GA.
\footnotesize{
\begin{longtblr}[
 caption = {County-by-county ad effect with the OR estimator in GA under $(\Gamma_0,\Gamma_1)=(1,1)$, $(\Gamma_0,\Gamma_1)=(0.990,1.006)$, and $(\Gamma_0,\Gamma_1) = (1.006,0.990)$. Each cells lists the TATE with $95\%$ CI in parentheses.},
  label = {tab:ga.county.ate},
]{
  colspec = {X| X XX},
}
\hline
   County & $\Gamma_0=\Gamma_1=1$
    & $\Gamma_0=0.990, \Gamma_1=1.006$ & $\Gamma_0=1.006, \Gamma_1=0.990$
    \\
    \hline
   Appling   & -0.03 (-0.33, 0.27)    &  0.35 (0.05, 0.65)    & -0.41 (-0.71, -0.11)  \\
        Atkinson   & -0.06 (-0.32, 0.21)    &  0.33 (0.06, 0.59)    & -0.44 (-0.71, -0.17)  \\
           Bacon   & -0.02 (-0.33, 0.29)    &  0.36 (0.05, 0.67)    &  -0.4 (-0.71, -0.09)  \\
           Baker   & -0.03 (-0.34, 0.27)    &  0.35 (0.04, 0.65)    & -0.42 (-0.72, -0.11)  \\
         Baldwin   & -0.05 (-0.36, 0.27)    &  0.34 (0.02, 0.65)    & -0.43 (-0.74, -0.11)  \\
           Banks   & -0.01 (-0.37, 0.34)    &  0.37 (0.02, 0.72)    &  -0.4 (-0.75, -0.05)  \\
          Barrow   &  -0.05 (-0.4, 0.31)    & 0.34 (-0.02, 0.69)    & -0.43 (-0.78, -0.08)  \\
          Bartow   & -0.03 (-0.36, 0.29)    &  0.35 (0.02, 0.68)    & -0.42 (-0.74, -0.09)  \\
       Ben Hill   & -0.04 (-0.33, 0.25)    &  0.34 (0.05, 0.63)    & -0.42 (-0.71, -0.13)  \\
        Berrien   &  -0.02 (-0.33, 0.3)    &  0.37 (0.05, 0.68)    &  -0.4 (-0.72, -0.08)  \\
          Bibb   & -0.06 (-0.41, 0.29)    & 0.32 (-0.02, 0.67)    & -0.44 (-0.79, -0.09)  \\
       Bleckley   & -0.02 (-0.32, 0.27)    &  0.36 (0.06, 0.65)    &   -0.4 (-0.7, -0.11)  \\
       Brantley   & -0.01 (-0.35, 0.34)    &  0.38 (0.03, 0.72)    & -0.39 (-0.73, -0.05)  \\
       Brooks   & -0.04 (-0.33, 0.26)    &  0.34 (0.05, 0.64)    & -0.42 (-0.71, -0.12)  \\
        Bryan   & -0.03 (-0.41, 0.34)    & 0.35 (-0.03, 0.73)    & -0.42 (-0.79, -0.04)  \\
        Bulloch   & -0.04 (-0.36, 0.27)    &  0.34 (0.03, 0.66)    & -0.42 (-0.74, -0.11)  \\
         Burke   & -0.04 (-0.37, 0.28)    &  0.34 (0.02, 0.66)    &  -0.43 (-0.75, -0.1)  \\
        Butts   & -0.03 (-0.34, 0.28)    &  0.35 (0.04, 0.66)    &  -0.41 (-0.72, -0.1)  \\
       Calhoun   &  -0.05 (-0.39, 0.3)    & 0.33 (-0.01, 0.68)    & -0.43 (-0.77, -0.08)  \\
    Camden   & -0.06 (-0.47, 0.36)    & 0.33 (-0.08, 0.74)    & -0.44 (-0.85, -0.03)  \\
      Candler   & -0.04 (-0.36, 0.27)    &  0.34 (0.03, 0.65)    & -0.42 (-0.74, -0.11)  \\
      Carroll   & -0.04 (-0.35, 0.28)    &  0.35 (0.03, 0.66)    & -0.42 (-0.73, -0.11)  \\
        Catoosa    & -0.02 (-0.37, 0.33)    &  0.36 (0.01, 0.71)    & -0.41 (-0.75, -0.06)  \\
        Charlton   &  -0.04 (-0.37, 0.3)    &  0.34 (0.01, 0.68)    & -0.42 (-0.76, -0.08)  \\
       Chatham   &  -0.06 (-0.4, 0.28)    & 0.32 (-0.02, 0.66)    &  -0.44 (-0.78, -0.1)  \\
  Chattahoochee   & -0.08 (-0.48, 0.31)    &   0.3 (-0.09, 0.7)    & -0.47 (-0.86, -0.07)  \\
     Chattooga   & -0.02 (-0.37, 0.33)    &  0.36 (0.01, 0.71)    & -0.41 (-0.75, -0.06)  \\
       Cherokee   & -0.03 (-0.41, 0.34)    & 0.35 (-0.03, 0.72)    & -0.42 (-0.79, -0.04)  \\
        Clarke   & -0.06 (-0.38, 0.26)    &     0.32 (0, 0.64)    & -0.45 (-0.77, -0.12)  \\
         Clay   & -0.05 (-0.41, 0.32)    &  0.33 (-0.03, 0.7)    & -0.43 (-0.79, -0.06)  \\
       Clayton   &  -0.1 (-0.55, 0.36)    & 0.29 (-0.16, 0.74)    & -0.48 (-0.93, -0.03)  \\
       Clinch   & -0.03 (-0.32, 0.26)    &  0.35 (0.06, 0.65)    &  -0.41 (-0.7, -0.12)  \\
          Cobb   & -0.07 (-0.44, 0.31)    & 0.32 (-0.06, 0.69)    & -0.45 (-0.82, -0.08)  \\
        Coffee   & -0.05 (-0.32, 0.23)    &  0.34 (0.06, 0.61)    &  -0.43 (-0.7, -0.15)  \\
      Colquitt   & -0.05 (-0.33, 0.23)    &  0.33 (0.05, 0.61)    & -0.43 (-0.71, -0.15)  \\
      Columbia   & -0.04 (-0.41, 0.32)    &  0.34 (-0.02, 0.7)    & -0.43 (-0.79, -0.06)  \\
          Cook   &  -0.03 (-0.3, 0.25)    &  0.35 (0.08, 0.63)    & -0.41 (-0.68, -0.13)  \\
        Coweta   & -0.04 (-0.41, 0.33)    & 0.34 (-0.03, 0.71)    & -0.42 (-0.79, -0.05)  \\
       Crawford   &  -0.02 (-0.34, 0.3)    &  0.36 (0.04, 0.68)    &  -0.4 (-0.72, -0.08)  \\
        Crisp   & -0.05 (-0.37, 0.27)    &  0.33 (0.01, 0.66)    & -0.43 (-0.75, -0.11)  \\
         Dade   & -0.01 (-0.36, 0.33)    &  0.37 (0.03, 0.72)    &  -0.4 (-0.74, -0.05)  \\
      Dawson   & -0.01 (-0.37, 0.34)    &  0.37 (0.01, 0.73)    &  -0.4 (-0.75, -0.04)  \\
      DeKalb   & -0.07 (-0.47, 0.33)    & 0.31 (-0.09, 0.71)    & -0.45 (-0.85, -0.05)  \\
      Decatur   & -0.05 (-0.34, 0.24)    &  0.33 (0.04, 0.63)    & -0.43 (-0.72, -0.14)  \\
         Dodge   & -0.03 (-0.32, 0.27)    &  0.36 (0.06, 0.65)    &  -0.41 (-0.7, -0.11)  \\
         Dooly   & -0.05 (-0.37, 0.27)    &  0.33 (0.01, 0.65)    & -0.43 (-0.75, -0.11)  \\
     Dougherty   & -0.07 (-0.46, 0.33)    & 0.31 (-0.08, 0.71)    & -0.45 (-0.84, -0.05)  \\
       Douglas   & -0.07 (-0.46, 0.31)    & 0.31 (-0.07, 0.69)    & -0.46 (-0.84, -0.07)  \\
         Early   & -0.04 (-0.36, 0.27)    &  0.34 (0.02, 0.65)    & -0.42 (-0.74, -0.11)  \\
         Echols   & -0.04 (-0.35, 0.27)    &  0.34 (0.04, 0.65)    & -0.42 (-0.73, -0.12)  \\
     Effingham   & -0.03 (-0.39, 0.33)    & 0.35 (-0.01, 0.71)    & -0.41 (-0.77, -0.05)  \\
         Elbert   & -0.03 (-0.33, 0.27)    &  0.35 (0.05, 0.65)    & -0.41 (-0.71, -0.11)  \\
       Emanuel   & -0.04 (-0.34, 0.27)    &  0.35 (0.04, 0.65)    & -0.42 (-0.72, -0.12)  \\
         Evans   &  -0.03 (-0.3, 0.24)    &  0.35 (0.08, 0.62)    & -0.41 (-0.68, -0.15)  \\
         Fannin   & -0.01 (-0.36, 0.35)    &  0.38 (0.02, 0.73)    & -0.39 (-0.75, -0.03)  \\
       Fayette   & -0.05 (-0.46, 0.35)    & 0.33 (-0.07, 0.73)    & -0.44 (-0.84, -0.03)  \\
          Floyd   & -0.04 (-0.36, 0.28)    &  0.34 (0.02, 0.66)    &  -0.42 (-0.74, -0.1)  \\
        Forsyth   & -0.06 (-0.66, 0.54)    & 0.33 (-0.27, 0.93)    &  -0.44 (-1.04, 0.16)  \\
      Franklin   & -0.02 (-0.35, 0.32)    &   0.37 (0.03, 0.7)    &  -0.4 (-0.73, -0.06)  \\
      Fulton   & -0.07 (-0.47, 0.33)    & 0.31 (-0.09, 0.72)    & -0.45 (-0.85, -0.05)  \\
        Gilmer   & -0.02 (-0.37, 0.32)    &  0.36 (0.02, 0.71)    &  -0.4 (-0.75, -0.06)  \\
       Glascock   &     0 (-0.35, 0.35)    &  0.38 (0.03, 0.73)    & -0.38 (-0.74, -0.03)  \\
         Glynn   & -0.04 (-0.36, 0.28)    &  0.34 (0.02, 0.66)    & -0.42 (-0.74, -0.11)  \\
         Gordon   & -0.04 (-0.37, 0.29)    &  0.34 (0.02, 0.67)    &  -0.42 (-0.75, -0.1)  \\
         Grady   & -0.04 (-0.33, 0.26)    &  0.34 (0.05, 0.64)    & -0.42 (-0.71, -0.13)  \\
         Greene   & -0.04 (-0.35, 0.27)    &  0.34 (0.03, 0.65)    & -0.42 (-0.73, -0.11)  \\
      Gwinnett   &  -0.1 (-0.59, 0.39)    & 0.29 (-0.21, 0.78)    &  -0.48 (-0.98, 0.01)  \\
      Habersham   & -0.03 (-0.38, 0.31)    &  0.35 (0.01, 0.69)    & -0.42 (-0.76, -0.08)  \\
          Hall   & -0.07 (-0.41, 0.26)    & 0.31 (-0.02, 0.64)    & -0.46 (-0.79, -0.12)  \\
     Hancock   & -0.06 (-0.46, 0.35)    & 0.32 (-0.08, 0.73)    & -0.44 (-0.84, -0.03)  \\
       Haralson   & -0.01 (-0.35, 0.33)    &  0.37 (0.03, 0.72)    & -0.39 (-0.73, -0.05)  \\
        Harris   & -0.02 (-0.37, 0.32)    &   0.36 (0.02, 0.7)    &  -0.4 (-0.75, -0.06)  \\
          Hart   & -0.02 (-0.34, 0.29)    &  0.36 (0.04, 0.68)    &  -0.4 (-0.72, -0.09)  \\
         Heard   & -0.01 (-0.35, 0.32)    &   0.37 (0.03, 0.7)    & -0.39 (-0.73, -0.06)  \\
         Henry   & -0.07 (-0.46, 0.32)    &  0.31 (-0.08, 0.7)    & -0.45 (-0.84, -0.06)  \\
     Houston   & -0.05 (-0.38, 0.28)    &     0.33 (0, 0.67)    &  -0.43 (-0.77, -0.1)  \\
         Irwin   & -0.02 (-0.32, 0.28)    &  0.36 (0.06, 0.66)    &   -0.4 (-0.7, -0.11)  \\
       Jackson   & -0.03 (-0.39, 0.33)    &     0.36 (0, 0.72)    & -0.41 (-0.77, -0.05)  \\
         Jasper   & -0.01 (-0.31, 0.29)    &  0.37 (0.07, 0.67)    & -0.39 (-0.69, -0.09)  \\
     Jeff Davis   & -0.03 (-0.32, 0.26)    &  0.35 (0.06, 0.65)    & -0.41 (-0.71, -0.12)  \\
     Jefferson   & -0.05 (-0.38, 0.29)    &     0.33 (0, 0.67)    & -0.43 (-0.76, -0.09)  \\
       Jenkins   & -0.04 (-0.34, 0.27)    &  0.34 (0.04, 0.65)    & -0.42 (-0.73, -0.11)  \\
      Johnson   & -0.02 (-0.32, 0.27)    &  0.36 (0.06, 0.65)    &  -0.41 (-0.7, -0.11)  \\
        Jones   & -0.02 (-0.32, 0.28)    &  0.36 (0.06, 0.66)    &    -0.4 (-0.7, -0.1)  \\
          Lamar   & -0.03 (-0.34, 0.29)    &  0.35 (0.04, 0.67)    & -0.41 (-0.72, -0.09)  \\
         Lanier   & -0.03 (-0.32, 0.27)    &  0.35 (0.06, 0.65)    &  -0.41 (-0.7, -0.12)  \\
      Laurens   & -0.04 (-0.33, 0.26)    &  0.35 (0.05, 0.64)    & -0.42 (-0.71, -0.12)  \\
            Lee   & -0.03 (-0.36, 0.31)    &  0.36 (0.02, 0.69)    & -0.41 (-0.74, -0.07)  \\
        Liberty   & -0.09 (-0.48, 0.31)    &  0.29 (-0.1, 0.69)    & -0.47 (-0.87, -0.08)  \\
       Lincoln   & -0.03 (-0.34, 0.29)    &  0.35 (0.04, 0.67)    & -0.41 (-0.72, -0.09)  \\
         Long   &  -0.06 (-0.42, 0.3)    & 0.32 (-0.03, 0.68)    &  -0.44 (-0.8, -0.08)  \\
        Lowndes   & -0.05 (-0.35, 0.25)    &  0.33 (0.04, 0.63)    & -0.43 (-0.73, -0.14)  \\
       Lumpkin   & -0.02 (-0.37, 0.33)    &  0.37 (0.02, 0.72)    &  -0.4 (-0.75, -0.05)  \\
         Macon   &  -0.06 (-0.4, 0.29)    & 0.32 (-0.02, 0.67)    & -0.44 (-0.78, -0.09)  \\
        Madison   & -0.03 (-0.36, 0.31)    &  0.36 (0.02, 0.69)    & -0.41 (-0.74, -0.07)  \\
         Marion   & -0.04 (-0.33, 0.26)    &  0.34 (0.05, 0.64)    & -0.42 (-0.71, -0.13)  \\
       Mcduffie   & -0.03 (-0.32, 0.26)    &  0.35 (0.06, 0.64)    &  -0.41 (-0.7, -0.12)  \\
      Mcintosh   & -0.03 (-0.34, 0.27)    &  0.35 (0.04, 0.65)    & -0.41 (-0.72, -0.11)  \\
     Meriwether   & -0.04 (-0.35, 0.28)    &  0.34 (0.03, 0.66)    &  -0.42 (-0.73, -0.1)  \\
       Miller   & -0.02 (-0.31, 0.27)    &  0.36 (0.07, 0.65)    &   -0.4 (-0.7, -0.11)  \\
      Mitchell   & -0.04 (-0.35, 0.26)    &  0.34 (0.03, 0.64)    & -0.42 (-0.73, -0.12)  \\
       Monroe   & -0.02 (-0.35, 0.31)    &  0.36 (0.03, 0.69)    &  -0.4 (-0.73, -0.07)  \\
   Montgomery   & -0.02 (-0.33, 0.28)    &  0.36 (0.05, 0.66)    &   -0.4 (-0.71, -0.1)  \\
        Morgan   & -0.02 (-0.33, 0.29)    &  0.36 (0.06, 0.67)    &  -0.4 (-0.71, -0.09)  \\
        Murray   &  -0.03 (-0.36, 0.3)    &  0.35 (0.02, 0.68)    & -0.41 (-0.74, -0.08)  \\
     Muscogee   & -0.07 (-0.42, 0.28)    & 0.31 (-0.04, 0.66)    &   -0.45 (-0.8, -0.1)  \\
       Newton   &  -0.06 (-0.4, 0.29)    & 0.32 (-0.02, 0.67)    &  -0.44 (-0.79, -0.1)  \\
       Oconee   & -0.01 (-0.41, 0.38)    & 0.37 (-0.03, 0.76)    &      -0.4 (-0.79, 0)  \\
   Oglethorpe   & -0.02 (-0.35, 0.31)    &  0.36 (0.03, 0.69)    &  -0.4 (-0.73, -0.08)  \\
     Paulding   &  -0.04 (-0.37, 0.3)    &  0.34 (0.01, 0.68)    & -0.42 (-0.76, -0.09)  \\
        Peach   & -0.05 (-0.36, 0.25)    &  0.33 (0.02, 0.64)    & -0.43 (-0.74, -0.13)  \\
     Pickens   & -0.01 (-0.39, 0.36)    &     0.37 (0, 0.74)    &  -0.4 (-0.77, -0.02)  \\
       Pierce   & -0.02 (-0.36, 0.33)    &  0.37 (0.02, 0.71)    &  -0.4 (-0.75, -0.05)  \\
        Pike   & -0.01 (-0.35, 0.34)    &  0.37 (0.03, 0.72)    & -0.39 (-0.74, -0.04)  \\
          Polk   & -0.03 (-0.34, 0.28)    &  0.35 (0.04, 0.66)    &  -0.41 (-0.72, -0.1)  \\
      Pulaski   & -0.03 (-0.32, 0.27)    &  0.35 (0.06, 0.65)    &  -0.41 (-0.7, -0.11)  \\
       Putnam   & -0.03 (-0.34, 0.27)    &  0.35 (0.05, 0.65)    & -0.41 (-0.72, -0.11)  \\
      Quitman   & -0.05 (-0.38, 0.28)    &     0.33 (0, 0.65)    &  -0.43 (-0.75, -0.1)  \\
         Rabun   & -0.02 (-0.36, 0.32)    &  0.36 (0.02, 0.71)    &  -0.4 (-0.75, -0.06)  \\
      Randolph   &  -0.05 (-0.39, 0.3)    & 0.33 (-0.01, 0.68)    & -0.43 (-0.77, -0.08)  \\
     Richmond   &  -0.07 (-0.44, 0.3)    & 0.31 (-0.06, 0.69)    & -0.45 (-0.82, -0.08)  \\
     Rockdale   &  -0.08 (-0.5, 0.34)    &  0.3 (-0.12, 0.72)    & -0.46 (-0.88, -0.05)  \\
        Schley   & -0.01 (-0.33, 0.31)    &  0.37 (0.05, 0.69)    & -0.39 (-0.71, -0.07)  \\
      Screven   & -0.04 (-0.35, 0.28)    &  0.34 (0.03, 0.66)    & -0.42 (-0.73, -0.11)  \\
     Seminole   & -0.03 (-0.33, 0.26)    &  0.35 (0.05, 0.65)    & -0.42 (-0.71, -0.12)  \\
   Spalding   & -0.05 (-0.35, 0.26)    &  0.34 (0.03, 0.64)    & -0.43 (-0.73, -0.12)  \\
  Stephens   &  -0.02 (-0.34, 0.3)    &  0.36 (0.04, 0.69)    &  -0.4 (-0.73, -0.08)  \\
   Stewart   & -0.04 (-0.39, 0.31)    & 0.34 (-0.01, 0.69)    & -0.43 (-0.78, -0.07)  \\
   Sumter   & -0.05 (-0.38, 0.27)    &  0.33 (0.01, 0.65)    & -0.43 (-0.76, -0.11)  \\
    Talbot   &  -0.05 (-0.39, 0.3)    & 0.34 (-0.01, 0.68)    & -0.43 (-0.77, -0.08)  \\
 Taliaferro   & -0.04 (-0.39, 0.32)    &  0.34 (-0.01, 0.7)    & -0.41 (-0.77, -0.06)  \\
   Tattnall   & -0.04 (-0.36, 0.28)    &  0.34 (0.02, 0.66)    & -0.42 (-0.74, -0.11)  \\
    Taylor   & -0.03 (-0.33, 0.26)    &  0.35 (0.05, 0.65)    & -0.41 (-0.71, -0.12)  \\
   Telfair   & -0.03 (-0.33, 0.27)    &  0.35 (0.05, 0.65)    & -0.41 (-0.71, -0.11)  \\
   Terrell   &   -0.05 (-0.4, 0.3)    & 0.33 (-0.02, 0.68)    & -0.43 (-0.78, -0.08)  \\
    Thomas   & -0.04 (-0.34, 0.27)    &  0.35 (0.04, 0.65)    & -0.42 (-0.72, -0.11)  \\
      Tift   & -0.05 (-0.34, 0.24)    &  0.33 (0.05, 0.62)    & -0.43 (-0.72, -0.14)  \\
    Toombs   & -0.05 (-0.35, 0.26)    &  0.34 (0.04, 0.64)    & -0.43 (-0.73, -0.13)  \\
     Towns   & -0.01 (-0.38, 0.36)    &  0.37 (0.01, 0.74)    & -0.39 (-0.76, -0.03)  \\
   Treutlen   & -0.02 (-0.32, 0.28)    &  0.36 (0.06, 0.66)    &    -0.4 (-0.7, -0.1)  \\
     Troup   & -0.04 (-0.34, 0.26)    &  0.34 (0.04, 0.64)    & -0.42 (-0.73, -0.12)  \\
    Turner   & -0.04 (-0.34, 0.26)    &  0.34 (0.04, 0.64)    & -0.42 (-0.72, -0.12)  \\
    Twiggs   & -0.03 (-0.35, 0.28)    &  0.35 (0.03, 0.66)    &  -0.41 (-0.73, -0.1)  \\
     Union   & -0.01 (-0.37, 0.35)    &  0.37 (0.01, 0.73)    & -0.39 (-0.75, -0.03)  \\
      Upson   & -0.03 (-0.33, 0.26)    &  0.35 (0.05, 0.65)    & -0.42 (-0.71, -0.12)  \\
    Walker   & -0.02 (-0.37, 0.34)    &  0.36 (0.01, 0.72)    &  -0.4 (-0.76, -0.05)  \\
     Walton   & -0.03 (-0.39, 0.32)    &      0.35 (0, 0.7)    & -0.42 (-0.77, -0.07)  \\
     Ware   & -0.04 (-0.35, 0.27)    &  0.34 (0.03, 0.66)    & -0.42 (-0.73, -0.11)  \\
     Warren   &  -0.05 (-0.39, 0.3)    & 0.33 (-0.01, 0.68)    & -0.43 (-0.77, -0.08)  \\
 Washington   & -0.04 (-0.36, 0.28)    &  0.34 (0.02, 0.66)    &  -0.42 (-0.74, -0.1)  \\
      Wayne   & -0.03 (-0.34, 0.28)    &  0.35 (0.05, 0.66)    &  -0.41 (-0.72, -0.1)  \\
   Webster   & -0.03 (-0.34, 0.27)    &  0.35 (0.04, 0.65)    & -0.42 (-0.72, -0.11)  \\
   Wheeler   & -0.03 (-0.33, 0.28)    &  0.36 (0.05, 0.66)    &  -0.41 (-0.71, -0.1)  \\
    White   & -0.01 (-0.37, 0.35)    &  0.37 (0.01, 0.73)    & -0.39 (-0.75, -0.04)  \\
 Whitfield   &  -0.09 (-0.4, 0.22)    & 0.29 (-0.02, 0.61)    & -0.47 (-0.79, -0.16)  \\
     Wilcox   & -0.03 (-0.33, 0.27)    &  0.35 (0.06, 0.65)    & -0.41 (-0.71, -0.12)  \\
    Wilkes   & -0.04 (-0.35, 0.27)    &  0.34 (0.04, 0.65)    & -0.42 (-0.73, -0.11)  \\
    Wilkinson   & -0.03 (-0.33, 0.27)    &  0.35 (0.05, 0.65)    & -0.41 (-0.71, -0.11)  \\
     Worth   & -0.03 (-0.32, 0.27)    &  0.36 (0.06, 0.65)    &  -0.41 (-0.7, -0.11)  \\
\hline
\end{longtblr}
}

\normalsize

\subsection{Subgroup Analysis}
We present a subgroup analysis that was discussed briefly in Section 6.6 of the main text.   Voters are defined into subgroups by a three-way interaction between gender (female versus not female), race (Black or Latinx versus Others), and recent voting history (new voters versus returning voters). Details of data pre-processing are provided in Section \ref{subsec:preprocess.subgroup}.

\begin{figure}[!h]
    \centering
    \includegraphics[width=1\linewidth]{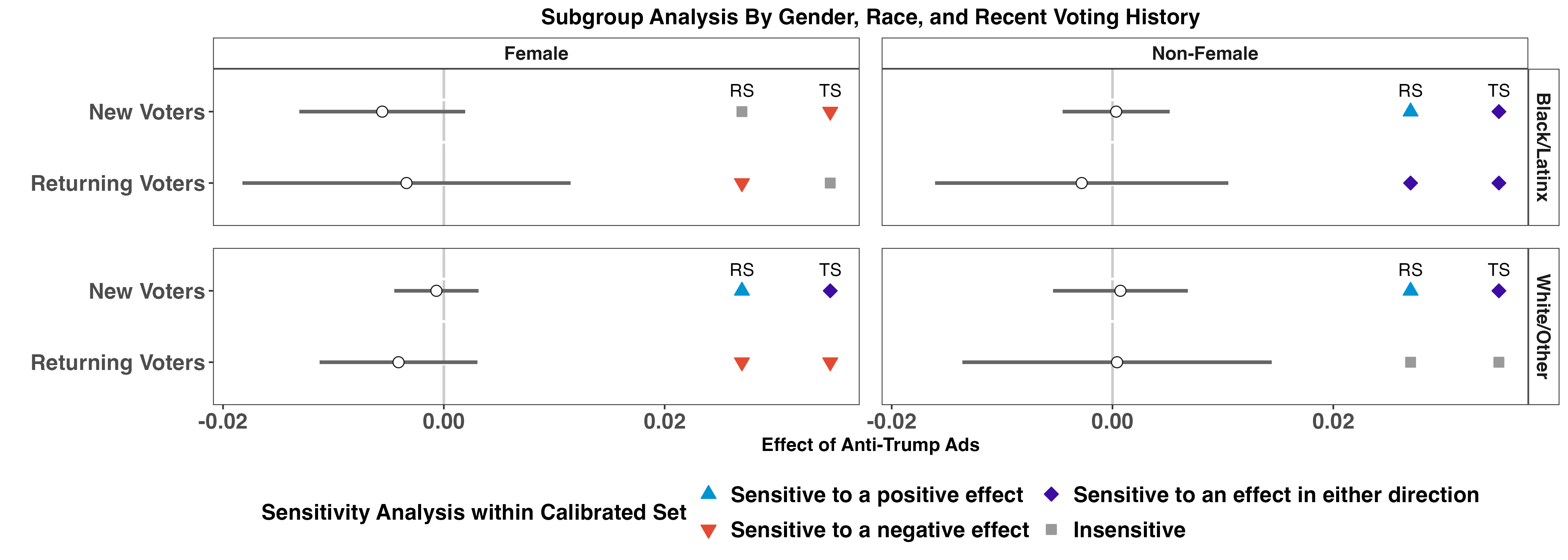}
    \caption{
    Subgroup analysis by gender, race, and recent voting history.
    The horizontal bar represents the 95\% CI of the ad effect under transportability.
    The colored boxes represent the results of the sensitivity analysis with the calibrated sets based on RS and TS partitions.
}
    \label{fig:ga.cate}
\end{figure}

Figure \ref{fig:ga.cate} summarizes the results. Under transportability,
we find some variations in the ad effect among different subgroups of Georgia voters,  but none of the  effects are statistically significant.
all subgroup effects are insignificant. When transportability is violated, we conduct a sensitivity analysis with the two calibrated sets based on RS and TS partitions. For both RS and TS partitions, the ad effect is sensitive in six out of eight subgroups in Georgia.
Notably, the two partitioning agrees that the ad is
(i) sensitive to a positive effect  among female, returning voters that are not Black or Latinx,
(ii) sensitive to an effect in either direction for returning, non-female voters who are Black or Latinx, and
(iii) insensitive for returning, non-female voters who are not Black or Latinx.
For the other subgroups, the conclusions of the two partitions are not identical and is a reflection of the differences in the two calibrated sets.

We briefly remark that under transportability, if $\calV = \calX$ and the subgroups are defined by every value of $\V$, the subgroup effect in the source population, i.e., $\E\{Y_i(1)-Y_i(0)\mid \V_i=\v, S_i=1\}$, is equivalent to the subgroup effect in the target population, i.e., $\E\{Y_i(1)-Y_i(0)\mid \V_i=\v, S_i=0\}$. However, when transportability is violated and $\calV\subset\calX$, the subgroup effect in the target population is a more complicated weighted average of the subgroup effect in the source population.

\subsection{Results under Bonferroni Correction}
 In this section, we provide the analysis result with a simple Bonferroni correction. Recall that under transportability, all effects (in both county-by-county analysis and in the subgroup analysis) are insignificant, thereby they remain insignificant after the Bonferroni correction. We next discuss result under the sensitivity analysis with the proposed calibration procedure.

 For the county-by-county analysis, the ad effect is insensitive in all counties, with either TS or RS  partitioning; recall that without Bonferroni correction, 6-155 counties are sensitive after calibration.
 For the subgroup analysis, effects in some subgroups are sensitive; a summary of the calibrated result is shown in the Table \ref{tab:subgroup.multiple}.
 \begin{table}[!h]
   \caption{Calibrated subgroup analysis with Bonferroni correction.}
       \label{tab:subgroup.multiple}
       \centering
       \begin{tabular}{l | l}
          \hline
      \multicolumn{2}{c}{Calibration based on TS Partitioning}  \\
      \hline
          Sensitive to a positive effect & Non-Female $\times$  Black/Latinx $\times$ New Voters\\
           Sensitive to a negative effect & Female $\times$ Not Black/Latinx $\times$ Returning Voters\\
           & Non-Female $\times$ Not Black/Latinx $\times$ New Voters\\
           Sensitive to either direction & Female $\times$ Not Black/Latinx  $\times$ New Voters\\
           Insensitive
            & Four Groups \\
            \hline
            \multicolumn{2}{c}{Calibration based on RS Partitioning}  \\
            \hline
          Sensitive to a positive effect & Female  $\times$ Not Black/Latinx  $\times$ New Voters\\
          & Non-Female  $\times$ Black/Latinx  $\times$ New Voters\\
          & Non-Female  $\times$ Not Black/Latinx  $\times$ New Voters\\
           Sensitive to a negative effect & Female  $\times$ Black/Latinx  $\times$ Returning Voters\\
           & Female  $\times$ Not Black/Latinx  $\times$ Returning Voters\\
           Sensitive to either direction & Non-Female  $\times$ Black/Latinx  $\times$ Returning Voters\\
           Insensitive
            & Two groups\\
            \hline
       \end{tabular}

   \end{table}

\subsection{Details on Data Pre-Processing for Subgroup Analysis}\label{subsec:preprocess.subgroup}
This section details the construction of variables regarding urbanicity and education attainment in the subgroup analysis presented in the previous subsection.
\subsubsection{Percentage of Bachelor's Degree Or Higher in ZIP Codes}
To construct a variable as a proxy for a voter's education attainment,
we leverage the ZIP code information from the PA voter database.
In specific, for every ZIP code in the PA voter database, we calculate the percentage of receiving a Bachelor's degree or higher from the 2022 American Community Survey (ACS), which is a comprehensive census that represents the U.S. population.
To preserve privacy, we excluded ZIP codes with fewer than 20 voters from the PA voter database or from the ACS data. This step removed 146 ZIP codes and 2149 voters.
As a result, for each voter, we have the percentage of Bachelor's degree or higher in their ZIP-code area. And for analysis, we  divided the percentages into five groups by every 20 percent.

\subsubsection{Urbanicty in Census Tracts}
For urbanicity, we mapped a voter's address with the 2020 U.S. census which classifies a census tract as urban or rural (i.e., not urban) based on characteristics including population, housing, and land area among others. We refer readers to the U.S. Census Bureau's urban-rural classification for the criterion of classifying a census tract as urban or rural. Among all 4,880,729 voters, the addresses of 176,866 (0.04\%) cannot be matched with a census tract. Their urbanicity was imputed by the proportion of urban voters with the same ZIP code (if the proportion is less than 50\%, we imputed the urbanicity to be rural and vice versa), except for 1,147 whose urbanicity cannot be imputed because their ZIP codes are either missing or do not match with ZIP codes of other voters. These voters take 0.02\% of the original voters and have been excluded from the analysis in Section 6.3.

\newpage
\section{Simulations}\label{supp.sec:sim}
In this section, we validate asymptotic properties of our proposed estimators on simulated datasets generated according to the 2020 RCT data.

In order to generate data that mimics the 2020 RCT data, we let the source covariate  $\X_i$ be gender, race, and age groups and set its distribution $\X_i\mid S_i=1$ to be the empirical distribution of these covariates in the 2020 RCT data.
Given $\x\in\calX$, the treatment is randomized within 18 strata mimicking the design in \cite{aggarwal20232}. The $\mu_1(\x)$ and $\mu_0(\x)$ are  generated in two scenarios. In Scenario (A), they differ by $0.005$ or $-0.005$ whereas the overall average effect is close to zero, mimicking the real data where the overall ad effect is negligible despite small, heterogeneous effects in subgroups.
In Scenario (B), the difference between $\mu_1(\x)$ and $\mu_0(\x)$ is larger in magnitude and more heterogeneous.
The covariate distribution on the target population, $p_{\X\mid S=0}$ is generated such that $p_{\X\mid S=0}(\x)/p_{\X\mid S=1}(\x)$ is between $0.9$ and $1.1$. Table \ref{tab:simulation.dgp} presents the values of this generation.
The target covariate $\V_i$ is set to be the gender variable alone.
The sensitivity parameter $\gamma_0$ is set to zero and $\gamma_1$ varies. The source sample size $\nt$ and target sample size $\nt$ are set equal.
\begin{table}[!h]
    \centering
\caption{Data generation in simulated datasets.}
\label{tab:simulation.dgp}
\footnotesize
    \begin{tabular}{ccc | c c c |cc| cc}
\hline
\multirow{2}{*}{Gender} &
\multirow{2}{*}{Race}&
\multirow{2}{*}{Age Group}&
\multirow{2}{*}{$p_{\X\mid S=1}(\x)$}&
\multirow{2}{*}{$p_{\X\mid S=0}(\x)$}&
\multirow{2}{*}{$\pi(\x)$}
 &\multicolumn{2}{|c|}{Scenario (A)}
  &\multicolumn{2}{|c}{Scenario (B)}\\
& & && & &$\mu_0(\x)$& $\mu_1(\x)$ &$\mu_0(\x)$& $\mu_1(\x)$\\
    \hline
    Female    &  Black    &   18-24    & 0.0061    & 0.0055    & 0.6    & 0.4    & 0.35 & 0.2 & 0.6 \\
  Female    &  Black    &   25-34    & 0.0077    & 0.0071    & 0.7    & 0.4    & 0.35  & 0.2 & 0.6 \\
  Female    &  Black    &   Other    & 0.0157    & 0.0150    & 0.8    & 0.5    & 0.45  & 0.7 & 0.2 \\
  Female    & Latinx    &   18-24    & 0.0073    & 0.0066    & 0.6    & 0.5    & 0.45 & 0.7 & 0.2  \\
  Female    & Latinx    &   25-34    & 0.0089    & 0.0083    & 0.8    & 0.4    & 0.35  & 0.3 & 0.3 \\
  Female    & Latinx    &   Other    & 0.0147    & 0.0139    & 0.9    & 0.5    & 0.45 & 0.7 & 0.2 \\
  Female    &  Other    &   18-24    & 0.1001    & 0.1042    & 0.6    & 0.6    & 0.55 & 0.3 & 0.5 \\
  Female    &  Other    &   25-34    & 0.1271    & 0.1353    & 0.8    & 0.5    & 0.45 & 0.6 & 0.2 \\
  Female    &  Other    &   Other    & 0.2016    & 0.2218    & 0.9    & 0.6    & 0.55  &0.3 & 0.5 \\
 Other    &  Black    &   18-24    & 0.0197    & 0.0193    & 0.6    & 0.3    & 0.35  &0.2 & 0.6\\
  Other    &  Black    &   25-34    & 0.0280    & 0.0285    & 0.8    & 0.2    & 0.25 &0.2 & 0.6 \\
 Other    &  Black    &   Other    & 0.0397    & 0.0409    & 0.8    & 0.3    & 0.35 &0.2 & 0.6 \\
  Other    & Latinx    &   18-24    & 0.0174    & 0.0169    & 0.6    & 0.3    & 0.35 &0.25 & 0.55 \\
  Other    & Latinx    &   25-34    & 0.0201    & 0.0200    & 0.8    & 0.3    & 0.35  & 0.25  & 0.55\\
  Other    & Latinx    &   Other    & 0.0211    & 0.0212    & 0.9    & 0.4    & 0.45  & 0.25 & 0.55 \\
  Other    &  Other    &   18-24    & 0.1061    & 0.1118    & 0.7    & 0.5    & 0.55 &7 & 0.2 \\
  Other    &  Other    &   25-34    & 0.1277    & 0.1375    & 0.8    & 0.4    & 0.45&0.25 & 0.55  \\
  Other    &  Other    &   Other    & 0.1310    & 0.1425    & 0.9    & 0.5    & 0.55  & 0.7 & 0.2 \\
\hline
    \end{tabular}
\end{table}

After generating datasets, the propensity score $\pi(\x)$ is estimated with the average proportion of treated units within each. The outcome regression functions $\mu_a(\x)$ and $\rho_a(\v)$ are estimated by  reweighing samples with $S_i=1$ and $A_i=a$ as in (5). 
The density ratio $w(\v)$ is estimated with (\ref{eq:wv.est.discrete}). For the OR estimator, the inference is based on 1000 bootstrap iterations. For the EIF-based estimator, the inference is based on the cross-fitting procedure with $K=2$ splits. The confidence level is set to $1-\alpha = 0.95$. Simulation results are based on 1000 replicates.

From results in Table \ref{tab:sim.gender}, both estimators are consistent and their empirical standard deviation (SD) decays with $\sqrt{n}$. The estimated SEs are close to the empirical SDs and the coverage rate nears the nominal level $0.95$. These results validate bootstrap CI consistency in  Theorem 4.1
as well as the asymptotic Normality of the EIF-based cross-fitting estimator in Theorem 4.2. 

\begin{table}[h]
    \caption{Simulation results. Bias, RMSE, empirical standard deviation (Emp.SD) and estimated standard error (Est.SE) have been multiplied with $1000$.}
    \label{tab:sim.gender}
    \centering
    \footnotesize
    \begin{tabular}{ll | lllll| lllll}
    \hline
    \multicolumn{2}{c|}{$\gamma_1=1$}&\multicolumn{5}{c|}{Scenario (A)}&\multicolumn{5}{c}{Scenario (B)}\\
    \hline
    Estimator& $\ns(=\nt)$ & Bias & RMSE & Emp.SD & Est.SE & Rate  & Bias & RMSE & Emp.SD & Est.SE & Rate \\
    \hline
 OR   &$10^5$   &  -0.135    &       4.317    &         4.317    &         4.275    &           0.943  &0.076    &       4.169    &         4.171    &         4.123    &           0.952  \\
       OR  & 2$\times10^5 $  &   -0.126    &       3.047    &         3.046    &         3.018    &           0.953 &  0.030   &       2.951   &         2.952    &         2.913    &           0.939  \\
    \hline
   EIF   & $10^5$ &  0.004    &       4.307    &         4.309    &         4.283    &           0.953 &  -0.0.082   &       3.943    &        3.944   &         4.135   &           0.955  \\
        EIF & 2$\times10^5 $ & -0.008    &      3.029    &         3.030   &         3.024    &           0.953  & -0.549    &       2.996    &         2.947    &         2.920    &           0.945  \\
    \hline
    \multicolumn{2}{c|}{$\gamma_1=1.05$}&\multicolumn{5}{c|}{Scenario (A)}&\multicolumn{5}{c}{Scenario (B)}\\
    \hline
 OR   &$10^5$   &    -0.136    &       4.318    &         4.318    &         4.276    &           0.945 &  0.076    &       4.178    &         4.180    &         4.130    &           0.953  \\
       OR  & 2$\times10^5 $  &  -0.126    &       3.047    &         3.046    &         3.019    &           0.954 &0.029    &       2.957    &         2.958    &         2.920    &           0.940  \\
    \hline
   EIF   & $10^5$ &  0.225    &       4.356    &         4.357    &         4.283    &           0.947 &  -0.265   &       4.152   &         4.145    &         4.142    &           0.948 \\
        EIF & 2$\times10^5 $ &   0.095  &       3.028    &         3.028    &         3.024    &           0.943&   -0.483    &       2.984    &         2.946    &         2.925    &           0.941  \\
  \hline
    \end{tabular}
\end{table}